\documentclass[a4paper,11pt,preprintnumbers]{article}
\pdfoutput=1
\usepackage{jheppub}

\usepackage{mathrsfs}
\usepackage{amsfonts}
\usepackage{amsmath,amssymb,bm}
\usepackage{array}
\usepackage{verbatim}
\usepackage{epsfig}
\usepackage{graphicx}
\usepackage[colorlinks=true,citecolor=blue,linkcolor=blue,urlcolor=blue]{hyperref}
\usepackage[normalem]{ulem}
\usepackage{xcolor}
\usepackage{ulem}
\usepackage{multirow}

\usepackage{graphicx}% Include figure files
\graphicspath{{./figures/}}
\usepackage{dcolumn}% Align table columns on decimal point
\usepackage{bm}% bold math
\usepackage{slashed}
\usepackage{siunitx}
\usepackage{ulem,xpatch}
\usepackage{hyperref}% add hypertext capabilities
\usepackage[mathlines]{lineno}% Enable numbering of text and display math
\usepackage{comment}
\usepackage{soul}
\usepackage{xcolor}
\usepackage{xfrac}
\newcommand{\inspire}[1]{[\href{https://inspirehep.net/literature?q=#1}{\sc inSPIRE}]}

\newcommand{\diff}[1]{\mathrm{d}#1}
\newcommand{\xb}{x_{\mbox{\tiny\!$B$}}}
\newcommand{\hxb}{\hat{x}_{\mbox{\tiny\!$B$}}}

\newcommand{\Rqed}{R_{\mbox{\tiny \rm QED}}}
\newcommand{\xbtrue}{x_{\mbox{\tiny\!$B$}, \rm true}}
\renewcommand{\t}[1]{\widetilde{#1}}
\renewcommand{\l}{\ell}
\newcommand{\lp}{{\ell'{}}}
\newcommand{\tlp}{\t{\ell}'{}}
\newcommand{\tl}{\t{\ell}}
\newcommand{\tg}{\t{g}}
\newcommand{\hq}{\hat{q}}
\newcommand{\llbrack}{[\![}
\newcommand{\rrbrack}{]\!]}
\newcommand{\Lqcd}{\Lambda_{\mbox{\tiny QCD}}}

\newcommand{\alfa}{\alpha}
\newcommand{\pTbar}{\overline{\rm p}_T}
\newcommand{\PTbar}{\overline{\rm P}_T}
\def\bea#1\eea{\begin{align}#1\end{align}}

\newcommand{\bef}{\begin{figure}[h!tb]\centering}
\newcommand{\beft}{\begin{figure}[t]\centering}
\newcommand{\eef}{\end{figure}}
\def\qeq{\mathrel{%
    \mathchoice{\QEQ}{\QEQ}{\scriptsize\QEQ}{\tiny\QEQ}%
}}
\def\QEQ{{%
    \setbox0\hbox{=}%
    \rlap{\hbox to \wd0{\hss?\hss}}\box0
}}

\newcommand{\eref}[1]{eq.~(\ref{e.#1})}

\newcommand{\fref}[1]{figure~\ref{f.#1}}

\newcommand{\ssref}[1]{section~\ref{ss.#1}}

\title{A new approach to semi-inclusive deep-inelastic scattering with QED and QCD factorization}

\author[a]{Tianbo~Liu,}
\author[b]{W.~Melnitchouk,}
\author[b,c]{Jian-Wei~Qiu}
\author[b]{and~N.~Sato}

\affiliation[a]{Key Laboratory of Particle Physics and Particle Irradiation (MOE),\\
Institute of Frontier and Interdisciplinary Science, Shandong University,\\
Qingdao, Shandong 266237, China}
\affiliation[b]{Theory Center, Jefferson Lab,\\
Newport News, Virginia 23606, USA}
\affiliation[c]{Department of Physics, William \& Mary,\\
Williamsburg, Virginia 23187, USA}
\emailAdd{liutb@sdu.edu.cn}
\emailAdd{wmelnitc@jlab.org}
\emailAdd{jqiu@jlab.org}
\emailAdd{nsato@jlab.org}

\date{\today}
\preprint{JLAB-THY-21-3489}

\abstract{
We present the details of a new factorized approach to semi-inclusive deep-inelastic scattering which treats QED and QCD radiation on equal footing, and provides a systematically improvable approximation to the extraction of transverse momentum dependent parton distributions. We demonstrate how the QED contributions can be well approximated by collinear factorization, and illustrate the application of the factorized approach to QED radiation in inclusive scattering. For semi-inclusive processes, we show how radiation effects prevent a well-defined ``photon-nucleon'' frame, forcing one to use a two-step process to account for the radiation. We illustrate the utility of the new method by explicit application to the spin-dependent Sivers and Collins asymmetries.}


\arxivnumber{2108.13371}

\begin{document}
\maketitle
\flushbottom

%%%%%%%%%%%%%%%%%%%%%%%%%%%%%%%%%%%%%%%%%%%%%%%%%%%%%%%%%%
\section{Introduction}

From the pioneering elastic scattering experiments in the 1950s that revealed the finite size of the proton~\cite{Hofstadter:1955ae}, to the classic deep-inelastic measurements that observed the first glimpses of the proton's pointlike substructure~\cite{Bloom:1969kc, Breidenbach:1969kd} in the 1960s and 1970s, electron scattering has been an indispensable tool for hadron structure studies.
These experiments established critical milestones that ultimately paved the way to the fundamental theory of the strong interactions, Quantum Chromodynamics (QCD).
The theoretical formalism of collinear factorization, developed in the 1980s in the context of perturbative QCD~\cite{Collins:1981uw, Collins:1984kg, Collins:1989gx}, provided a rigorous and systematic path between high-energy scattering observables, such as deep-inelastic scattering (DIS) cross sections, and the quark and gluon (or parton) longitudinal momentum distributions that characterize the proton's internal structure.
Over the course of the last few decades, a wealth of experimental data has been accumulated on proton and nuclear targets which has revealed intriguing features of the flavor and spin dependence of the parton distribution functions (PDFs) in nucleons and nuclei~\cite{Jimenez-Delgado:2013sma, Aidala:2012mv, Ethier:2020way}.

More recently, it has been recognized that one can access also the transverse momentum distributions of quarks and gluons, which, when combined with the longitudinal information, holds the promise of systematically mapping out the full three-dimensional partonic structure of the nucleon in momentum space~\cite{Bacchetta:2006tn, Angeles-Martinez:2015sea, Collins:2016hqq, Diehl:2015uka, Gutierrez-Reyes:2019vbx, Liu:2018trl}.
Charting the distributions in the transverse plane is naturally more involved, however, requiring the development of transverse momentum dependent (TMD) factorization in appropriate regions of kinematics~\cite{Collins:2011zzd}.
Tremendous interest has been generated in recent years~\cite{Ji:2004wu, Aybat:2011zv, Aybat:2011ge, Bacchetta:2004jz, Collins:2017oxh, Gamberg:2017jha, Anselmino:2013lza} by the prospects of extracting TMD information from experiments at existing facilities such as COMPASS at CERN and Jefferson Lab, which use lepton probes (muons and electrons, respectively), and RHIC at BNL, which utilizes proton beams.

One of the key processes that has been embraced as a potentially rich source of information about TMD PDFs is semi-inclusive DIS (SIDIS), where in addition to the scattered lepton, a high-momentum hadron (typically a pion) is detected in coincidence in the final state~\cite{Avakian:2016rst}.
With the incident $\ell$ and scattered $\ell'$ lepton four-momenta defining the leptonic plane, and the target nucleon and produced hadron four-momenta $P$ and $P_h$ defining the hadronic plane, specific angular modulations between these planes in SIDIS can allow the extraction of various types of TMDs that are not accessible from traditional inclusive observables.
(Note that for simplicity we will refer in this paper to scattering of leptons from nucleons, however, the results apply equally to scattering from any other hadron or nucleus.)
Here, as for inclusive scattering, the large four-momentum transfer $q \equiv \ell-\ell'$ between the leptons provides the hard scale, $Q^2 \equiv - q^2 \gg \Lqcd^2$, for the factorization of the SIDIS process.

A clear advantage of electrons (and other pointlike leptons, such as positrons and muons) is that they are much cleaner probes of nucleon structure than are hadron beams, whose internal partonic structure is typically intertwined with that of the probe and hence more difficult to disentangle.
At the same time, it has long been understood that electron scattering at large momentum transfer can be a source of considerable photon radiation, which can significantly distort the inferred nucleon structure if it is not properly accounted for.
In particular, the radiation can not only affect the momentum transfer $q$ from the lepton to the nucleon, it can also alter the angular modulation between the leptonic and hadronic planes, making it problematic to define the transverse momentum of the produced hadron, $P_{hT}$, in the true photon-nucleon frame.
This in turn can induce angular modulations which can mimic those arising from the true nucleon structure effects encoded by the TMDs.

In the literature, modifications to inclusive DIS~\cite{Mo:1968cg, Bardin:1989vz, Badelek:1994uq} and $e^+ e^-$ annihilation~\cite{Blumlein:2007kx, Bertone:2019hks, Ablinger:2020qvo, Frixione:2019lga} cross sections induced by electromagnetic radiation have been treated in the form of corrections to the tree-level cross sections, in some cases improved by resummation of logarithmic-enhanced radiative effects~\cite{Kripfganz:1990vm, Spiesberger:1994dm, Blumlein:2002fy}, what are then subtracted to reveal the true Born cross sections without radiation.
Unfortunately, without accounting for all radiated photons experimentally, some of the radiative corrections (RCs) rely on knowing the invariant mass of the hadronic final state and subsequent Monte Carlo simulation~\cite{Charchula:1994kf, Pierre:2019nry, Kwiatkowski:1990es, Arbuzov:1995id}, which inherently introduces a degree of model dependence in the procedure.
For processes beyond inclusive DIS the prescription of matching to the Born cross section by removing the radiation effects becomes increasingly difficult~\cite{Ent:2001hm, Afanasev:2002ee}.
For exclusive or semi-inclusive cross sections, which are parametrized by 18 structure functions, the procedure becomes effectively impractical without introducing severe approximations.

Despite the complications, several pioneering efforts have been made to address RCs in SIDIS reactions, most notably within the covariant approach of Bardin and Shumeiko~\cite{Bardin:1976qa}, in which infrared divergences from real and virtual photon emission are shown to cancel.
Compared with the Mo and Tsai approach~\cite{Mo:1968cg}, an advantage of the covariant method is that the expressions for the RCs do not depend on parameters introduced to separate the photon emission on the hard and soft parts of the amplitudes.
Using this approach, the corrections to the triply-differential (transverse momentum integrated) SIDIS cross section
    $\diff^3\sigma/\diff \xb\, \diff y\, \diff z_h$ 
were considered by Soroko and Shumeiko for unpolarized~\cite{Soroko:1989zt} as well as polarized~\cite{Soroko:1991zr} scattering, where $\xb = -q^2/2P\cdot q$ is the Bjorken scaling variable, $y = P \cdot q/P \cdot \ell$ is the energy loss of the incident electron, and $z_h = P \cdot P_h / P \cdot q$ is the longitudinal momentum fraction carried by the final state hadron.

This was extended by Akushevich {et al.}~\cite{Akushevich:1999hz, Akushevich:2007jc} to the case of the angular dependent cross section for unpolarized SIDIS, 
    $\diff^5\sigma/\diff \xb\, \diff y\, \diff z_h\, \diff P_{hT}^2\, \diff \phi_h$,
where $\phi_h$ is the azimuthal angle between the lepton and hadron production planes, including contributions from the exclusive radiative tail.
Ilyichev and Osipenko~\cite{Ilyichev:2013ega} considered a higher-order background to this five-fold unpolarized SIDIS cross section arising from exclusive lepton-pair production, which in the region $\phi_h=180^\circ$ can be comparable to the SIDIS signal.
Contributions from two-photon emission, which enter at the same order, were also considered in this work.
Most recently, Akushevich and Ilyichev~\cite{Akushevich:2019mbz} derived within the same approach analytical expressions for RCs to sixfold differential SIDIS cross sections for scattering longitudinally polarized leptons from nucleons with arbitrary polarization,
    $\diff^6\sigma / \diff \xb\, \diff y\, \diff z_h\, \diff P_{hT}^2\, \diff\phi_h\, \diff\psi$,
where $\psi$ is the azimuthal angle between the lepton scattering plane and the spin direction of the incident nucleon.
The calculations included the ``model-independent contributions,'' proportional to $\log(Q^2/m_e^2)$, where $m_e$ is the lepton mass, associated with the emission of real photons from leptons, along with leptonic vertex correction, and vacuum polarization. 
Not considered in the analysis were corrections from real and virtual photon emissions by hadrons, QED hadronic vertex corrections, or two-photon exchange contributions.
% By measuring the momentum transfer, $q \equiv \ell-\ell'$, from an incident lepton $\ell$ scattered to a lepton with momentum $\ell'$, and keeping $Q \equiv \sqrt{-q^2} \gg 1/R$, where $R$ is the hadron radius, the DIS experiments provided a short-distance electromagnetic probe of the point-like quarks inside hadrons, ultimately giving birth to QCD as the theory of strong interactions.
% Without observing specific final states other than the scattered  lepton, this modern version of Rutherford scattering provided the first glimpse of the hadrons' internal landscape of quarks and gluons (or collectively, partons), parametrized through the parton distribution functions (PDFs) as probability densities for finding a parton inside the hadron with momentum fraction $x$~\cite{Feynman:1973xc}.

In a recent paper~\cite{Liu:2020rvc}, we proposed a new factorized framework for SIDIS reactions, which simultaneously treats QED and QCD radiative effects on the same footing and in a systematically improvable manner. In this approach, the lepton-nucleon SIDIS cross section is effectively an inclusive cross section to observe one lepton and one hadron in the final state.  It is a well-defined two-scale observable when $Q^2$ is much larger than the momentum imbalance between the observed final-state lepton and hadron, and the imbalance is sensitive to the collision-induced QED and QCD radiation and transverse momentum of the active partons and leptons.  Our new factorized framework for SIDIS is effectively a {\it hybrid} factorization approach with {\it collinear} factorization for the two leptons and {\it TMD} factorization for two hadrons when the SIDIS cross section is in the two-scale regime.

In the present work, we provide the details that justify this hybrid factorization approach, and demonstrate why the collinear factorization is a good approximation for organizing all-order contributions of collision-induced radiation from the leptons in both lepton-nucleon DIS and SIDIS.
We illustrate this with explicit examples of applications of the factorization approach to QED radiation in inclusive DIS, and compare with existing RC calculations.
For SIDIS, we quantitatively demonstrate that the amount of transverse momentum broadening generated by the collision-induced QED and QCD radiation from a ``point-like'' lepton is much smaller than the typical transverse momentum of a colliding parton (which could be further enhanced by QCD radiation from its intrinsic value) for all foreseeable energies of lepton-nucleon scattering experiments. 
The momentum imbalance between the observed lepton and hadron in the final state is therefore dominated by the transverse momentum dependence of the nucleon TMDs, which makes SIDIS a particularly useful process for TMD extraction. This finding justifies our hybrid factorization approach to handling high-order QED and QCD contributions to SIDIS consistently, with collinear and TMD factorization for the leptons and hadrons describing their leading nonperturbative contributions via universal collinear and TMD distributions, respectively, when the SIDIS is in the two-scale regime.
In addition, our hybrid factorization approach to SIDIS in this two-scale regime avoids having to deal with a full TMD factorization for all four observed particles (the two leptons and two hadrons) in both QED and QCD~\cite{Collins:2007nk}.

The challenge for the traditional method of treating QED radiation as an RC factor applied to the QED Born cross section is the difficulty in controlling the ``true'' momentum transfer to the incident nucleon.
Even when the momentum transfer~$Q^2$ is sufficiently large for a perturbative hard scale, QED radiation can render the ``true'' momentum transfer $\widehat{Q}^2$ to the colliding nucleon, which has a minimum 
\begin{equation}
    \widehat{Q}^2_{\rm min} =\, Q^2\, \frac{(1-y)}{(1-\xb\,y)},
\label{e.Qhat2min}
\end{equation}
such that $\widehat{Q}^2 \ll Q^2$ when $\xb$ is small or $y$ is large and there is a large phase space for radiation.
When the ``true'' momentum transfer to the colliding nucleon $\widehat{Q}^2$ is not in the DIS regime because of QED radiation, high-twists and quasielastic or elastic tails could contribute to the lepton-nucleon cross section even when $Q^2$ is large.
This could naturally lead to model dependence of the RCs in order to remove or correct these non-DIS events, even for the inclusive DIS measurements.

Our proposed factorization approach for both QCD and QED contributions to DIS and SIDIS naturally maintains the ``true'' momentum transfer sufficiently large, $\widehat{Q}_{\rm min}^2 \gg \Lqcd^2$, to ensure factorization and avoid regimes where higher-twist and non-DIS events could be relevant.
To achieve this, we systematically separate the infrared-sensitive QED parts as $m_e\to 0$ from the infrared-safe QED terms in the same way as for QCD factorization.
We include all-order QED contributions to DIS and SIDIS cross sections by resumming infrared-sensitive terms into universal lepton distribution functions (LDFs) for the incident leptons, and lepton fragmentation functions (LFFs) of the observed leptons.
The infrared-safe contributions are calculated perturbatively in powers of $\alpha$, up to power corrections in powers of $m_e/\widehat{Q} \ll 1$.

The key impact of QED radiation on the SIDIS cross section is from the change of the momentum transfer to the colliding nucleon, in both its direction and  invariant mass, caused mainly by the logarithmic-enhanced collinear QED radiation.
In this paper, we extend the analysis~\cite{Liu:2020rvc} of the unpolarized SIDIS structure function to the spin-dependent case, for the specific examples of the Sivers and Collins asymmetries.
In particular, we demonstrate a ``no-go theorem'' for RCs in SIDIS, which arises from the dependence of the longitudinal and transverse polarization vectors $S_L$ and $S_T$ on the leptonic momentum fractions, and forces us to consider a two-step process to account for the radiation. 
We note that the same issue will affect the case of inclusive polarized DIS, in addition to SIDIS, and the extraction of the spin-dependent $g_1$ and $g_2$ structure functions.

We begin by reviewing in section~\ref{s.dis-fac} the factorized formalism for inclusive DIS in the presence of QED radiation, presenting the basic formulas for DIS cross sections in terms of universal LDFs and LFFs.
We assess the importance of the QED radiation numerically at various kinematics relevant to Jefferson Lab and EIC energies, and compare the results of our factorized approach with the traditional formulation of RCs in the literature.
In section~\ref{s.sidis-fac} we generalize the factorized formalism to the case of SIDIS processes, and discuss the specific collinear and TMD factorization for the leptonic tensor and structure functions relevant in different kinematics.
The numerical impact of QED effects on SIDIS observables is described in section~\ref{s.pheno}.
To demonstrate the practicality of our approach, we illustrate the formalism applied to unpolarized SIDIS structure functions, as well as to the azimuthal modulations for transversely polarized nucleons associated with the Sivers and Collins asymmetries.
In particular, we quantify the effect of the mismatch between the total four-momentum transferred from the incident lepton and the QED Born approximation on the problem of defining a unique photon-nucleon frame, and the resulting mixing induced between the different angular modulations. 
Finally, in section~\ref{s.conclusion} we summarize our main conclusions and discuss possible future extensions of this work.
Several appendices give additional details of the calculation of the NLO perturbative coefficients for the leptonic tensor (Appendix~\ref{s.app-nlo}), together with a few useful formulas (Appendix~\ref{s.app-int}), and a set of QED dependent kinematic expressions relevant for SIDIS calculations (Appendix~\ref{s.sidis-kin}).

%%%%%%%%%%%%%%%%%%%%%%%%%%%%%%%%%%%%%%%%%%%%%%%%%%%%%%%%%%
\section{Factorized formalism for inclusive DIS with QED}
\label{s.dis-fac}

\begin{figure}
\begin{center}
\includegraphics[width=0.35\textwidth,page=1]{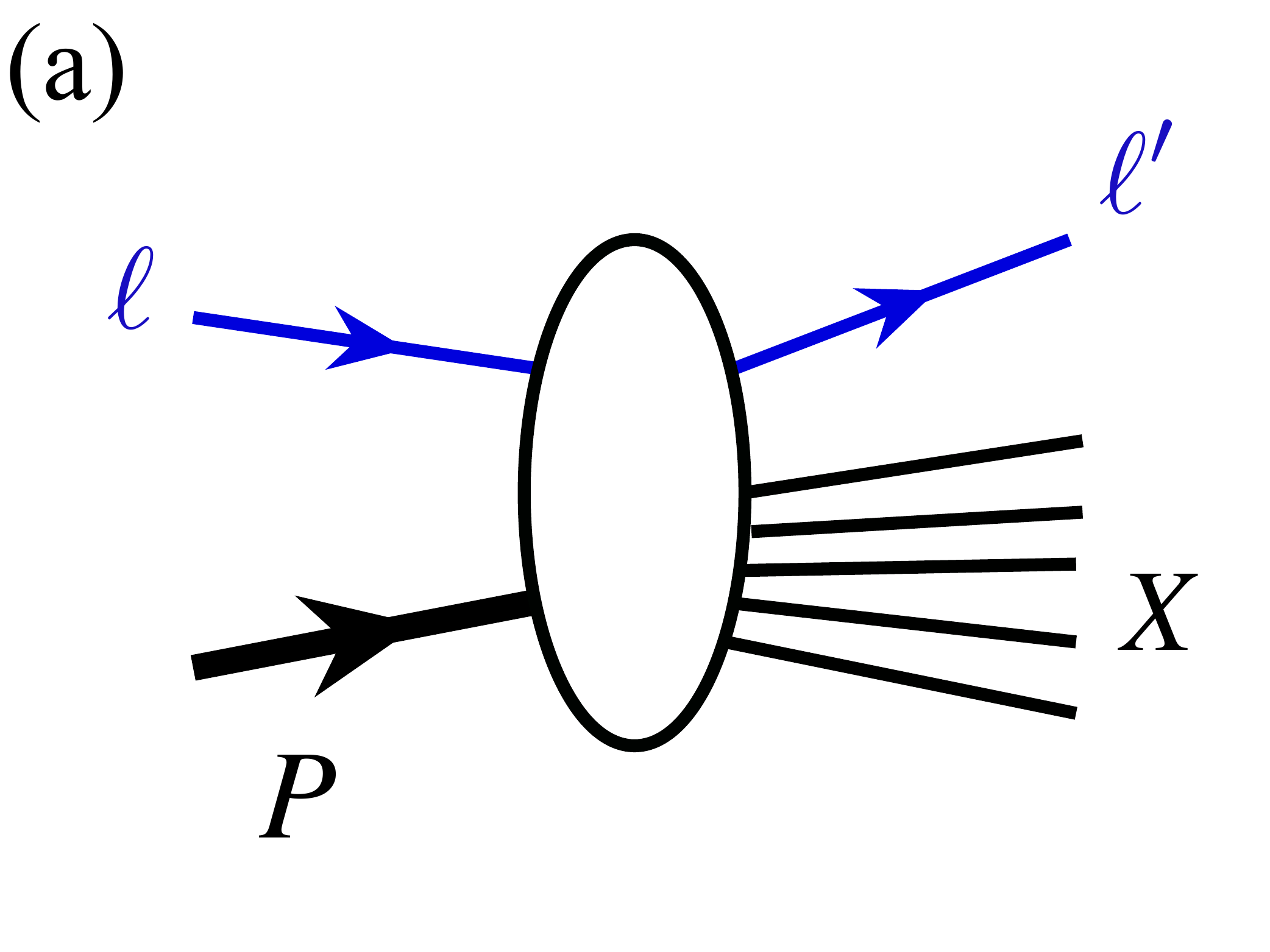} \qquad\qquad
\includegraphics[width=0.35\textwidth,page=2]{dis.pdf}
\vspace*{-0.3cm}
\caption{{\bf (a)}~Inelastic scattering amplitude for a lepton ($\l,\lambda_\l$) from a nucleon ($P,S$) to a scattered lepton ($\lp$) with inclusive final states $X$, and {\bf (b)}~the same amplitude via an exchange of one virtual photon ($q$).}
\label{f.dis}
\end{center}
\end{figure}

We begin our discussion by reviewing the more familiar case of inclusive DIS, where we demonstrate the factorized formulation for the QED radiative effects in terms of universal LDFs and LFFs and infrared-safe higher-order QED corrections.
Most generally, the cross section for inelastic scattering of a lepton $e$ of four-momentum $\ell$ and helicity $\lambda_\l$ from a nucleon $N$ of four-momentum $P$ and spin $S$ to a scattered lepton $e$ of four-momentum $\ell'$ with inclusive final states~$X$, $e(\l,\lambda_\l) + N(P,S) \to e(\lp) + X$, can be formally written in terms of the square of its scattering amplitude $M_{\l(\lambda_\l)P(S)\to \lp X}$, as sketched in figure~\ref{f.dis}(a),
\begin{eqnarray}
\label{e.dis} 
\diff\sigma_{\l(\lambda_\l) P(S)\to \lp X}
= \frac{1}{2s} \left| M_{\l(\lambda_\l) P(S)\to \lp X} \right|^2
\diff{\rm PS},
\end{eqnarray}
where $s=(\l + P)^2 \approx 2\l \cdot P$ is the total collision energy, and $\diff{\rm PS}$ indicates the differential phase space of the given final state, which will be specified below. 
Using the fact that the QED fine structure constant $\alpha=e^2/(4\pi)$ is small in the energy regime of interest, the DIS amplitude is often approximated by the amplitude with one-photon exchange, as shown in figure~\ref{f.dis}(b).
%
%%%%%%%%%%%%%%%%%%%%%%
%\beft
%\eef
%%%%%%%%%%%%%%%%%%%%%%%
%
The inclusive DIS cross section in this case can be written as
\begin{eqnarray}
\label{e.dis-1p} 
E_{\lp} \frac{\diff^3\sigma_{\l(\lambda_\l) P(S) \to \lp X}}{\diff^3\lp}
&\approx& \frac{2\alpha^2}{s\, Q^4} 
L_{\mu\nu}^{(0)}(\l,\lp,\lambda_\l)\,
W^{\mu\nu}(q,P,S),
\end{eqnarray}
where the zeroth-order leptonic tensor given by
\begin{eqnarray}
\label{e.dis-lmn} 
L_{\mu\nu}^{(0)}(\l,\lp,\lambda_\l)
&=&{\rm Tr}\Big[
\gamma_\nu\, 
\frac12 \left(1+\lambda_\l \gamma_5\right) 
\gamma\cdot \l\, \gamma_\mu\, \gamma\cdot \lp\Big]
\nonumber\\
&=&
2\big(
\l_\mu\, \ell'_\nu + \l_\nu\, \ell'_\mu - \l\cdot\lp g_{\mu\nu}
+ i\lambda_\l\, {\epsilon}_{\mu\nu\alpha\beta}\, \l^\alpha\,\lp^\beta
\big)\, ,
\end{eqnarray}
and ${\epsilon}_{\mu\nu\rho\sigma}$ is the totally antisymmetric tensor with ${\epsilon}^{0123}=1$.
The hadronic tensor is defined as
\begin{eqnarray}
\label{e.dis-wmn} 
W^{\mu\nu}(q,P,S)
&=&\frac{1}{4\pi} \sum_{X}
\int
\prod_{i\in X}\frac{\diff^3p_i}{(2\pi)^3 2E_i}\,
(2\pi)^4 \delta^{(4)}\Big(q+P-\sum_{i\in X} p_i\Big)
\nonumber\\
& & \hspace*{0cm}
\times
\langle P,S| J^\mu(0) | X\rangle
\langle X | J^\nu(0) | P,S \rangle,
\end{eqnarray}
where $J^\mu(0)$ is the electromagnetic current coupling to quarks.
In general, the hadronic tensor can be expanded in terms of spin-averaged $F_{1,2}(\xb,Q^2)$ and spin-dependent $g_{1,2}(\xb,Q^2)$ structure functions,
\begin{eqnarray}
\label{e.dis-wmnSFs} 
W^{\mu\nu}(q,P,S)
&=&
-\widetilde{g}^{\mu\nu}(q)\, F_1(\xb,Q^2) 
+\frac{1}{P\cdot q} \widetilde{P}^\mu \widetilde{P}^\nu\, 
F_2(\xb,Q^2)
\nonumber\\
& &+
\frac{i M}{P\cdot q}  {\epsilon}^{\mu\nu\alpha\beta}\, q_\alpha
\left[ S_\beta\, g_1(\xb,Q^2) 
+ \Big(S_\beta - \frac{S\cdot q}{P\cdot q} P_\beta \Big)\, g_2(\xb,Q^2) \right],
\end{eqnarray}
where the current conserving tensor $\widetilde{g}^{\mu\nu}$ and vector $\widetilde{P}^\mu$ are defined as
\begin{eqnarray}
\label{e.currentcon} 
\widetilde{g}^{\mu\nu}(q)
&\equiv & g^{\mu\nu} - \frac{q^\mu q^\nu}{q^2},
\qquad
\widetilde{P}^\mu
\equiv P^\mu - \frac{P \cdot q}{q^2} q^\mu,
\end{eqnarray}
such that 
    $ q_\mu\, \widetilde{g}^{\mu\nu}(q) 
    = q_\mu   \widetilde{P}^{\mu}
    = 0 $.
The target nucleon spin four-vector can be written in terms of the polarization vector ${\bm S}$, $S^\mu = (S^0,\,{\bm S})$, with $P \cdot S = 0$ and normalized such that $S^2 = -1$.

In the one-photon exchange approximation, and in the absence of photon radiation from leptons ({i.e.}, the QED Born approximation), the inclusive DIS cross section in eq.~\eqref{e.dis} can be expressed in terms of the spin-averaged and spin-dependent structure functions by using the leptonic and hadronic tensors in eqs.~\eqref{e.dis-lmn} and~\eqref{e.dis-wmn}, respectively.
For example, the unpolarized lepton-nucleon DIS cross section is given by
\begin{eqnarray}
\label{e.dis-1p-avg} 
E_{\lp}  
 \frac{\diff^3\sigma_{\l P\to \lp X}}{\diff^3\lp}
&\approx& 
\frac{4 \alfa^2}{s \xb y^2 Q^2}
\Big[ \xb y^2 F_1(\xb,Q^2)
    + \Big( 1-y-\frac{1}{4}\gamma^2 y^2 \Big) F_2(\xb,Q^2)
\Big],
\end{eqnarray}
where $\gamma = 2M \xb/Q$, and we neglect hadron masses relative to the center of mass energy $\sqrt{s}$, but keep finite mass terms with respect to $Q \equiv \sqrt{Q^2}$.
The one-photon exchange expression for the cross section (\ref{e.dis-1p-avg}) indicates that the nucleon structure functions $F_1$ and $F_2$ can be extracted directly from inclusive DIS data, and traditionally have often been considered as ``direct'' physical observables. 
With a large four-momentum transfer, $Q^2 \gg \Lqcd^2$, these structure functions can be further factorized in terms of quark and gluon PDFs \cite{Collins:1989gx}; for example, for the $F_2$ structure function,
\begin{eqnarray}
\label{e.dis-f2} 
F_2(\xb,Q^2) 
= \sum_a \int_{\xb}^1 \diff x\, C_{2a}\bigg(\frac{x}{\xb},\frac{Q^2}{\mu^2},\alpha_s\bigg)\,
f_a(x,\mu^2) + {\cal O}\bigg(\frac{\Lqcd^2}{Q^2}\bigg),
\end{eqnarray}
where the sum runs over all parton flavors $a$ ($=q,\bar{q},g$), $C_{2a}$ are coefficient functions calculable in QCD perturbation theory order-by-order in powers of the strong coupling $\alpha_s$, and $f_a(x,\mu^2)$ are universal PDFs of flavor $a$ probed with active parton momentum fraction $x$ and factorization scale~$\mu$.

In principle, any cross section with an identified hadron (in the initial or final state), such as the inclusive DIS cross section, cannot be fully calculated within QCD perturbation theory due to its dependence on the hadronic scale of the identified hadron.
The factorization formalism, as in eq.~\eqref{e.dis-f2}, is an approximation with the correction suppressed by inverse powers of the large momentum transfer $Q$ of the collision.
Similarly, other structure functions in eq.~\eqref{e.dis-wmnSFs} can also be factorized in terms of universal PDFs~\cite{Collins:2011zzd}.
If the factorized coefficients are calculated at leading order (LO) in $\alpha_s$, the two spin-averaged structure functions are related via the Callan-Gross relation~\cite{Callan:1969uq},
    $F_2(\xb,Q^2) = 2\xb F_1(\xb,Q^2) = \sum_a e_a^2\, \xb\, f_a(\xb,Q^2)$.
With the perturbatively calculated coefficient functions at next-to-leading order (NLO) and next-to-next-to leading order (NNLO) in $\alpha_s$, precise data from inclusive DIS have provided important constraints on QCD global analysis of PDFs~\cite{Lin:2017snn}.

%%%%%%%%%%%%%%%%%%%%%%
\beft
\includegraphics[width=0.6\textwidth]{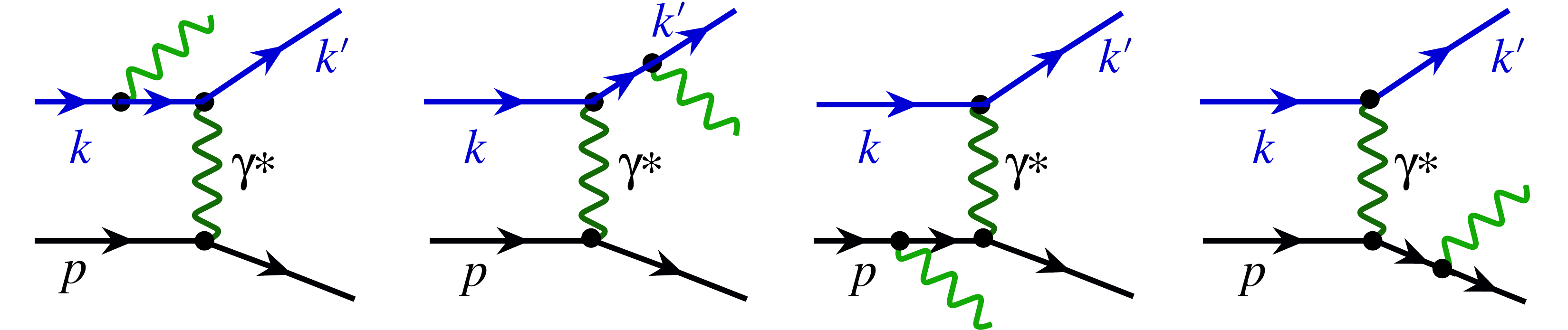} \\
\vspace{0.1in}
\includegraphics[width=0.7\textwidth]{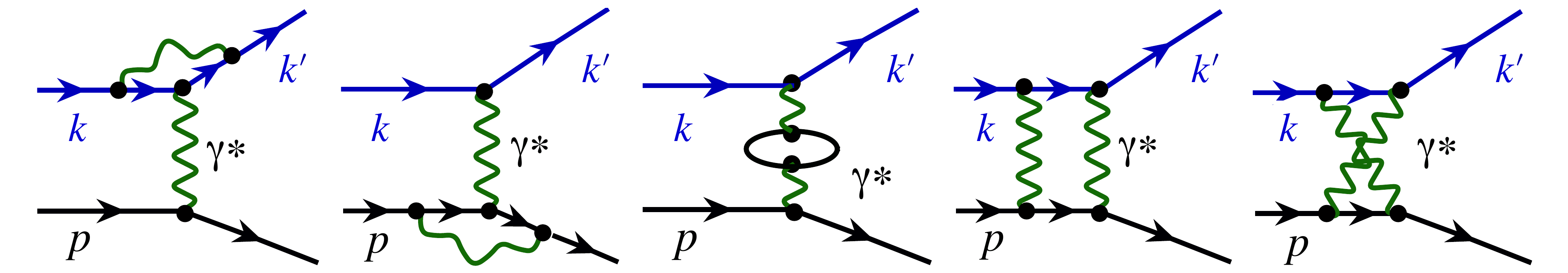}
\caption{Diagrams for the first real (top row) and virtual (bottom row) QED radiative contribution to scattering of a lepton (momentum $k$) from a quark ($p$) to a lepton ($k'$) and recoiling quark.}
\label{f.dis-qed2}
\eef
%%%%%%%%%%%%%%%%%%%%%%%

%%%%%%%%%%%%%%%%%%%%%%%%%%%%%%%%%%%%%%%%
\subsection{Inclusive DIS with QED radiative contributions}
\label{ss.qed-rc}

With the large momentum transfer, $Q^2 \gg \Lqcd^2$, lepton-nucleon scattering naturally triggers radiation of photons (photon showers), such as those from the incident and scattered charged leptons and quarks illustrated in figure~\ref{f.dis-qed2} at NLO in $\alpha$. 
Without being able to account for all radiated photons experimentally, this collision-induced QED radiation not only changes the momentum transfer $q$ between the incident lepton and nucleon, but also requires diagrams beyond the one-photon exchange approximation to maintain the gauge invariance of QED (or in general electroweak) contributions to the inclusive lepton-nucleon DIS cross section.
Beyond the one-photon exchange approximation, the structure functions, along with the PDFs from eq.~\eqref{e.dis-f2}, cannot be uniquely determined from inclusive DIS data without accounting for all QED radiative contributions to the measured cross section.

The traditional method to include all QED radiative contributions to the lepton-nucleon DIS cross sections is to introduce an RC factor to the Born cross section, so that one can still extract the structure functions from inclusive DIS data.
However, this approach necessarily introduces uncertainties in handling the contributions of QED diagrams beyond one-photon exchange, such as the virtual diagrams with two-photon exchange contributions in the second row in figure~\ref{f.dis-qed2} at NLO, and similar diagrams at higher orders.
Consistent treatment of such QED (or electroweak) contributions to the lepton-nucleon DIS cross sections is important for precision extraction of PDFs, and especially for searches of new physics beyond the standard model in processes such as parity-violating DIS.

%%%%%%%%%%%%%%%%%%%%%%
\beft
\includegraphics[height=0.2\textwidth]{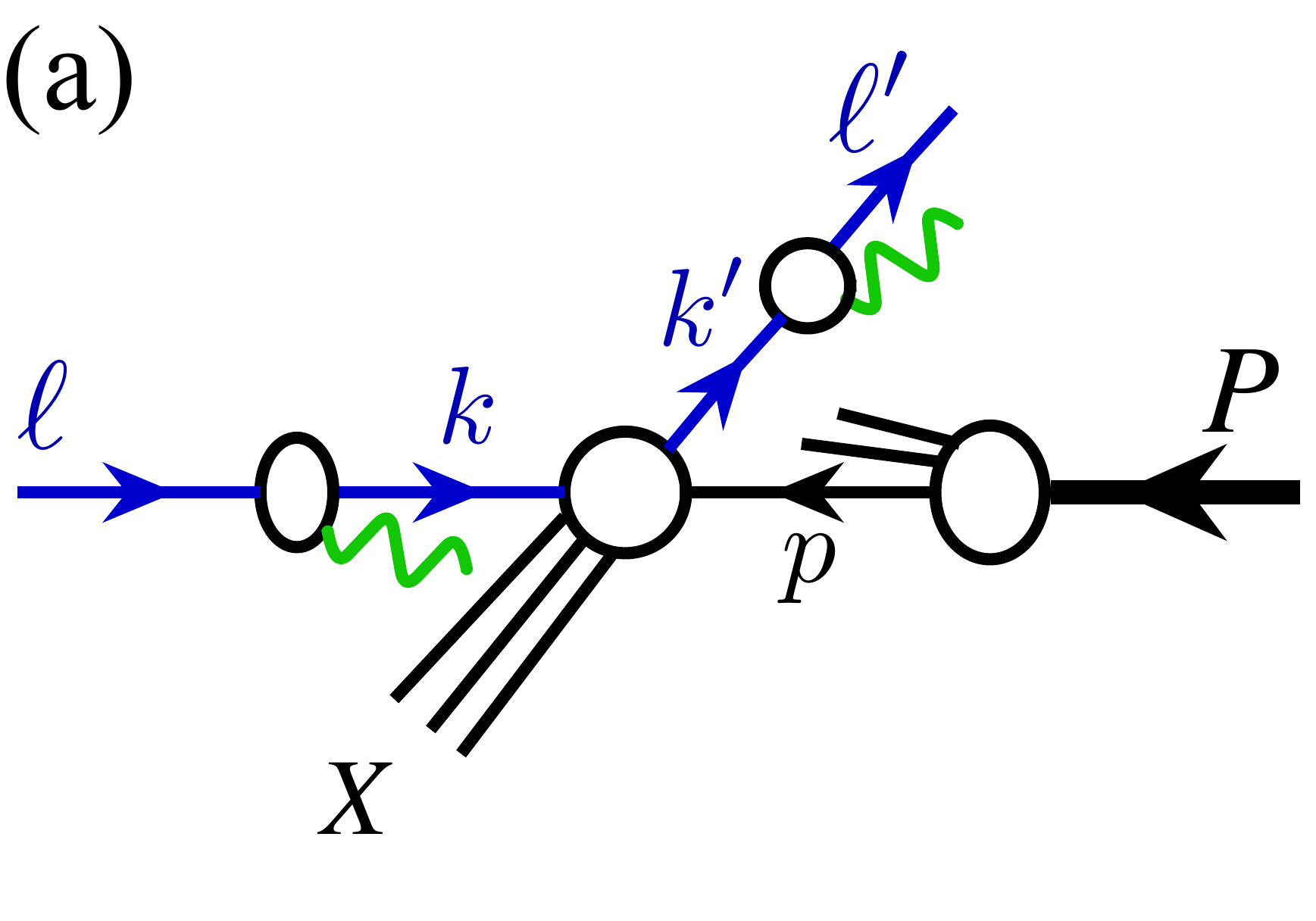} 
\hspace{1cm}
\includegraphics[height=0.2\textwidth]{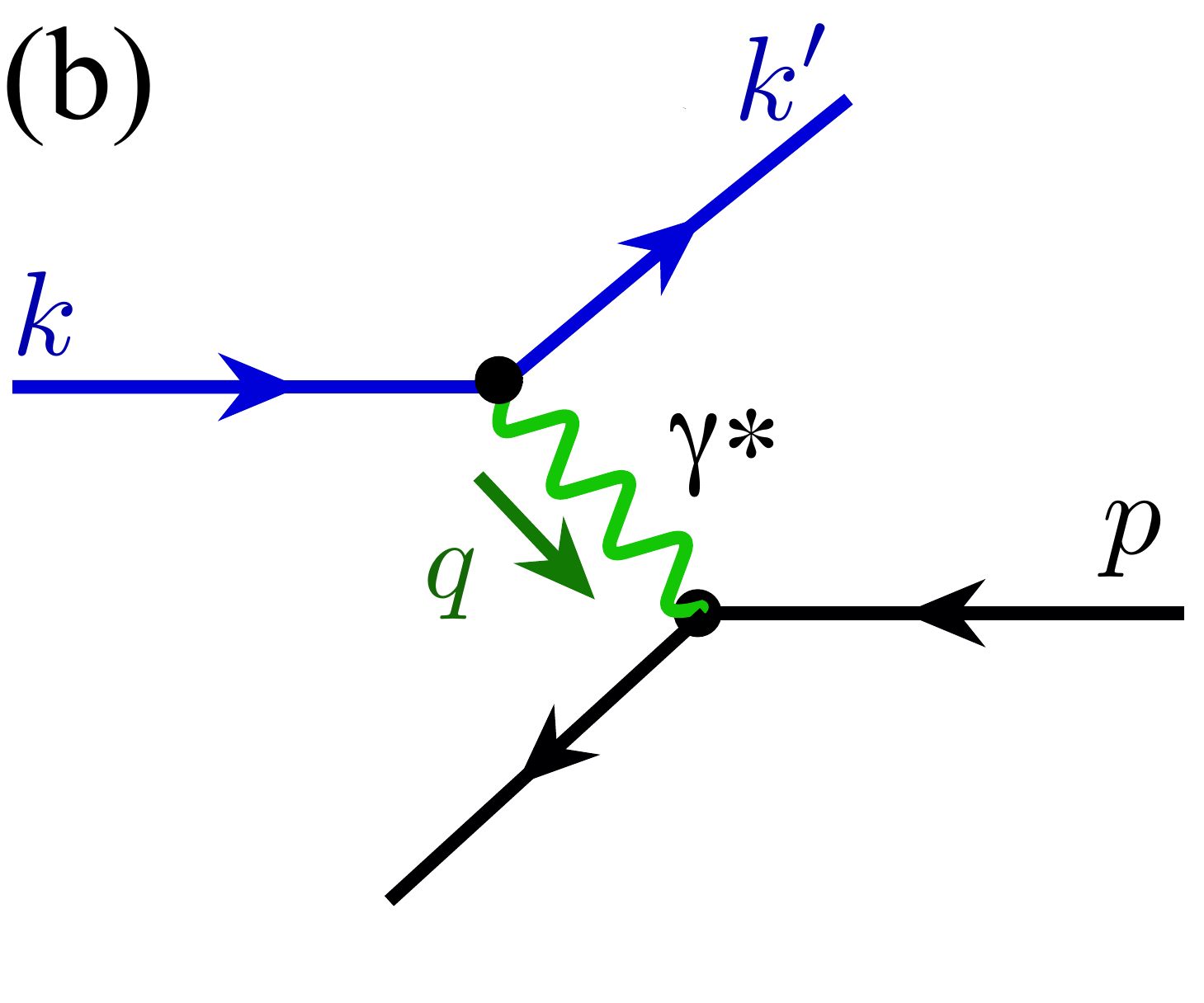}
\caption{Sketch of scattering amplitudes for {\bf (a)}~the factorized DIS process in eq.~\eqref{e.dis-fac}, and {\bf (b)}~lowest order lepton-quark scattering.}
\label{f.dis-fac}
\eef
%%%%%%%%%%%%%%%%%%%%%%%

Instead of treating QED radiation as a correction to the Born process, here we unify the QED and QCD contributions to the lepton-nucleon scattering cross section in a consistent factorization formalism. 
We consider the lepton-nucleon inclusive DIS in eq.~\eqref{e.dis} as an inclusive production of a scattered lepton of four-momentum $\ell'_\mu$ with a transverse component $\ell'_T \gg \Lqcd$ in the lepton-nucleon frame, where the colliding lepton and nucleon are head-on, as sketched in figure~\ref{f.dis-fac}(a).
Applying the factorization formalism previously developed for single-hadron production at large transverse momentum in hadronic collisions~\cite{Nayak:2005rt} to lepton-nucleon scattering, the factorized cross section for the unpolarized inclusive DIS reaction $e(\l) + N(P) \to e(\lp) + X$ can be written as
\begin{eqnarray}
E'\frac{\diff \sigma_{\l P\to \lp X}}{\diff^3 \lp}
&=& \frac{1}{2s} \sum_{i j a} 
\int_{\zeta_{\rm min}}^1 \frac{\diff\zeta}{\zeta^2} 
\int_{\xi_{\rm min}}^1 \frac{\diff\xi}{\xi}\, D_{e/j}(\zeta,\mu^2)\, f_{i/e}(\xi,\mu^2)
\notag\\
& & 
\times
\int_{x_{\rm min}}^1\frac{\diff x}{x}\, f_{a/N}(x,\mu^2)\,
\widehat{H}_{ia\to jX}(\xi \l,x P,\lp/\zeta,\mu^2)\
+\ \cdots ,
\label{e.dis-fac}
\end{eqnarray}
where $i$, $j$, $a$ include all QED and QCD particles, and the ellipsis represents corrections suppressed by inverse powers of $\l_T'$.
The lower limits of the integrations in eq.~\eqref{e.dis-fac} depend on external kinematics as specified in eq.~\eqref{e.minvalues} below, and $f_{a/N}(x,\mu^2)$ is the PDF of the colliding nucleon $N$ with momentum fraction $x=p^-/P^-$ carried by the active parton of flavor $a$ (either a quark, antiquark or gluon in QCD, or a lepton or photon in QED)~\cite{Collins:1981uw}, where we use the light-cone vector notation
    $v^\pm=(v^0 \pm v^3)/\sqrt{2}$ 
for any four-vector $v^\mu$ with the plus direction defined along the lepton momentum $\l$.
(Note that we take the nucleon to be moving along the $-z$ direction, with the incident lepton along the $+z$ direction.)

In eq.~\eqref{e.dis-fac}, the LDF $f_{i/e}(\xi,\mu^2)$ gives the probability to find a lepton (or parton) of flavor~$i$ with momentum $k \sim \xi \l$ in the incident lepton of flavor~$e$, defined analogously to the PDF of a hadron~\cite{Collins:1981uw}, but with the hadron state replaced by an asymptotic lepton state $| e \rangle$.
Explicitly, for a lepton (or quark) distribution in a lepton $e$ with momentum $\l$, the LDF is defined as
\begin{equation}
{\hskip -0.09in}
f_{i/e}(\xi,\mu^2) = \int\frac{\diff z^-}{4\pi}\, e^{i \xi \l^+ z^-}
\langle e(\l) |\,
  \overline{\psi}_i(0) \gamma^+ \Phi_{[0,z^-]}\, \psi_i(z^-)
| e(\l) \rangle,
\label{e.lpdf}
\end{equation}
where $\xi=k^+/\l^+$ is the light-cone momentum fraction carried by the active lepton (or quark) of momentum $k$ and flavor $i$, as sketched in figure~\ref{f.lpdf}, $\mu$ is a scale to renormalize the nonlocal fermion operator, and
    $\Phi_{[0,z^-]} = \exp[-ie\int_0^{z^-}\!\!d\eta^- A^+(\eta^-)]$
is the gauge link with a photon (or gluon) field $A^\mu$.
Similarly, the photon (or gluon) distribution function of a lepton can be defined in the same way as the gluon distribution of a hadron, except replacing the hadron state by a lepton state, and the gluon field by corresponding photon field for the photon distribution function~\cite{Collins:1981uw}.

%%%%%%%%%%%%%%%%%%%%%%
\beft
\includegraphics[width=0.4\textwidth]{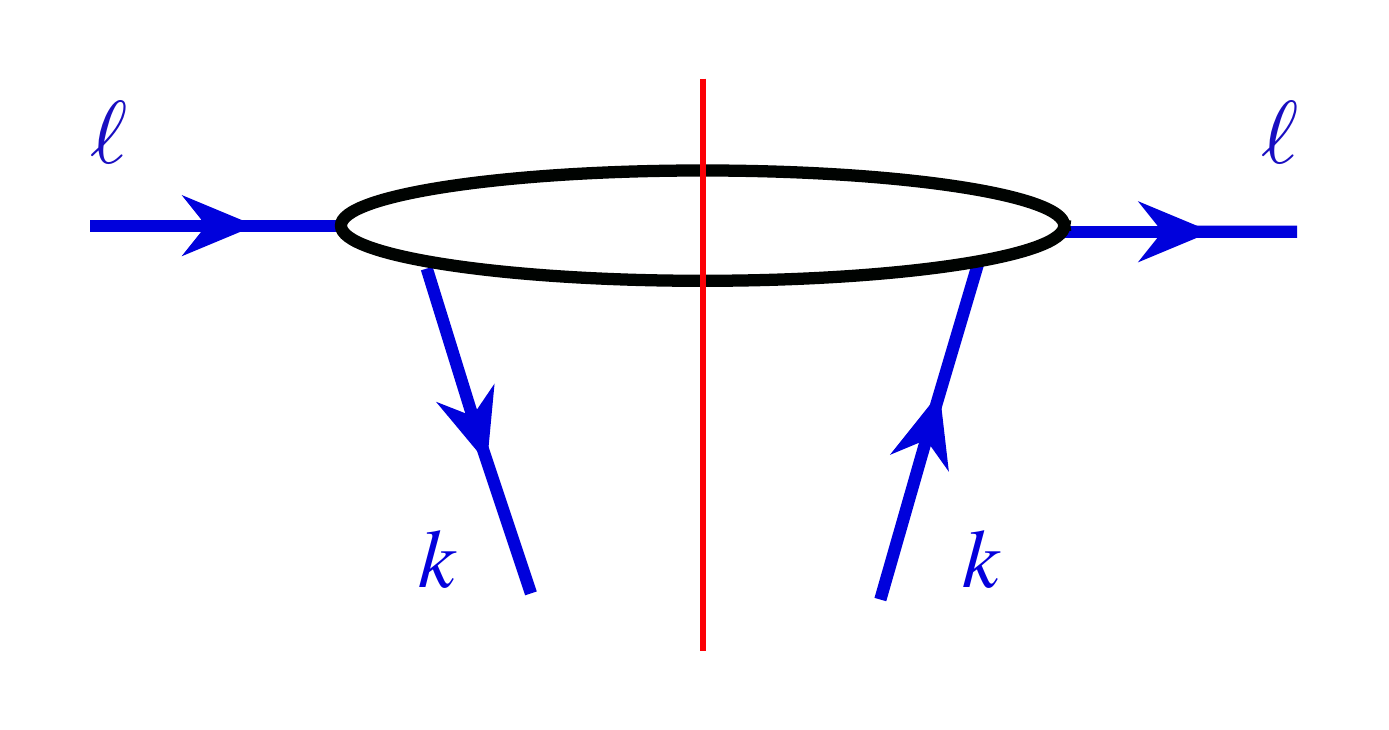} 
\vspace*{-0.4cm}
\caption{Sketch of the lepton distribution function for finding a fermion (quark or lepton) inside a colliding lepton of momentum $\l$, where the active fermion of momentum $k$ in the amplitude and its complex conjugate is contracted by the cut-vertex, $\gamma^+/(2\l^+)\, \delta(\xi-k^+/\l^+)$.}
\label{f.lpdf}
\eef
%%%%%%%%%%%%%%%%%%%%%%%

In analogy with the LDF, the LFF $D_{e/j}(\zeta,\mu^2)$ in eq.~(\ref{e.dis-fac}) describes the emergence of the final lepton $e$ with momentum $\ell'$ from a lepton (or parton) of flavor $j$ with momentum~\mbox{$k'\sim \lp/\zeta$}. 
Formally, the LFF for a fermion (lepton or quark) of flavor $j$ to decay into the observed lepton $e$ is defined as 
\begin{eqnarray}
\label{e.lff}
D_{e/j}(\zeta,\mu^2)
&=& \frac{\zeta}{2}\, \sum_X \int \frac{\diff z^-}{4\pi}\,
e^{i \ell'^+ z^-/\zeta} \nonumber\\
& & \times\,
{\rm Tr}
\Big[ 
\gamma^+
\langle 0|\, \overline{\psi}_j(0) \Phi_{[0,\infty]\,} |e(\lp), X\rangle
\langle e(\lp), X| \psi_j(z^-) \Phi_{[z^-,\infty]} |0\rangle 
\Big],
\end{eqnarray}
where the plus direction is taken along the observed lepton momentum 
    $\lp^\mu = (\lp^+, 0^-, \bm{0}_T)$
[note that the plus direction for the LDF in eq.~\eqref{e.lpdf} was defined along $\ell$].
The LFF from a photon (or gluon) is defined analogously to the gluon fragmentation function to a hadron, with the hadron state replaced by the observed lepton state and gluon field replaced by the corresponding photon field for the photon fragmentation function~\cite{Collins:1981uw}.
Both the LDF \eqref{e.lpdf} and LFF \eqref{e.lff} are defined in analogy with the quark PDF in the nucleon, $f_{a/N}(x,\mu^2)$, and quark to hadron fragmentation function~\cite{Collins:1981uw}, with the quark and gluon fields replaced by lepton and photon fields, and the hadron state by a lepton state.

In eq.~\eqref{e.dis-fac}, the function $\widehat{H}_{ia \to jX}$ is the lepton-parton (or parton-parton) scattering cross section, with all logarithmic collinear sensitivities along the direction of observed momenta, $\l, \lp$ and $P$, removed, and is therefore infrared safe and insensitive to taking the $m_e \to 0$ or $m_q\to 0$ limits.
The infrared-safe $\widehat{H}_{ia \to jX}$ can be perturbatively calculated by expanding the factorized formula~\eqref{e.dis-fac} order-by-order in powers of $\alpha$ and $\alpha_s$, with $\widehat{H}_{ia \to j}^{(m,n)}$ denoting the contribution at ${\cal O}(\alpha^m \alpha_s^n)$.

The factorized inclusive DIS cross section in eq.~\eqref{e.dis-fac} resums all logarithmic enhanced contributions from collision-induced radiation collinearly sensitive to the incident lepton into LDFs $f_{i/e}$, radiation that is collinearly sensitive to the scattered lepton into LFFs $D_{e/j}$, and radiation collinear to the colliding nucleon into the nucleon PDFs $f_{a/N}$. 
We stress that our factorization approach to inclusive DIS does not require the approximation of one-photon exchange.  
The factorization formula~\eqref{e.dis-fac} does provide a perturbatively stable basis for the reliable extraction of the nucleon PDFs, $f_{a/N}$, from inclusive DIS cross sections, along with the universal LDFs and LFFs, without the need for introducing the concept of hadron structure functions.
In this approach the structure functions are not direct physical observables, but are in practice tied to the one-photon exchange approximation.
In addition, the factorized formalism in eq.~\eqref{e.dis-fac} naturally accounts for all leading power QED contributions in the $m_e/Q$ expansion of the inclusive DIS cross section, order-by-order in powers of $\alpha$, such as those in figure~\ref{f.dis-qed2}, as well as the resummation of logarithmically enhanced collinear radiative contributions into LDFs and LFFs.

\begin{figure}[t]
\centering
    \includegraphics[width=0.6\textwidth]{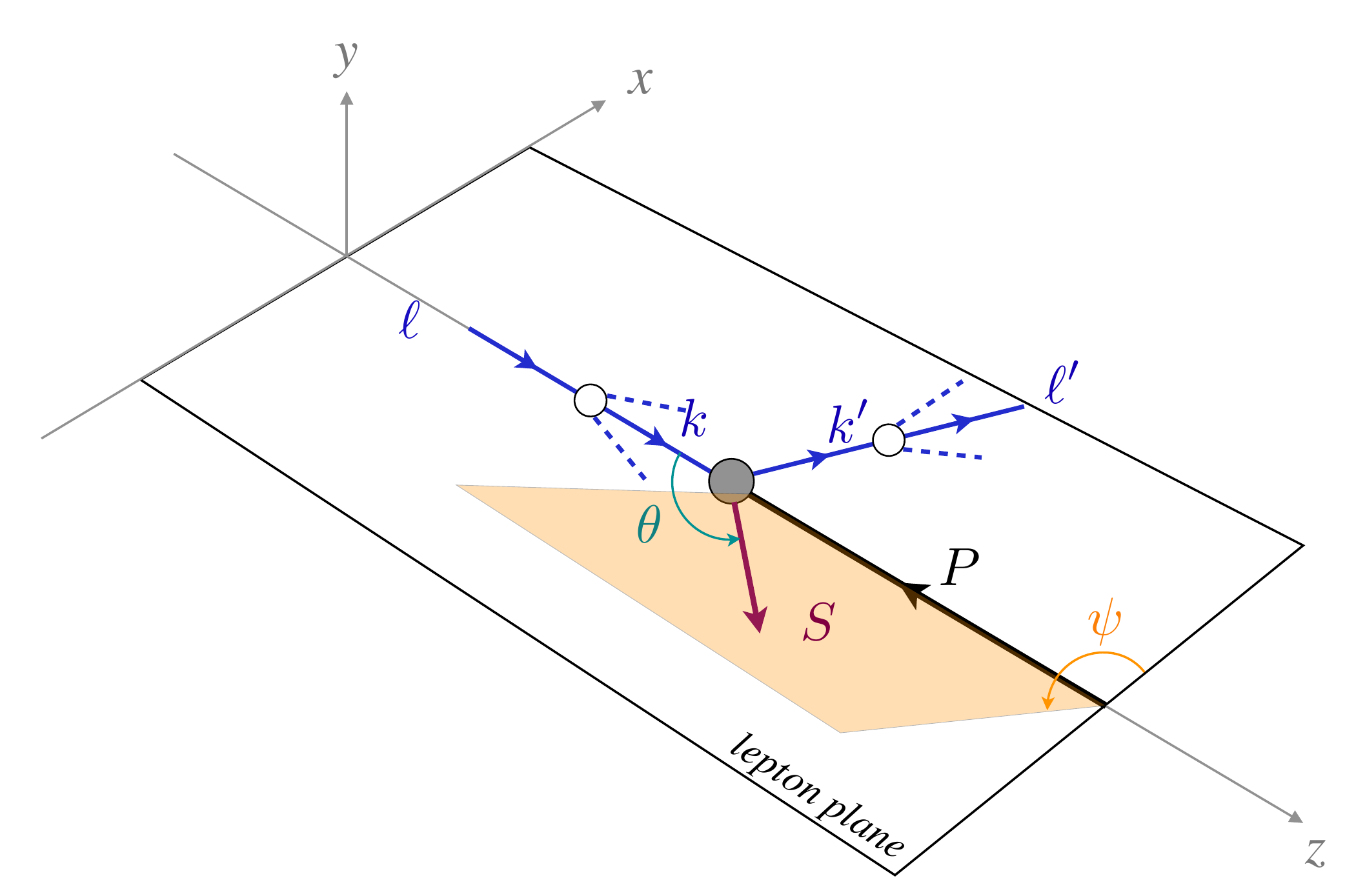}
    \caption{
    Sketch of the kinematical variables describing inclusive DIS from a nucleon (with momentum $P$ and spin $S$), with the incident ($\l$) and scattered ($\lp$) leptons defining the lepton plane.}
\label{f.dis-frame}
\end{figure}

As an additional approximation, if one can justify that the inclusive DIS cross section for a lepton of momentum $\l$ and helicity $\lambda_\l$ colliding with a nucleon of momentum $P$ and spin $S$, as sketched in figure~\ref{f.dis-frame}, is dominated by the subprocesses consistent with the one-photon approximation, which is equivalent to setting $i=j=e$ for the $\sum_{ij}$ in eq.~\eqref{e.dis-fac}, then the factorization formula in eq.~\eqref{e.dis-fac} can be further simplified to
\begin{align}
\label{e.dis-fac-qed0} 
E_{\ell'}
 \frac{\diff^3\sigma_{\l(\lambda_\l) P(S)\to \lp X}}{\diff^3\lp}
 &\approx
    \sum_{\lambda_k}
    \int^1_{\zeta_{\rm min}} \frac{\diff\zeta}{\zeta^2} \, D_{e/e}(\zeta,\mu^2)
    \int^1_{\xi_{\rm min}} \diff\xi \, f_{e(\lambda_k)/e(\lambda_\l)}(\xi,\mu^2)
\nonumber\\
 & {\hskip 0.2in} \times
    \left[
E_{k'}
 \frac{\diff^3\hat{\sigma}_{k(\lambda_k) P(S) \to k' X}}{\diff^3 k'}
 \right]_{k=\xi\l,\, k'=\lp/\zeta},
\end{align}
where $\lambda_k$ is the helicity of the lepton of momentum $k$ that collides with the nucleon.
The cross section $\hat{\sigma}_{k(\lambda_k) P(S)\to k' X}$ is infrared-safe as $m_e \to 0$, with all collinear sensitive QED radiative contributions along the lepton momenta $\l$ and $\lp$ resummed into $f_{e/e}$ and $D_{e/e}$, respectively.
At lowest order in powers of $\alpha$, effectively with one-photon exchange, the cross section can be written as
\begin{eqnarray}
\label{e.dis-qed-1p} 
E_{k'} \frac{\diff^3\hat{\sigma}_{k(\lambda_k) P(S)\to k' X}}{\diff^3k'}
&\approx& 
\frac{2\alpha^2}{\hat{s}\, \widehat{Q}^4} 
L_{\mu\nu}^{(0)}(k,k',\lambda_k)\,
W^{\mu\nu}(\hat{q},P,S),
\end{eqnarray}
with $L_{\mu\nu}^{(0)}(k,k',\lambda_k)$ and $W^{\mu\nu}(\hat{q},P,S)$ defined in eqs.~(\ref{e.dis-lmn}) and (\ref{e.dis-wmnSFs}), respectively.
We can express the phase space of the scattered lepton $\lp$ in terms of more commonly used variables,
\begin{align}
\label{e.dis-variables}
\frac{\diff^3\lp}{E_\lp} 
&= \Big(\frac{y}{2\xb}  \Big)\, \diff\xb\, \diff Q^2\, \diff\psi 
 = \Big(\frac{Q^2}{2\xb}\Big)\, \diff\xb\, \diff y\,   \diff\psi,
\end{align}
where $\psi$ is an angle between the leptonic plane and the nucleon spin plane defined by vectors ${\bm P}$ and ${\bm S}$, as shown in figure~\ref{f.dis-frame}, with integration over $\diff\psi$ giving a factor $2\pi$ for unpolarized or longitudinally polarized DIS.
Substituting the tensors in eqs.~(\ref{e.dis-lmn}) and (\ref{e.dis-wmnSFs}) into eq.~(\ref{e.dis-qed-1p}), and then substituting (\ref{e.dis-qed-1p}) into eq.~(\ref{e.dis-fac-qed0}),
we can express the spin-averaged lepton-nucleon DIS cross section in terms of structure functions evaluated at the shifted variables $\xb \to \hxb$ and $Q^2 \to \widehat{Q}^2$,
\begin{eqnarray}
\frac{\diff^2 \sigma_{\l P\to \lp X}}{\diff \xb \diff y}
&\approx&
\int^1_{\zeta_{\rm min}}\, \frac{\diff\zeta}{\zeta^2}\,
\int^1_{\xi_{\rm min}}\, \diff\xi\, D_{e/e}(\zeta,\mu^2)\, f_{e/e}(\xi,\mu^2)
\left[ \frac{Q^2}{\xb}\, \frac{\hxb}{\widehat{Q}^2} \right]
\nonumber\\
& &  
\times\
\frac{4\pi \alfa^2}{\hxb\, \hat{y}\, \widehat{Q}^2}
\Big[ \hxb \hat{y}^2\, F_1(\hxb,\widehat{Q}^2)
    + \Big( 1-\hat{y}-\frac{1}{4}\hat{y}^2\hat{\gamma}^2 \Big)\, F_2(\hxb,\widehat{Q}^2)
\Big].
\label{e.dis-fac-1p}
\end{eqnarray}
Here the factor $\big[(Q^2/\xb)\, (\hxb/\widehat{Q}^2)\big]$ is the Jacobian from eq.~(\ref{e.dis-variables}), and the variables with carets ``\, $\widehat{}$\, '' are defined with respect to a virtual photon with momentum 
    $\hat{q}^\mu = \xi \l^\mu - \lp^\mu/\zeta$, 
\begin{eqnarray}
\widehat{Q}^2 = -\hat{q}^2 = \frac{\xi}{\zeta} Q^2,
\qquad
\hxb = \frac{\widehat{Q}^2}{2P\cdot\hat{q}}\,,
\qquad
\hat{y} = \frac{P\cdot \hat{q}}{P\cdot k}\, ,
\qquad
\hat{\gamma} = \frac{2M\hxb}{\widehat{Q}}\, ,
\label{e.hat-variables}
\end{eqnarray}
with 
    $\widehat{Q}^2 = \hxb\,\hat{y}\,\hat{s}$ and 
    $\hat{s} = (k+P)^2 \approx \xi s$.
The factorization formalism with the one-photon exchange in eq.~(\ref{e.dis-fac-1p}) resums all logarithmic enhanced QED radiative contributions to the inclusive DIS cross section into the universal LDFs and LFFs.

We stress that the result in eq.~\eqref{e.dis-fac-1p} is derived from~\eqref{e.dis-fac-qed0} with the approximation of eq.~(\ref{e.dis-qed-1p}), and should be valid so long as QED power corrections, proportional to powers of $m_e/\widehat{Q}$, are small, without assuming any QCD factorization of the nonperturbative $F_1$ and $F_2$ structure functions.
The QCD factorization of $F_1$ and $F_2$ into expressions involving PDFs may indeed not be valid if the ``true'' hard scale $\widehat{Q}^2$ is not in the DIS regime, or if $\hxb$ is too close to 1 when the power corrections are large.
On the other hand, eq.~\eqref{e.dis-fac-1p} does express a valid QED factorization formalism that preserves the concept of the $F_1$ and $F_2$ DIS structure functions in the one-photon exchange scenario.
With knowledge of the LDFs and LFFs, eq.~(\ref{e.dis-fac-1p}) allows the extraction of $F_1$ and $F_2$ as functions of $\hxb$ and $\widehat{Q}^2$ via global analysis of all DIS cross section data at measured $\xb$ and $Q^2$ values, without necessarily addressing whether they can be factorized into PDFs.

It is important also to note, as we discuss in more detail in section~\ref{ss.connection} below, that the integration over the leptonic momentum fractions $\xi$ and $\zeta$ in eq.~(\ref{e.dis-fac-1p}), resulting from the induced QED radiation, allows the ``true'' Bjorken variable experienced by the colliding nucleon, $\hxb$, to take any value between $\xb$ and 1 for any measured $\xb$.
Namely, if one insisted on deriving an RC factor to mimic the impact of all (or the dominant) induced QED radiation, one would require knowledge of the structure functions for all possible values of the Bjorken variable between $\xb$ and $1$, which is the quantity that we are trying to measure in the first place.
Such an RC factor, therefore, is necessarily model dependent.
Furthermore, the structure functions are nonperturbative quantities and the validity of their factorization into PDFs requires power corrections $\propto 1/(1-\hxb)\widehat{Q}^2$ to be small and controllable. 
However, modeling the structure functions with known PDFs in order to derive the RC factor could lead to uncontrollable systematic uncertainties, since such power correction could be enhanced by not only $\hxb \to 1$ but also the fact that $\widehat{Q}_{\rm min}^2$ as given in eq.~(\ref{e.Qhat2min}) is $\leq Q^2$.

Under the collinear factorization approach to inclusive DIS in eq.~(\ref{e.dis-fac}), the active leptons of momentum $k$ and $k'$ are in the same plane as the incoming and scattered leptons of momentum $\l$ and $\lp$.
In the one-photon exchange approximation, therefore, the factorization formalism in eq.~(\ref{e.dis-fac-qed0}) also applies to the polarized inclusive DIS cross section, 
\begin{eqnarray}
&&
\frac{\diff \sigma_{\l(\lambda_\l) P(S)\to \lp X}}{\diff \xb\, \diff y\, \diff \psi}
-
\frac{\diff \sigma_{\l(\lambda_\l) P(-S)\to \lp X}}{\diff \xb\, \diff y\, \diff \psi}
\approx
\int^1_{\zeta_{\rm min}}\, \frac{\diff\zeta}{\zeta^2}\,
\int^1_{\xi_{\rm min}}\, \diff\xi\, D_{e/e}(\zeta,\mu^2)\, {\Delta f_{e/e(\lambda_\l)}}(\xi,\mu^2)
\nonumber\\
& & \hspace{2cm}
\times
\left[ \frac{Q^2}{\xb}\, \frac{\hxb}{\widehat{Q}^2} \right]
\frac{4\lambda_\l\, \alfa^2}{\widehat{Q}^2}
\Bigg\{ \cos\theta 
\bigg[
    \Big( 1 - \frac{\hat{y}}{2} - \frac{1}{4}\hat{\gamma}^2\hat{y}^2 \Big) 
    g_1(\hxb,\widehat{Q}^2)
  - \frac{1}{2} \hat{\gamma}^2\hat{y}\, 
    g_2(\hxb,\widehat{Q}^2)
\bigg]
        \label{e.dis-fac-pol}
\nonumber\\
& & \hspace*{3.3cm} 
-\sin\theta \cos\psi\ \hat{\gamma}\sqrt{1-\hat{y}-\frac{1}{4}\hat{\gamma}^2\hat{y}^2}
\bigg[ \frac{\hat{y}}{2}\, g_1(\hxb,\widehat{Q}^2)
     + g_2(\hxb,\widehat{Q}^2) 
\bigg]
\Bigg\}\, ,
\end{eqnarray}
where 
    $\Delta f_{e/e(\lambda_\l)} 
    \equiv [f_{e(\lambda_k=1)/e(\lambda_\l)} - f_{e(\lambda_k=-1)/e(\lambda_\l)}]/2
    = [f_{e(\lambda_k=1)/e(\lambda_\l)} - f_{e(\lambda_k=1)/e(-\lambda_\l)}]/2$
denotes the lepton helicity distribution,
and $\theta$ is the angle between the colliding lepton of 3-momentum ${\bm \l}$ and the direction of the nucleon spin ${\bm S}$ [$\cos\theta = M (\l\cdot S)/(\l\cdot P)$], as shown in figure~\ref{f.dis-frame}.

If one wishes to include higher order QED radiative contributions that are not resummed into the LDFs and LFFs, one should use the more general factorization formalism in eq.~(\ref{e.dis-fac}) [or in eq.~(\ref{e.dis-fac-qed0}) under the one-photon approximation] for the inclusive DIS cross section.
In this case all QED radiative contributions are systematically included into the infrared-safe hard part $\widehat{H}_{ia \to jX}$, order-by-order in powers of $\alpha$, and the universal LDFs and LFFs.

% .......................................................................
\subsection{Universal LDFs and LFFs}
\label{ss.ldf-lff}

\begin{figure}[t]
\centering
\includegraphics[width=0.3\textwidth,page=1]{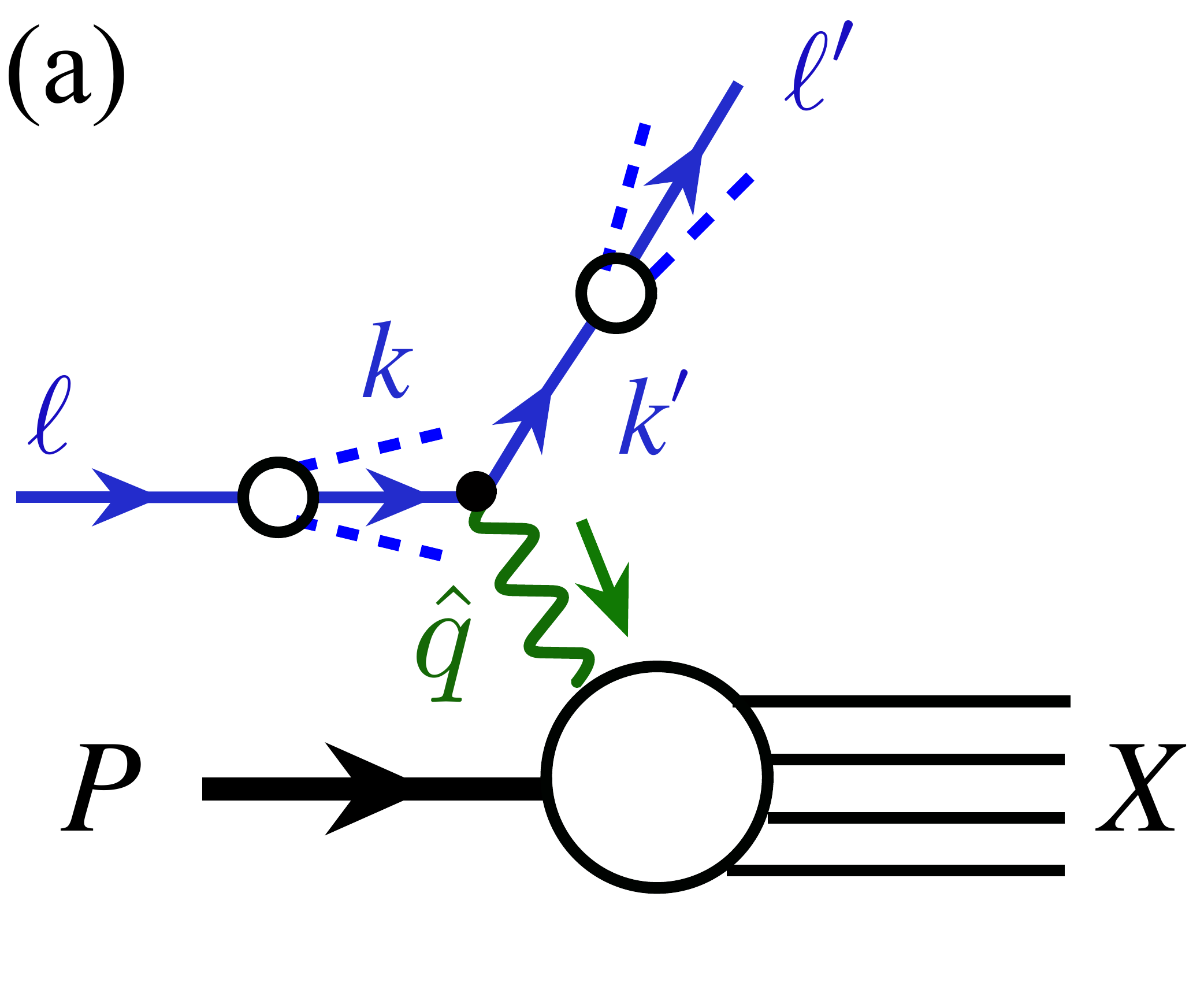}\qquad\qquad
\includegraphics[width=0.3\textwidth,page=2]{dis-1boson.pdf} 
\hspace*{-0.3cm}
\caption{Sketch of sample scattering amplitudes for inclusive DIS with
{\bf (a)}~one-photon exchange, and {\bf (b)}~one-gluon exchange.}
\label{f.dis-amp}
\end{figure}

The universal LDFs and LFFs share the same operator definitions with the hadron PDFs and FFs, as discussed above.  
Like the hadron PDFs and FFs, in principle the LDFs and LFFs are nonperturbative due to the fact that they can have hadronic components from high-order processes.
For example, the colliding electron could radiate a photon, the photon could split into quark-antiquark pair, and the quark could initiate a hard scattering to produce the observed lepton of momentum $\lp$, leading to a factorized nonperturbative term proportional to the LDF $f_{q/e}(\xi,\mu^2)$.
In this case one has contributions to the factorized inclusive DIS cross section in eq.~(\ref{e.dis-fac}) in terms of nonperturbative quark or gluon PDFs of a colliding electron, as well as quark or gluon FFs to the observed lepton, as illustrated in figure~\ref{f.dis-amp}.
In addition, even the LDF $f_{e/e}(\xi,\mu^2)$ may have nonperturbative hadronic component from high-order processes, although the impact of its hadronic components may be very small in the energy regime of interest.

If we could restrict the events where there is effectively no hadronic activity along the direction of the observed lepton, we could suppress the contributions from the types of subprocesses in Fig.~\ref{f.dis-amp}(b), even though these are expected to be small.
Such restriction could be imposed on the measurements to identify ``isolated lepton'' events, similar to the ``isolated photon'' events in hadronic collision~\cite{Harris:1990if}. 
However, such isolation could reduce the phase space for real gluon emission to break the perturbative infrared cancellation between the real and virtual diagrams, since the phase space for the virtual gluon loop is not affected by the isolation requirement.
This was recognized in the ``isolated photon" case~\cite{Berger:1990es,Berger:1995cc}, and has being consistently taken care of by the proposed implementation of the isolation~\cite{Frixione:1998jh}.  
Photon isolation is very important and needed for high energy photon production as a hard probe of short-distance dynamics, since there is a large background of high energy photons from the decay of an energetic $\pi^0$. 
However, as discussed in section~\ref{sss.ltmd-fac} below, the hadronic content of the collision-induced shower along the direction of the lepton is strongly suppressed by powers of $\alpha \sim 10^{-2}$, and we expect that the leading lepton is almost ``isolated''.
A detailed implementation of such ``isolated leptons'' is beyond the scope of current paper.

Neglecting the hadronic contribution, the LDF $f_{e/e}(\xi,\mu^2)$ can be calculated perturbatively in QED with a properly defined renormalization for the nonlocal operators.
Denoting by 
    $f^{(m)}_{e/e}(\xi,\mu^2)$
the LDF evaluated perturbatively to order ${\cal O}(\alpha^m)$, we have, for example, the LO LDF given by
    $f^{(0)}_{e/e}(\xi) = \delta(\xi-1)$.  
At NLO, the leading logarithmically enhanced real and virtual contribution in the light-cone gauge are given by the diagrams in figure~\ref{fig:lpdf1}(a)~and~(b), respectively, leading to the result in the $\overline{\rm MS}$ scheme,
\begin{eqnarray}
f_{e/e}^{(1)}(\xi,\mu^2)
= \frac{\alpha}{2\pi}
\left[
    \frac{1+\xi^2}{1-\xi}
    \ln\frac{\mu^2}{(1-\xi)^2\, m^2_e}
\right]_+ , 
\label{e.lpdf1}
\end{eqnarray}
where the standard ``+'' prescription is used.
As expected, the perturbatively calculated LDF,
    $f_{e/e}(\xi,\mu^2) 
    \approx f_{e/e}^{(0)}(\xi,\mu^2) + f_{e/e}^{(1)}(\xi,\mu^2)$,
preserves lepton number, 
    \mbox{$\int_0^1 d\xi\, f_{e/e}(\xi,\mu^2) = 1$}.
A more comprehensive derivation of this LDF, beyond our NLO QED calculation, can be found in Ref.~\cite{Frixione:2021wzh}.
As for the contributions to hadron PDFs, high-order logarithmically enhanced contributions to LDFs can be systematically resummed by solving the evolution equations for these collinearly factorized distributions, including PDFs of the lepton if we apply collinear factorization for the collision-induced QED and QCD radiation from the leptons~\cite{Williams:1934ad, vonWeizsacker:1934nji, Dokshitzer:1977sg, Gribov:1972ri, Lipatov:1974qm, Altarelli:1977zs}.
For the flavor non-singlet evolution, for example, one has
\begin{equation}
\mu^2\frac{\diff}{\diff\mu^2} f_{e/e}(\xi,\mu^2) 
= \int_{\xi}^1\frac{\diff\xi'}{\xi'}\,
P_{ee}\bigg(\frac{\xi}{\xi'},\alpha,\alpha_s\bigg) f_{e/e}(\xi',\mu^2),
\label{e.dglap}
\end{equation}
where the evolution kernel $P_{ee}$ is calculable perturbatively order-by-order in powers of $\alpha$ and $\alpha_s$.
At~${\cal O}(\alpha)$, from eq.~(\ref{e.lpdf1}) one has
    $P_{ee}^{(1)}(z,\alpha,\alpha_s) = (\alpha/2\pi) \left[(1+z^2)/(1-z)\right]_+$.

%%%%%%%%%%%%%%%%%%%%%%
\beft
\centering
\includegraphics[height=0.17\textwidth]{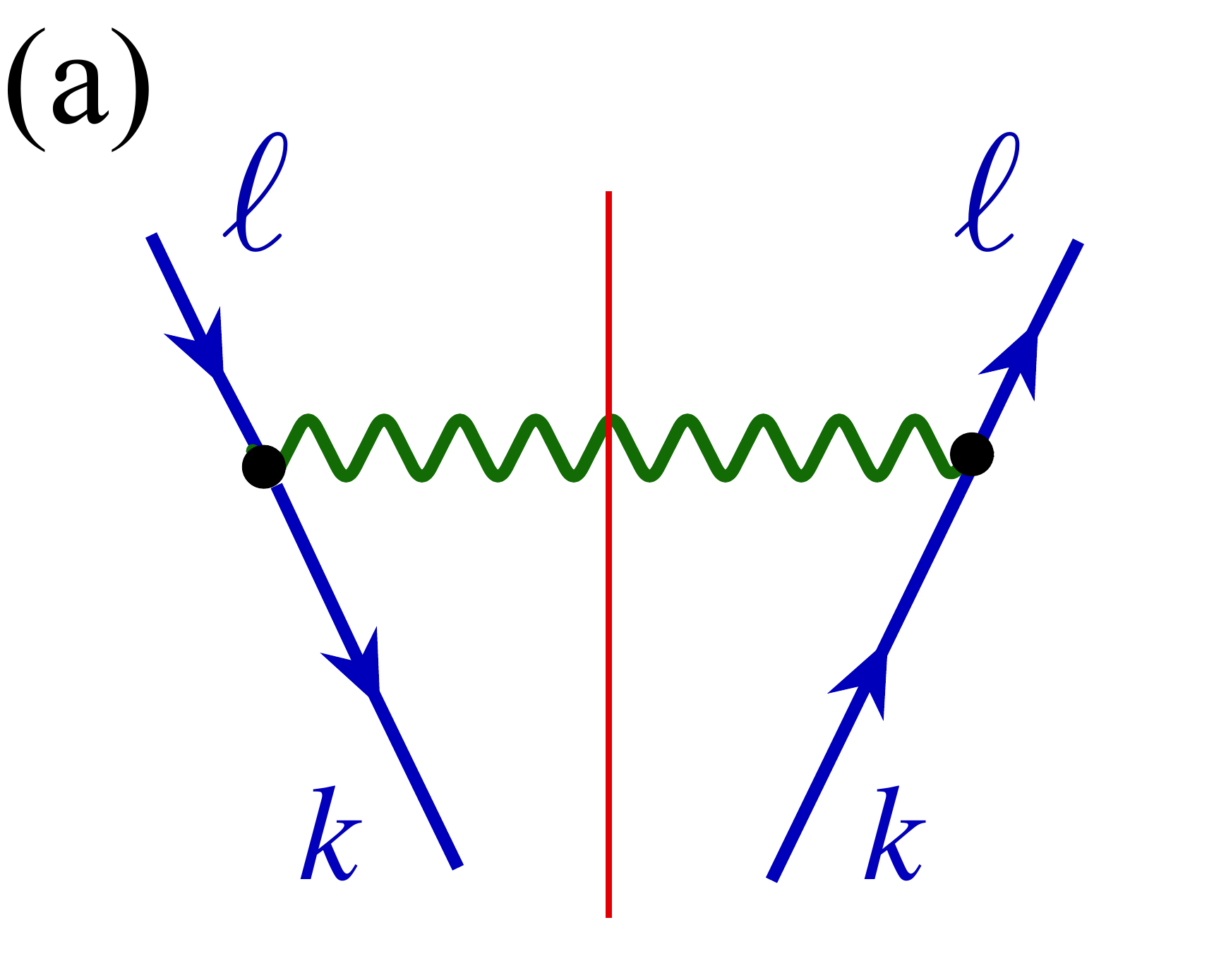}
\hspace{1cm}
\includegraphics[height=0.17\textwidth]{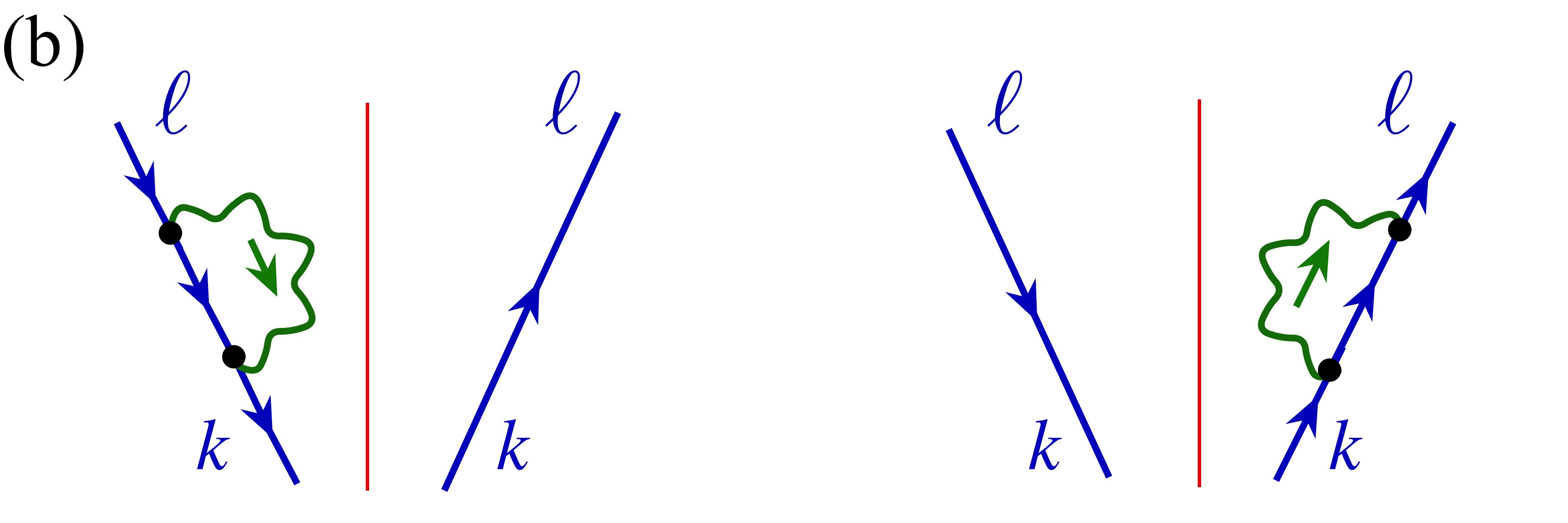}
\caption{Examples of {\bf (a)} real and {\bf (b)} virtual QED diagrams contributing to the NLO lepton distribution $f_{e/e}^{(1)}$.}
\label{fig:lpdf1}
\eef
%%%%%%%%%%%%%%%%%%%%%%%

Similarly, the LFFs can also be calculated perturbatively in QED, if we neglect their hadronic components.
At LO, the LFF is given by the trivial expression
    $D^{(0)}_{e/e}(\zeta) = \delta(\zeta - 1)$,
while at ${\cal O}(\alpha)$ we have analogous expression to that in eq.~(\ref{e.lpdf1}),
% for the logarithmically enhanced contributions,
%
\begin{eqnarray}
D_{e/e}^{(1)}(\zeta,\mu)
= \frac{\alpha}{2\pi}
\left[
    \frac{1+\zeta^2}{1-\zeta} 
    \ln \frac{\zeta^2\mu^2}{(1-\zeta)^2\, m_e^2}
\right]_+ \, .
\label{e.lff1}
\end{eqnarray}
As for the LDFs, the logarithmically enhanced high-order contributions to the LFFs can be resummd by solving the corresponding evolution equations.

Within the collinear factorization framework, in analogy with PDFs and FFs of hadrons, we can derive the LDFs and LFFs by solving their corresponding evolution equations with nonperturbative input distributions at an initial scale $\mu_0$.
Unlike PDFs or FFs of hadrons, however, which are completely nonperturbative, we could use the perturbatively calculated LDFs and LFFs in QED at $\mu_0$ as a reasonable model for the input distributions, neglecting their QCD contributions as an approximation.
In practice, the input distributions can always be improved by comparing with experimental data.
We also note that the choice of $\mu_0$ is not unique, which impacts the size of uncalculated higher order contributions to LDFs and LFFs in QED, as well as the size of neglected nonperturbative QCD contributions to LDFs and LFFs.
Our choice of $\mu_0$ will be specified in section~\ref{s.pheno}, and further discussion about this will be explored in future work.

% .......................................................................
\subsection{Short-distance partonic hard parts}
\label{ss.dis-hard}

As with all QCD factorization approaches, the partonic hard parts $\widehat{H}_{ia \to jX}$ in eq.~\eqref{e.dis-fac} are infrared safe and insensitive to taking the limits $m_e \to 0$ or $m_q \to 0$.
They can be calculated perturbatively by applying the factorized formula \eqref{e.dis-fac} to lepton-parton scattering order-by-order in powers of $\alpha$ and $\alpha_s$, and depend on the choice for the renormalization scheme of the LDFs and LFFs, in addition to the scheme that defines the PDFs.

To compute the leading order infrared-safe hard part in eq.~\eqref{e.dis-fac}, we can replace the target nucleon by a point-like quark target, $q$.
The lepton-quark cross section can then be expanded to a given order in $\alpha$ and $\alpha_s$, with the ${\cal O}(\alpha^m \alpha_s^n)$ contribution to the cross section denoted by
    $\sigma^{(m,n)}_{eq} 
    \equiv 2s E' \diff \sigma_{eq \to eX}^{(m,n)}/\diff^3 \ell'$.
[Note that the subscripts on the partonic cross section here refer to particle type, in contrast to the hadronic cross sections discussed above, as in eq.~(\ref{e.dis-fac-qed0}) and subsequently, which are labeled by the leptons' and hadrons' momenta.]
Expanding the partonic cross section to the lowest order, {i.e.,} ${\cal O}(\alpha^2 \alpha_s^0)$, we have
\begin{align}
\sigma_{eq}^{(2,0)}
&= D_{e/e}^{(0)} \otimes f_{e/e}^{(0)} \otimes f_{q/q}^{(0)} \otimes \widehat{H}_{eq \to eX}^{(2,0)}
 = \widehat{H}_{eq \to eX}^{(2,0)}\, ,
\end{align}
where $\otimes$ indicates the convolution of momentum fractions, and the ${\cal O}(\alpha_s^0)$ quark distribution 
    $f^{(0)}_{q/q}(x) = \delta(x-1)$
is also used.
Evaluating the lowest order lepton-quark scattering diagram, one finds for the hard part function
\begin{eqnarray}
\widehat{H}_{eq \to eX}^{(2,0)}
&=&
\frac{4 \alpha^2 e_q^2}{\zeta}
\bigg[ \frac{(\zeta\xi x s)^2 + (x u)^2}{(\xi t)^2} \bigg]\,
\delta\big(\zeta\xi x s + x u + \xi t\big),
\label{e.fac-h0}
\end{eqnarray}
with Mandelstam variables
$s = (\l + P)^2$,
$u = (\lp - P)^2 = (y-1) s$,
and
$t = (\l - \lp)^2 = - Q^2$.
Substituting the calculated $\widehat{H}^{(2,0)}_{eq \to eX}$ into eq.~\eqref{e.dis-fac-1p}, we then have 
\begin{align}
E_\lp \frac{\diff \sigma_{\l P\to \lp X}}{\diff^3 \lp}
&\approx \frac{2\alpha^2}{s} \sum_q 
\int_{\zeta_{\rm min}}^1 \frac{\diff\zeta}{\zeta^2} 
\int_{\xi_{\rm min}}^1 \frac{\diff\xi}{\xi} D_{e/e}(\zeta)\, f_{e/e}(\xi)
\notag\\
& \quad
\times
\int_{x_{\rm min}}^1 \frac{\diff x}{x}\, e_q^2\, f_{q/N}(x)
\frac{x^2\zeta \big[ (\zeta\xi s)^2 + u^2\big]}
     {(\xi t)^2 (\zeta\xi s + u)}\,
\delta\big( x - x_{\rm min} \big),
\label{e.fac0}
\end{align}
where the lower limits of the integrations are given by
\begin{subequations}
\label{e.minvalues}
\begin{eqnarray}
\zeta_{\rm min} &=& -\frac{t+u}{s}\,
                 =\, 1 - (1-\xb)\, y,
\\
\xi_{\rm min}   &=& -\frac{u}{\zeta s + t}\,
                 =\, \frac{1-y}{\zeta-\xb\, y},
\\
x_{\rm min}     &=& -\frac{\xi t}{\zeta\xi s + u}\,
                 =\, \frac{\xi\, \xb\, y}{\xi\zeta + y - 1}.
\end{eqnarray}
\end{subequations}
Choosing the leading order contributions
    $f_{e/e}(\xi) \approx f_{e/e}^{(0)}(\xi)$
and
    $D_{e/e}(\zeta) \approx D_{e/e}^{(0)}(\zeta)$
in eq.~\eqref{e.fac0}, and noting that to ${\cal O}(\alpha_s^0)$ the structure functions in eq.~\eqref{e.dis-fac-1p} are given by 
    $F_2(\xb) = 2 \xb F_1(\xb) = \sum_q e_q^2\, \xb f_{q/N}(\xb)$,
one can reproduce the lepton-nucleon cross section \eqref{e.dis-fac-1p} from eq.~\eqref{e.fac0}.
The key difference between eqs.~\eqref{e.dis-fac-1p} and \eqref{e.fac0}, apart from infrared-safe high order QED contribution, is the resummation of logarithmic-enhanced photon radiation for the colliding and scattered leptons into the LDFs and LFFs, respectively.

With the factorization formalism in eq.~\eqref{e.dis-fac}, one can systematically improve the ``RCs'' by calculating the infrared-safe hard parts $\widehat{H}_{eq \to eX}^{(m,n)}$ perturbatively for $m > 2$, and determining the lepton mass-sensitive, but universal, LDFs and LFFs.
For example, at $m=3$ one can write
\begin{eqnarray}
\widehat{H}_{eq \to eX}^{(3,0)}
&=& \sigma_{eq}^{(3,0)} 
 - D_{e/e}^{(1)} \otimes \widehat{H}_{eq \to eX}^{(2,0)} 
 - f_{e/e}^{(1)} \otimes \widehat{H}_{eq \to eX}^{(2,0)}
 - f_{q/q}^{(1)} \otimes \widehat{H}_{eq \to eX}^{(2,0)}\, ,
\label{e.fac-h1}
\end{eqnarray}
where $\sigma_{eq}^{(3,0)}$ is the NLO QED contribution to electron-quark scattering with a proper collinear regularization, and is given by the diagrams in figure~\ref{f.dis-qed2}.
The function $\widehat{H}_{eq \to eX}^{(2,0)}$ is given in eq.~\eqref{e.fac-h0}, and $f_{e/e}^{(1)}$ and $D_{e/e}^{(1)}$ are the NLO electron distribution and fragmentation functions of an electron in eqs.~\eqref{e.lpdf1} and \eqref{e.lff1}, respectively, if we regularize the perturbative collinear divergence by the electron mass. 
The NLO quark distribution function in a quark, $f_{q/q}^{(1)}$, is similar to $f_{e/e}^{(1)}$, and its exact expression depends on how the perturbative collinear divergence associated with massless quark is regularized~\cite{Collins:2011zzd}.
The three subtraction terms in eq.~\eqref{e.fac-h1} involve convolutions over different momentum fractions to remove the collinear-sensitive photon radiation from the scattered lepton, incident lepton, and incident quark, respectively.
With infrared safety, the perturbatively calculated RCs are completely perturbative-stable and insensitive to the lepton mass $m_e \to 0$, with all $m_e$ sensitive RCs resummed into universal LDFs and LFFs.

\begin{figure}[tp]
    \centering
    \includegraphics[width=0.75\textwidth]{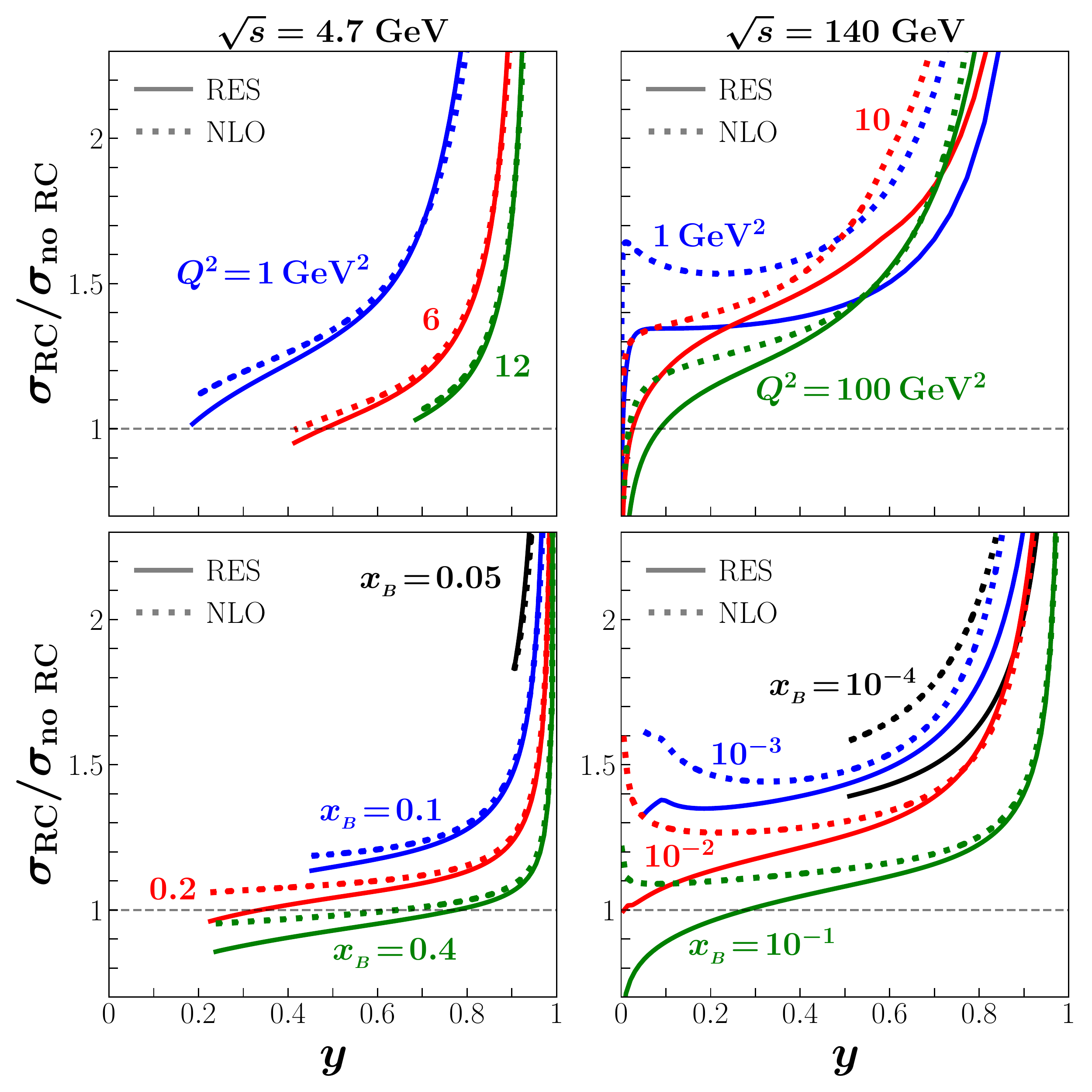}
    \caption{Ratio of inclusive $ep$ cross sections with QED radiation effects ($\sigma_{\rm RC}$) to those without radiation ($\sigma_{\rm no\, RC}$) versus $y$ at fixed values of $Q^2$ {\bf (top row)} and fixed $\xb$ {\bf (bottom row)} for Jefferson Lab energy $\sqrt{s}=4.7$~GeV {\bf (left)}, and EIC energy $\sqrt{s}=140$~GeV {\bf (right)}, for the resummed (``RES'', solid lines) and fixed-order (``NLO'', dashed lines) results.}
    \label{fig:idis-y}
\end{figure}

The effects of the QED radiation on the inclusive cross section are illustrated numerically in figure~\ref{fig:idis-y}, for typical Jefferson Lab ($\sqrt{s}=4.7$~GeV) and EIC ($\sqrt{s}=140$~GeV) center of mass energies.
The ratios of the full cross sections to the Born results show that the effects of the QED radiation can be quite large in some regions of kinematics, especially at larger values of $y$.
For the ratios with fixed $Q^2$ values, since $\xb\, y\, s = Q^2$ the large-$y$ region corresponds to small values of $\xb$, and lower $y$ values correspond to larger $\xb$.
The minimum value of $y$ accessible is restricted by the cut $W > W_{\rm min} = 2$~GeV, which excludes the nucleon resonance region, and corresponds to a maximum value of $\xb < \xb^{\rm max} = Q^2/(W_{\rm min}^2-M^2+Q^2)$. 
At Jefferson Lab energies this places a strong restriction on the range of $y$ values allowed, while at EIC energies the effect of the cut is less dramatic.
Note that for $Q^2=1$~GeV$^2$, for example, the resonance region cut corresponds to a maximum value $\xb^{\rm max} \approx 0.24$, while for $Q^2=10$~GeV$^2$, $\xb^{\rm max} \approx 0.76$, and for $Q^2=100$~GeV$^2$, $\xb^{\rm max} \approx 0.97$.

\begin{figure}[tp]
    \centering
    \includegraphics[width=0.42\textwidth]{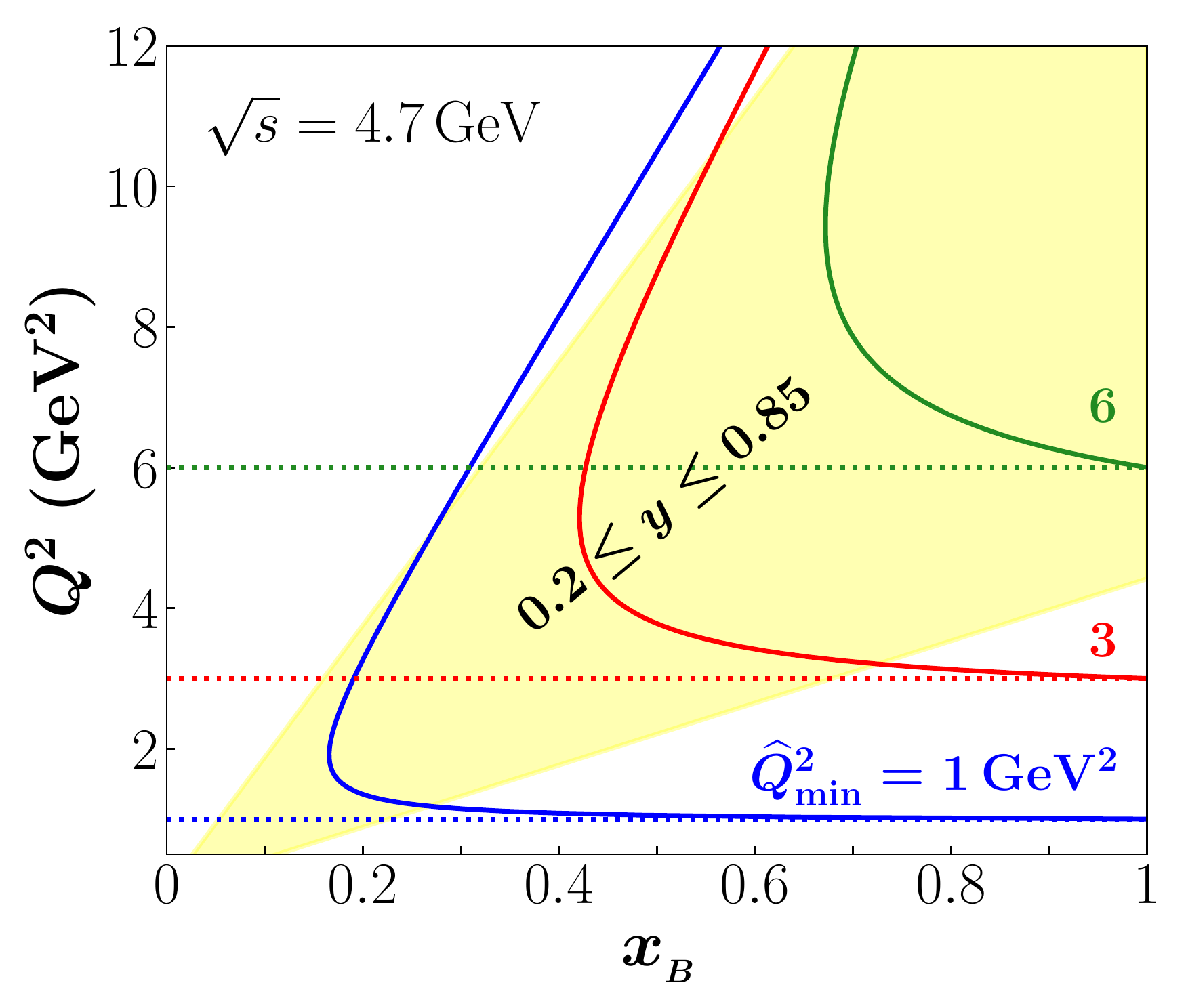}
    \includegraphics[width=0.42\textwidth]{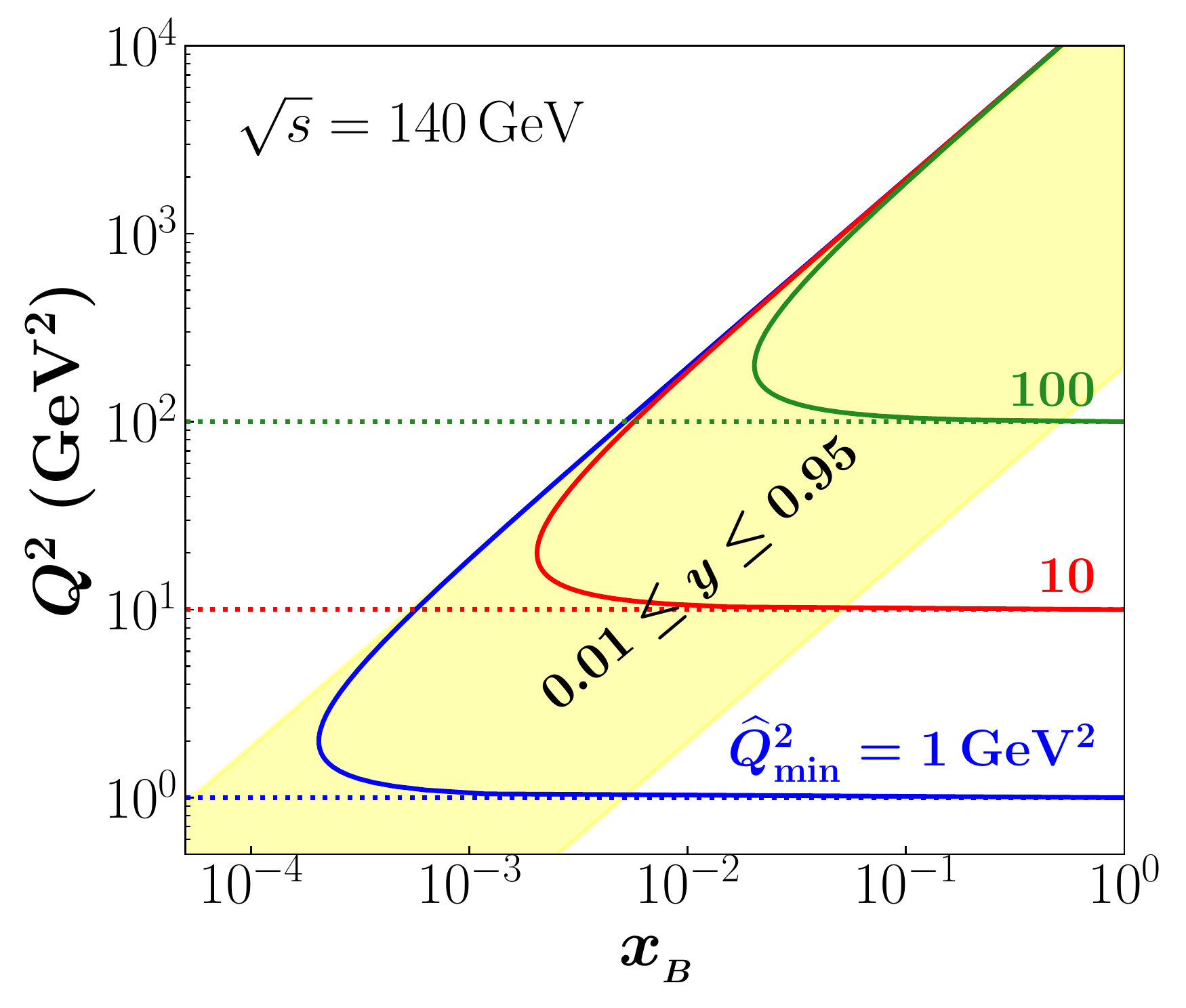}
    \caption{Available phase-space for lepton-nucleon DIS with collision-induced QED radiative contributions at Jefferson Lab ($\sqrt{s}=4.7$~GeV) {\bf (left)} and EIC ($\sqrt{s}=140$~GeV) {\bf (right)} kinematics. The colored lines denote regions of fixed $\widehat{Q}^2_{\rm min}$ and the diagonal yellow bands represent typical ranges of $y$ at those facilities.}
    \label{fig:ps}
\end{figure}

For the ratios at fixed values of $\xb$ in figure~\ref{fig:idis-y}, the effects also increase at larger~$y$, which corresponds to larger $Q^2$ values.
The minimum value of $y$ is restricted by the \mbox{$Q^2>1$~GeV$^2$} cut, which is imposed to exclude regions where the factorized framework would not be applicable.
This constraint becomes more evident at smaller $\xb$ values, which again is less dramatic at the higher EIC energies, where the limit on the $y$ range is visible for $\xb \lesssim 10^{-4}$.
In addition, with the collision-induced QED radiation, the hard scale of the collision (momentum transfer experienced by the colliding nucleon) changes from $Q^2$ to $\widehat{Q}^2=(\xi/\zeta)\, Q^2$, which has a minimum value of
        $\widehat{Q}^2_{\rm min} \leq Q^2$
given by eq.~(\ref{e.Qhat2min}).
The induced QED radiation could push the scattering between the virtual photon and the colliding nucleon out of the DIS regime when the ``true'' probing scale $\widehat{Q}^2$ is less than 1~GeV$^2$, even though $Q^2$ itself would be above the cut.
Instead of restricting $Q^2 > 1$~GeV$^2$, a requirement of $\widehat{Q}^2_{\rm min} > 1$~GeV$^2$ could impose a stronger constraint on the range of $\xb$ for a given value of $y$, as shown in figure~\ref{fig:ps}.

Overall, the radiative effects are positive over most kinematics, with the 
$\sigma_{\rm RC} / \sigma_{\rm no\ RC}$ ratio dropping below unity only at the lowest $y$ values, especially for larger $\xb$.
The effect of the resummation is generally a decrease in the magnitude of the radiative effects relative to the NLO calculation, except at the highest $y$ values where it enhances the corrections.
Clearly, the effects of the QED radiation are nontrivial and will have a significant impact on the extraction of PDF information from inclusive DIS experiments.
This is especially pertinent at large values of $y$ and small $\xb$, where more phase space is available for both QED and QCD radiation, and will be of particular interest at these kinematics in future EIC measurements.

% .......................................................................
\subsection{QED radiative contributions vs. radiative corrections}
\label{ss.connection}

Before moving to the more involved case of semi-inclusive lepton-nucleon scattering, we conclude the discussion of QED radiative effects in inclusive DIS by comparing our proposed factorization approach with existing approaches that isolate such contributions in the form of QED ``radiative corrections.''
With a large momentum transfer, the collision-induced QED radiation is an integral part of the experimentally measured cross section for deep-inelastic lepton-nucleon scattering.
Historically, tremendous efforts have been devoted to isolate and remove collision-induced QED contributions from measured cross sections that would enable one to focus purely on QCD effects in lepton-nucleon scattering.
The RCs can be represented schematically in the form
\begin{align}
    \sigma_{\rm obs}(\xb,Q^2)\,
    \qeq\, \Rqed(\xb,Q^2; \xbtrue,Q^2_{\rm true}) 
    \times 
    \sigma_{\rm Born}(\xbtrue,Q^2_{\rm true}) 
    + \sigma_X(\xb,Q^2),
    \label{e.standard}
\end{align}
where $\sigma_{\rm obs}$ is the physically measured cross section, $\sigma_{\rm Born}$ is the ideal lepton-nucleon cross section without the collision-induced QED radiation contamination, and $\Rqed$ and $\sigma_X$ are correction factors that are computed theoretically.
The variables $\xbtrue$ and $Q^2_{\rm true}$ represent the ``true'' or effective momentum scales that are experienced by the colliding nucleon, and differ from the corresponding experimental $\xb$ and $Q^2$ due to the induced QED radiation.  

For the expression in eq.~(\ref{e.standard}) to be a valid basis on which to quantitatively account for QED radiation, there must exist some controllable approximation scheme applicable for the full kinematic regime where the cross sections are measured.
More importantly, the following two conditions should be met in order to isolate QED contribution in terms of RCs:
\begin{enumerate}
\item[(1)]
the correction factors $\Rqed$ and $\sigma_X$ should not depend on the hadron structure that we wish to extract, and they can be systematically calculated in QED to high precision; 
\item[(2)]
the effective scale $Q^2_{\rm true}$ for the Born cross section $\sigma_{\rm Born}$ should be large enough to keep the ``true'' scattering within the DIS regime.
\end{enumerate}
In particular, with the one-photon approximation, the exchanged virtual photon (or vector boson, in general), with its fully determined four-momentum under the QED Born kinematics, would be able to serve as a localized and well-controlled hard probe to explore the partonic structure of the colliding nucleon.

In practice, however, the collision-induced QED radiation will change the momentum of the exchanged photon from $q$ to $\hat{q}$, as shown in figure~\ref{f.dis-amp}(a) under the one-photon approximation for the hard collision.
Since $\hat{q}$ is not fixed by the observed external momenta, it must be integrated over if we cannot account for all radiated photons.
The integration necessarily includes contributions from radiation that can distort $\hat{q}$ so much as to move the collision with the nucleon out of the desired DIS regime, when $\widehat{Q}^2 \equiv -\hat{q}^2 \lesssim 1$~GeV$^2$, and introduce contamination from elastic scattering events.
As indicated in figure~1 of ref.~\cite{Breidenbach:1969kd} from over 50 years ago, and verified by numerous experiments since, the event rate for inclusive lepton-nucleon DIS is expected to be much larger than the typical rate from elastic scattering when the probing scale is larger than $\approx 1$~GeV$^2$.
The collision-induced QED radiation could make the ``true'' probing scale $Q^2_{\rm true} = \widehat{Q}^2$ smaller, however, effectively enhancing the rate of non-DIS events and the size of non-factorized power corrections, even when $Q^2 = {\cal O}(1~{\rm GeV}^2)$ or larger.
Furthermore, QED radiation from final-state nucleons in elastic events requires a larger $Q^2$ to mimic DIS events.
Since these non-DIS events involve nonperturbative strong interaction physics that cannot be calculated reliably, QED RC factors that aim to ``correct'' for this QED contamination are necessarily model dependent. 
Some of these contaminations are sensitive to the very hadronic physics that we aim to explore in the DIS and SIDIS reactions.

\begin{figure}[t]
    \centering
    \includegraphics[width=0.42\textwidth]{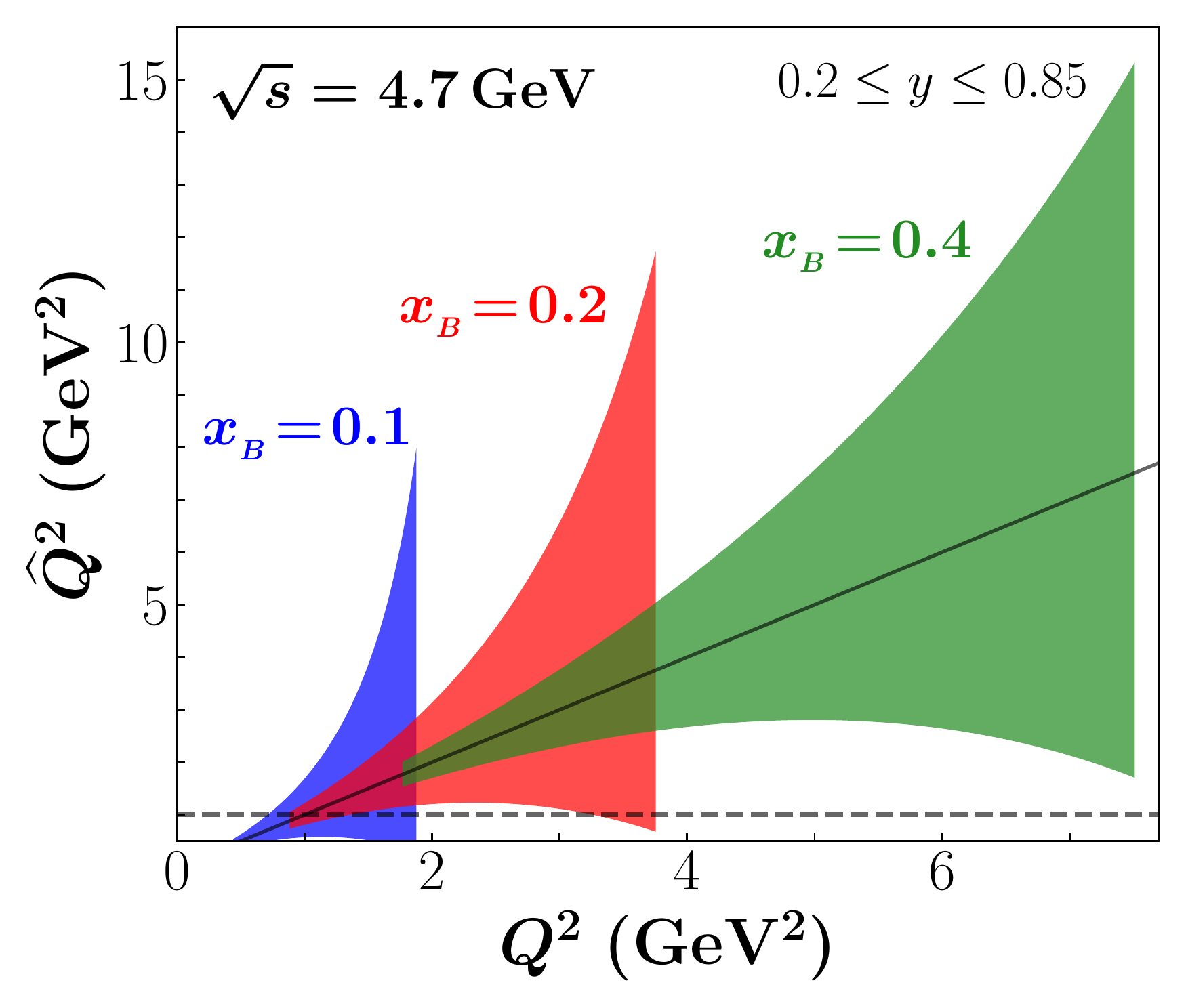}
    \includegraphics[width=0.42\textwidth]{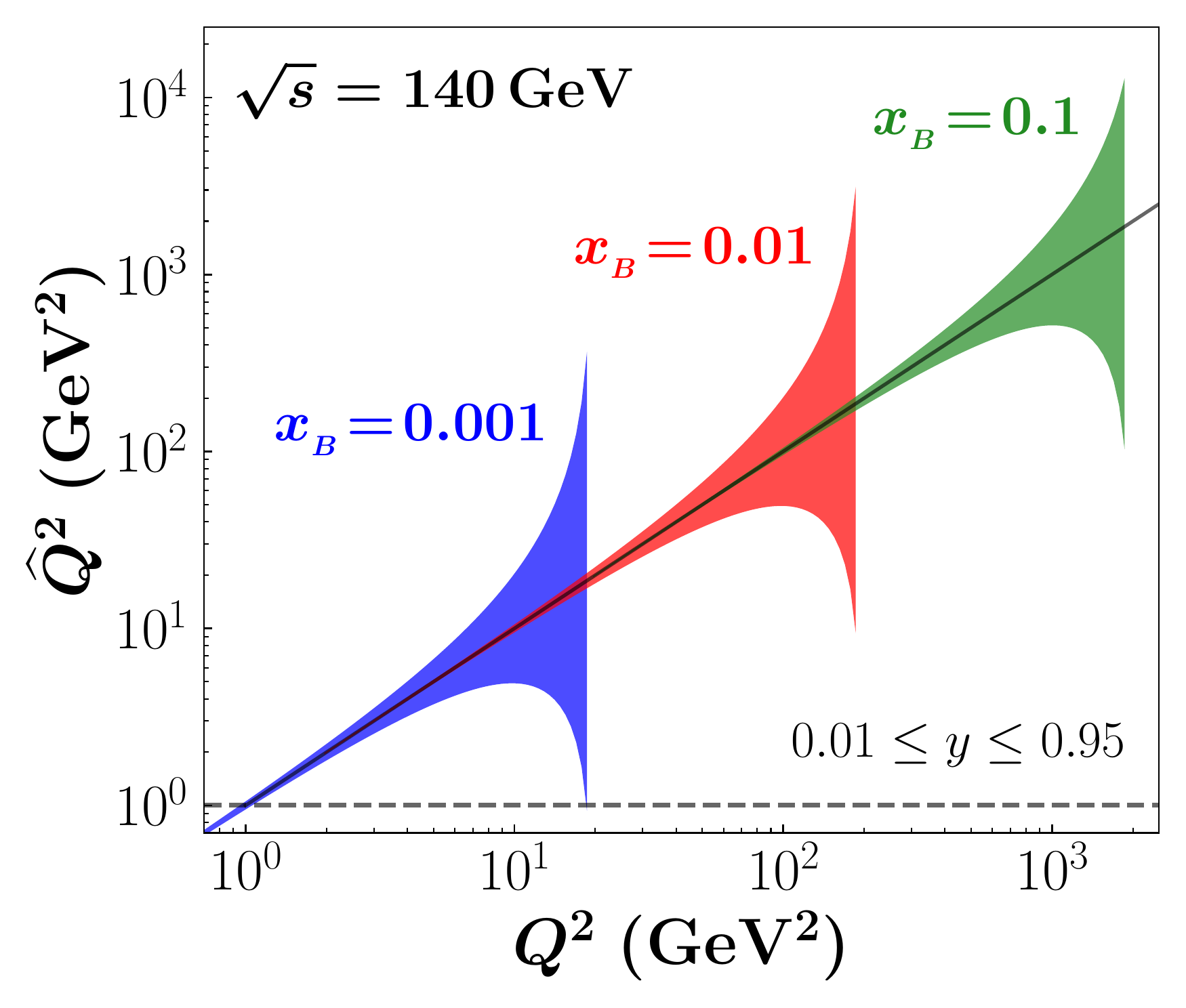}
    \caption{The range of the hard scale $\widehat{Q}^2$ experienced by the nucleon as a function of the measured scale $Q^2$, for fixed values of $\xb$, at Jefferson Lab ($\sqrt{s}=4.7$~GeV) {\bf (left)} and EIC ($\sqrt{s}=140$~GeV) {\bf (right)} kinematics. The straight black lines correspond to $\widehat{Q}^2=Q^2$.}
    \label{fig:qh2range}
\end{figure}

A further complication stems from the fact that photons are massless and the lepton mass is much smaller than the typical hard scale for QCD dynamics.
Consequently, RC factors based on fixed-order QED calculations are often infrared sensitive as $m_e \to 0$, involving infrared cutoff parameters, such as the total energy of soft photons in the treatment by Mo and Tsai~\cite{Mo:1968cg} or the minimum photon energy in the approach of Bardin and Shumeniko~\cite{Bardin:1976qa}. These parameters need to be tuned to the data.

As will be discussed in detail in the next section, the collision-induced QED radiation also leads to uncertainty in determining the photon-nucleon frame in which the produced hadron momentum, the hadronic plane, angular modulations and, most importantly, the TMD factorization of SIDIS, are defined.
Consequently, hadronic model dependence is inevitably introduced into attempts to derive RCs for SIDIS~\cite{Soroko:1989zt, Soroko:1991zr, Akushevich:2019mbz}.
In contrast, rather than searching for more reliable RC factors with which to extract the ideal $\sigma_{\rm Born}$ in eq.~(\ref{e.standard}) from the experimental cross section, $\sigma_{\rm obs}$, our proposed approach is a systematically improvable and reliable way to calculate the induced QED radiative contributions to all orders in powers of $\alpha$.
In analogy with the calculation of the induced QCD radiative contributions to the measured cross sections, our factorization approach organizes all-order contributions with respect to both QCD and QED, such as in \eref{dis-fac} for the inclusive DIS cross section. 
Instead of the RC approach of eq.~(\ref{e.standard}), our factorization approach can be schematically represented as
\begin{align}
    \sigma_{\rm obs}(\xb,Q^2) 
    &= \sigma_{\rm lep}^{\rm univ}(\mu^2;m_e^2)
    \otimes
    \sigma_{\rm had}^{\rm univ}(\mu^2;\Lqcd^2)
    \otimes
    \widehat{\sigma}_{\mbox{\scriptsize{IR-safe}}}(\hxb,\widehat{Q}^2,\mu^2) 
    \nonumber\\
    &\quad
    + {\cal O}\left(\frac{\Lqcd^2}{Q^2},\frac{m_e^2}{Q^2}\right),
\label{e.rc-fac}
\end{align}
where all infrared-sensitive contributions to the cross sections are either factorized into the universal leptonic and hadronic distribution or fragmentation functions, $\sigma_{\rm lep}^{\rm univ}$ and $\sigma_{\rm had}^{\rm univ}$, which are renormalization group improved with the factorization scale $\mu^2$, or neglected as power-suppressed corrections, and $\otimes$ represents the convolution over the respective leptonic and partonic momentum fractions.

The IR-safe and perturbatively calculable short-distance coefficient functions $\widehat{\sigma}_{\mbox{\scriptsize{IR-safe}}}$ depend on the ``true'' probing scales $\hxb$ and $\widehat{Q}^2$ for the colliding nucleon under the one-photon approximation, and can be systematically improved by higher-order contributions in powers of both $\alpha$ and $\alpha_s$.
As illustrated in figure~\ref{fig:qh2range}, for a given value of $Q^2$ the true probing scale $\widehat{Q}^2$ can be in the range
    $\widehat{Q}_{\rm min}^2 \leq \widehat{Q}^2 \leq \widehat{Q}_{\rm max}^2$,
where
\begin{equation}
    \widehat{Q}_{\rm min}^2 = Q^2\, \frac{(1-y)}{(1-\xb\, y)}
\qquad {\rm and} \qquad
    \widehat{Q}_{\rm max}^2 = Q^2\, \frac{1}{(1-y+\xb\, y)}
\end{equation}
are the minimum and maximum values.
To obtain a single $Q^2_{\rm true}$ value from the range of $\widehat{Q}^2$ that defines the QED RC factor $\Rqed$ in eq.~(\ref{e.standard}), one must model the colliding nucleon's response at different values of $\widehat{Q}^2$, and such modeling could impact the quantity itself that we wish to extract from the measured cross sections. 
In addition, the Bjorken scaling variable in eq.~(\ref{e.hat-variables}),
    $\hxb = \xb\, \xi\, y/(\xi\zeta + y -1)$,
ranges between its minimum value, $\hxb^{\rm min} = \xb$, and its maximum value, $\hxb^{\rm max} = 1$. 
With the collision-induced QED radiation, for given values of $\xb$ and $Q^2$ from the measured lepton and nucleon momenta $\l$, $\lp$, and $P$, we actually probe the colliding nucleon over a much wider kinematic region of
    $\hxb \in [\hxb^{\rm min}, \hxb^{\rm max}]$ and 
    $\widehat{Q}^2 \in [\widehat{Q}_{\rm min}^2, \widehat{Q}_{\rm max}^2]$.

%========================
\begin{figure}[t]
    \centering
    \includegraphics[width=0.6\textwidth]{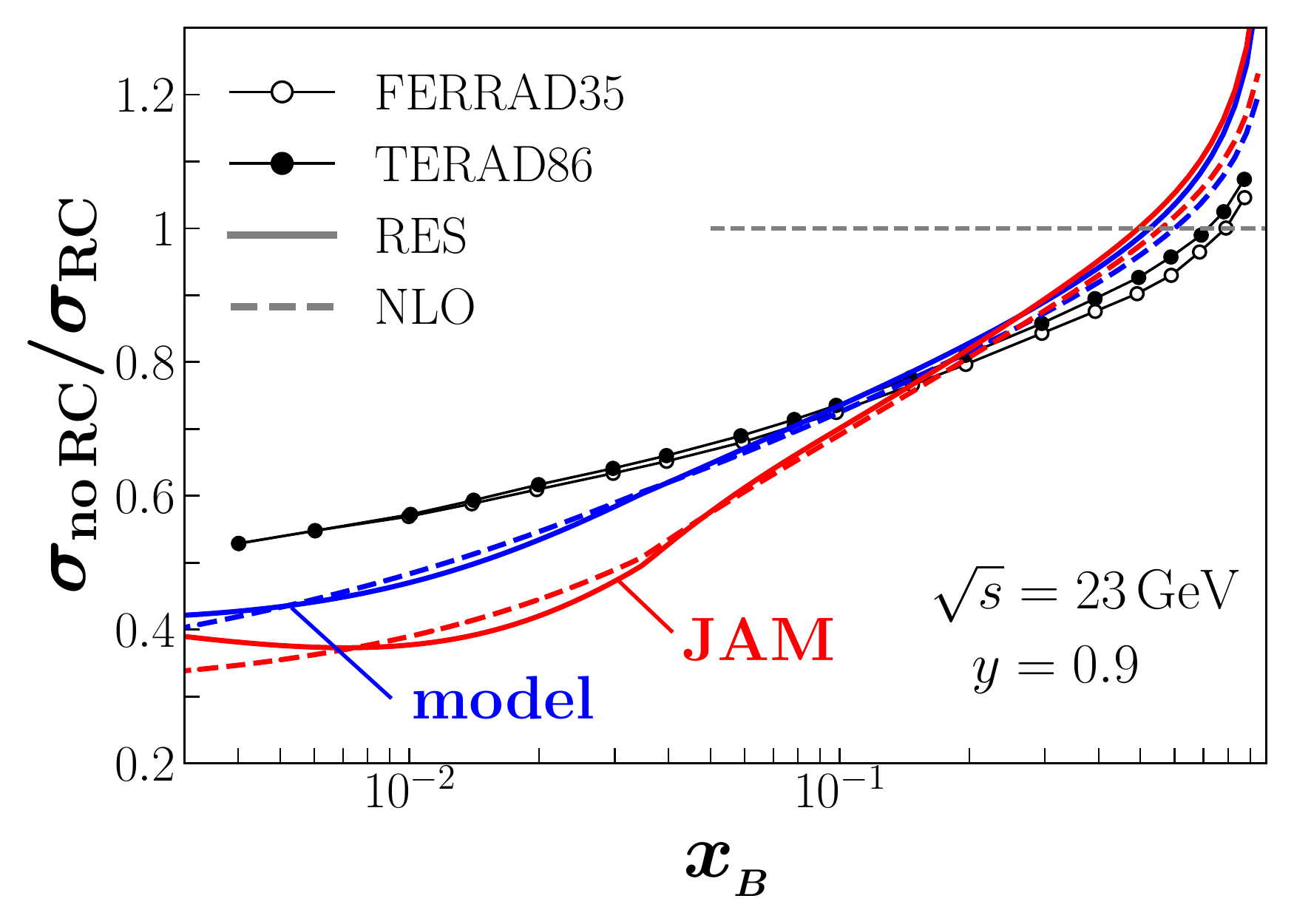}
    \caption{Comparison of our factorized results for the ratio $\sigma_{\rm no\, RC}/\sigma_{\rm RC}$ versus $\xb$ with those from Ref.~\cite{Badelek:1994uq} using the FERRAD35 and TERAD86 codes for the Mo-Tsai~\cite{Mo:1968cg} and Bardin {\it et al.}~\cite{Akhundov:1985qu} schemes, respectively, at matching kinematics ($\sqrt{s}=23$~GeV, $y=0.9$). Our factorized results for the resummed (RES, solid lines) and NLO (dashed lines) calculations are computed using PDFs from the JAM global QCD analysis~\cite{Sato:2019yez} (red lines) and using a simple model (blue lines, see text).}
    \label{fig:compare}
\end{figure}
%========================

In Fig.~\ref{fig:compare} we  show the effect of radiation on the ratio of inclusive $ep$ cross sections as in Fig.~\ref{fig:idis-y}, but inverted to match the definition used in Ref.~\cite{Badelek:1994uq}.  
To demonstrate the dependence of the traditional method of calculating QED contributions (as an RC factor applied to the Born term) on the hadron structure input, we show the inclusive ratio for two different sets of proton PDFs, namely, PDFs from the JAM global QCD analysis~\cite{Sato:2019yez} and using a simple model 
    $x f_{i/p}(x) \sim x^a (1-x)^b$.
For the latter, we choose the shape parameters 
    $a=0.5$ and $b=3$ for valence quarks, 
    $a=-0.08$ and $b=7$ for sea quarks, and 
    $a=-0.08$ and $b=5$ for gluons,
with the valence distributions normalized to 2 (1) for $u$ ($d$) quarks, and the sea quark and gluon distributions normalized with momentum fractions
    $\langle x \rangle_i = \{ 0.030, 0.036, 0.016, 0.005, 0.41 \}$
for $i = \{ \bar u, \bar d, s=\bar{s}, c=\bar{c}, g \}$.
Note that for the traditional method, the RC factor at ($x_B,Q^2)$ is sensitive to the input of hadron structure function for $x_B \in [x_B,1)$ and $Q^2\in [\widehat{Q}^2_{\rm min}, \widehat{Q}^2_{\rm max}]$.
For comparison, we also show in Fig.~\ref{fig:compare} the results from Ref.~\cite{Badelek:1994uq} using the Mo-Tsai~\cite{Mo:1968cg} and Bardin {\it et al.}~\cite{Akhundov:1985qu} schemes at matching kinematics ($\sqrt{s}=23$~GeV, $y=0.9$).
The comparison clearly shows the sensitivity of the ratio $\sigma_{\rm RC}/\sigma_{\rm no\, RC}$ to the hadronic structure input, which is problematic given that the aim is to extract this very structure from the data.

As discussed in \ssref{qed-rc}, the novelty of our approach is the fact that we do not need to assume any prior knowledge about the hadronic structures, provided that the power corrections are suppressed. 
The exact demarcation of the phase space where our proposed factorization approach is applicable cannot be determined {\it a priori}, but can be found through global analysis involving multiple high-energy reactions with overlapping partonic kinematics, which can ultimately confirm and validate the universality of the inferred structures.

In our proposed new approach to the QED radiation (\ref{e.rc-fac}), all collision-induced QED contributions to the measured cross sections are organized such that all leading power infrared-sensitive contributions are included into the universal LDFs and LFFs.
All leading power infrared-safe contributions are included in the calculable hard parts, and the rest can be neglected or further improved as power corrections. 
Although QED radiation changes the momentum of the exchanged hard photon and introduces uncertainty in controlling the ``true'' hard probe, our factorization formalism as in \eref{dis-fac} provides the minimum value of the probing scale, $\widehat{Q}_{\rm min}^2$. 
As shown in figure~\ref{fig:ps}, the collision-induced QED radiation does remove some phase space from the DIS regime, particularly when $\xb$ is small or $y$ is large, which corresponds to more phase space for radiation.

Most importantly, in our factorization approach to the collision-induced QED radiative contributions, neither the universal infrared-sensitive LDFs and LFFs, nor the calculable QED hard parts depend on the nonperturbative hadron structure, such as PDFs, fragmentation functions, or TMDs that we aim to extract.
That is, our factorization approach does not require any modeling of hadronic physics and is not sensitive to infrared cutoffs, which are the two main uncertainties of existing approaches to treating induced QED radiation via RC factors.

As with all factorization approaches, on the other hand, we do not know exactly the size of the power corrections or the precise functional forms of the universal infrared-sensitive LDFs and LFFs in \eref{dis-fac}.
Although we could have a better control on LDFs and LFFs in QED than for corresponding partonic functions in QCD, the global analysis of all possible data is still needed to identify regions where the process-dependent power corrections are small, and one can demonstrate the universality of the infrared-sensitive functions.

%%%%%%%%%%%%%%%%%%%%%%%%%%%%%%%%%%%%%%%%%%%%%%%%%%%%%%%%%%
\section{Factorized formalism for semi-inclusive DIS with QED}
\label{s.sidis-fac}

\begin{figure}[t]
    \centering
    \includegraphics[width=0.3\textwidth]{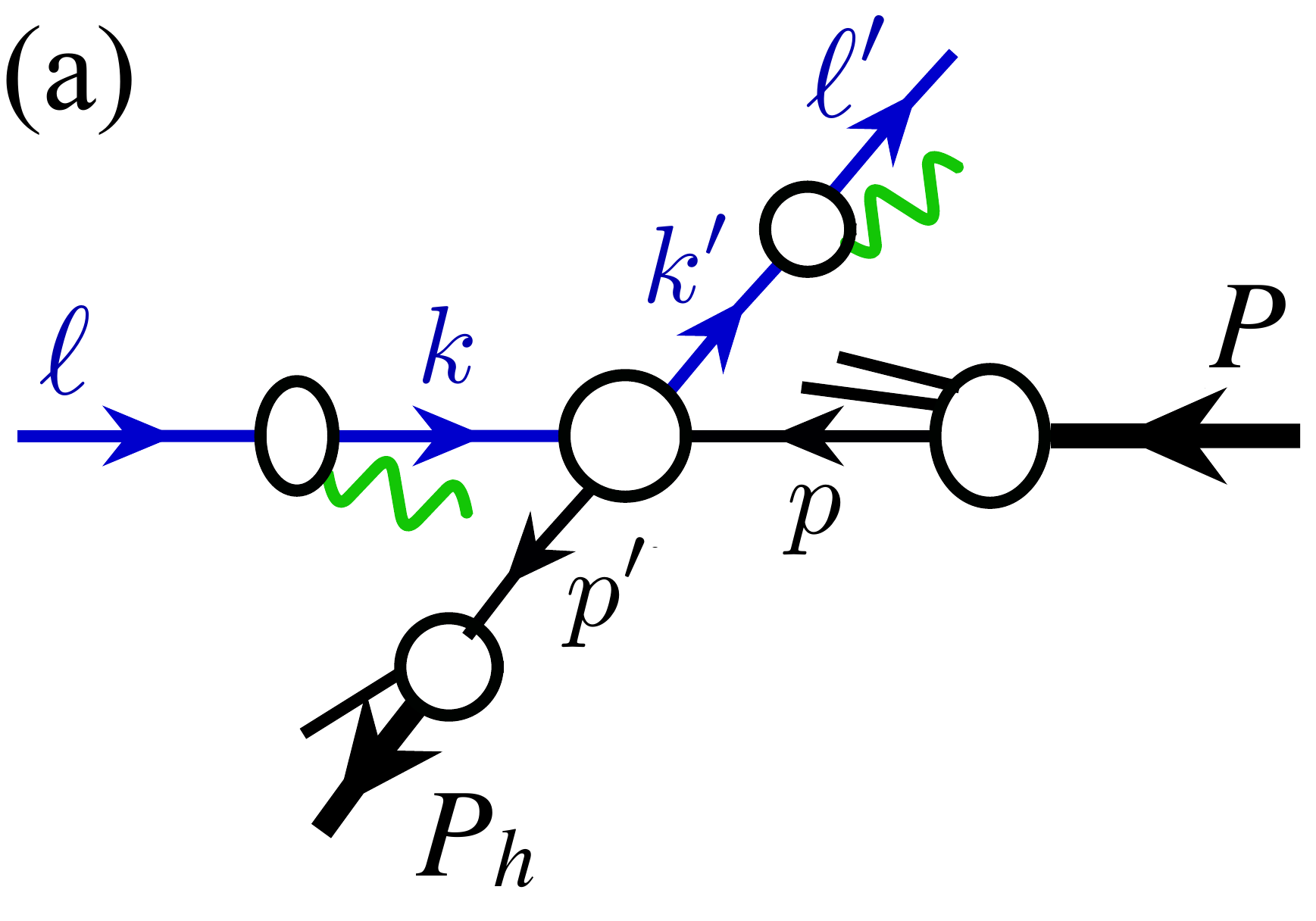}
    \hspace{1cm}
    \includegraphics[width=0.3\textwidth]{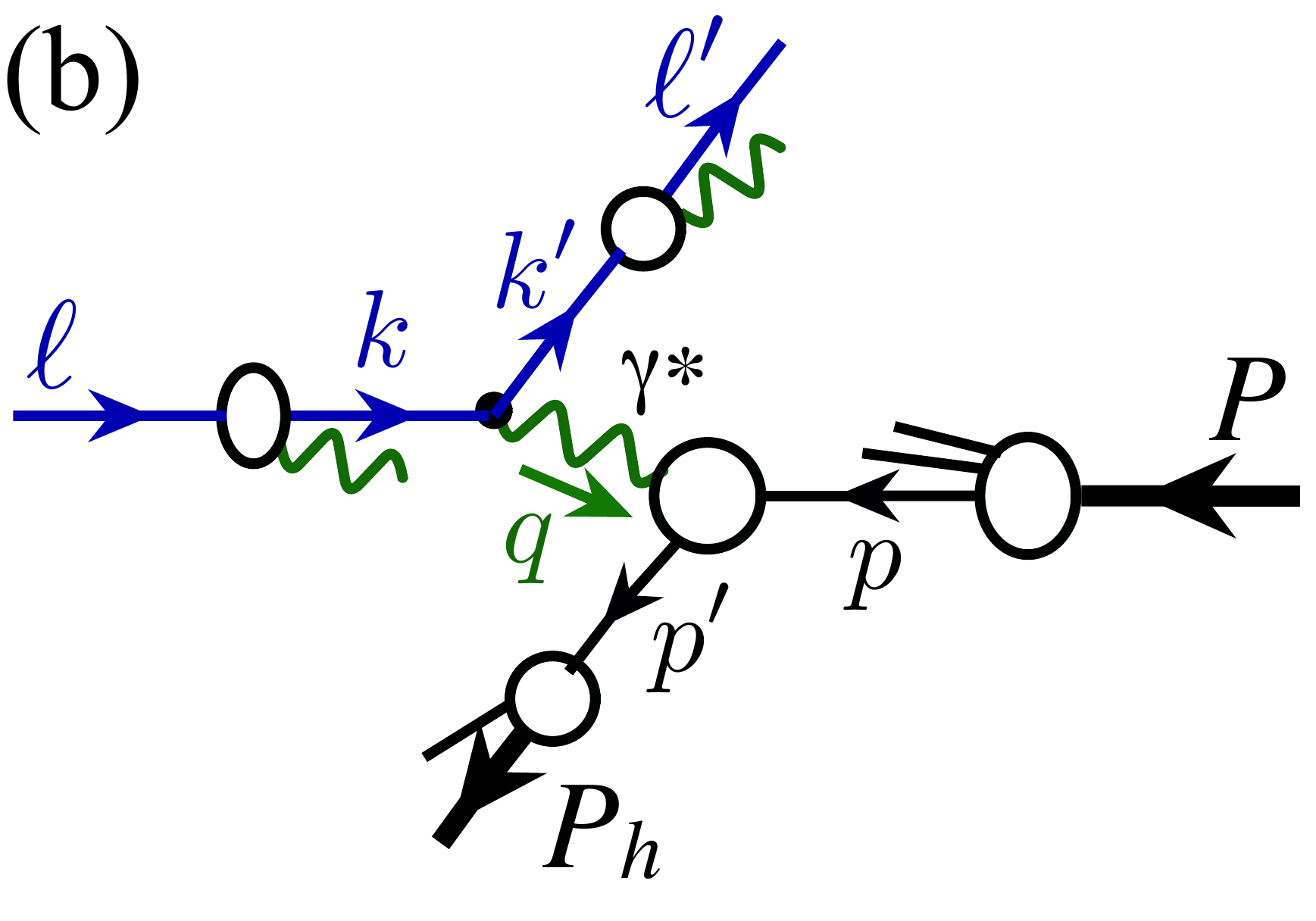} 
    \caption{Sketch of {\bf (a)} the SIDIS process $e(\ell)+N(P) \to e(\ell')+h(P_h)+X$, and {\bf (b)} SIDIS in the one-photon exchange approximation.}
\label{f.sidis}
\end{figure}

In this section, we expand our combined QED and QCD factorization approach to contributions to the cross section for the SIDIS process,
    $e(\l,\lambda_\l) + N(P,S) \to e(\lp) + h(P_h) + X$, 
for the semi-inclusive production of a hadron $h$ with four-momentum $P_h$ in coincidence with the scattered lepton $\lp$.
As for the case of inclusive DIS in eq.~(\ref{e.dis}) of section~\ref{s.dis-fac}, the SIDIS cross section can be formally written in terms of the square of its scattering amplitude, $M_{\l P\to \lp P_h X}$,
\begin{eqnarray}
\label{e.sidis} 
\diff\sigma_{\ell P \to \ell' P_h X}
= \frac{1}{2s} 
  \big| M_{\ell P\to \ell' P_hX} \big|^2\, 
  \diff{\rm PS},
\end{eqnarray}
where for convenience the dependence on the electron and nucleon polarization ($\lambda_\l$ and $S$, respectively) is suppressed.  
In analogy with the inclusive DIS case, we consider SIDIS as an inclusive production of a large-$\ell'_T$ lepton plus a large-$P_{hT}$ hadron (or jet) in lepton-nucleon collisions, as illustrated in figure~\ref{f.sidis}(a).
In the plane transverse to the lepton-nucleon collision axis, the regime where the transverse momenta $\bm{\ell}'_T$ and $\bm{P}_{hT}$ are almost back-to-back, namely,
    $\PTbar \equiv |\bm{\ell}'_T - \bm{P}_{hT}|/2 
    \gg |\bm{\ell}'_T + \bm{P}_{hT}| \equiv \pTbar$, 
is suited for TMD factorization, while the region where $\PTbar \sim \pTbar$ is suited for collinear factorization.

% .......................................................................
\subsection{Collinear factorization for semi-inclusive DIS with QED contributions}
\label{ss.sidis-fac-co}

With an exchange of a single hard scale, $\PTbar \sim \pTbar \gg \Lqcd$, the invariant mass of any pair of initial-state colliding particle and final-state observed particle momenta is a hard scale, whose absolute value is much larger than $\Lqcd$.
Applying the arguments in ref.~\cite{Collins:1981ta}, we can factorize the SIDIS cross section in the regime where $\PTbar \sim \pTbar \gg \Lqcd$ as
\begin{eqnarray}
E_{\lp} E_{P_h}\frac{\diff \sigma_{\l P\to \lp P_h X}}{\diff^3 \lp \diff^3 P_h }
&=& \frac{1}{2s} \sum_{ijab} 
\int_{\zeta_{\rm min}}^1 \frac{\diff\zeta}{\zeta^2} 
\int_{\xi_{\rm min}}^1 \frac{\diff\xi}{\xi}\, 
D_{e/j}(\zeta,\mu^2)\, f_{i/e}(\xi,\mu^2)
\notag\\
& &
\times
\int_{z_{\rm min}}^1 \frac{\diff z}{z^2} 
\int_{x_{\rm min}}^1 \frac{\diff x}{x}\, 
D_{h/b}(z,\mu^2)\, f_{a/N}(x,\mu^2)\,
\notag\\
& &
\times\,
\widehat{H}_{ia\to jbX}(\xi,x,\l_T'/\zeta,P_{hT}/z,\mu^2)\
+\ \cdots ,
\label{e.sidis-fac-co}
\end{eqnarray}
where the indices $i$, $j$, $a$, $b$ include all QED and QCD particles, and the ellipsis represents power corrections suppressed by inverse powers of $\ell_T'$ and $P_{hT}$, or $\PTbar \sim \pTbar$, defined in the lepton-nucleon frame. 
The lower limits of the integrations depend on the collision energy $\sqrt{s}$ and the observed lepton and hadron momenta, $\lp$ and $P_h$, respectively, and are given in eqs.~(\ref{e.minvalues}) in the previous section.

The functions $f_{i/e}(\xi,\mu^2)$, $D_{e/j}(\zeta,\mu^2)$ and $f_{a/N}(x,\mu^2)$ in eq.~(\ref{e.sidis-fac-co}) are LDFs, LFFs and PDFs, respectively, and are the same universal functions as those in eq.~(\ref{e.dis-fac}) for inclusive DIS. 
The function $D_{h/b}(z,\mu^2)$ in eq.~(\ref{e.sidis-fac-co}) is the collinear fragmentation function (FF) to the observed hadron $h$ of momentum $P_h$ from a parton $b$ of momentum $p'$, which is defined in ref.~\cite{Collins:1981uw} for $b = q, \bar{q}, g$ as a function of momentum fraction $z = P_h^+/p'^+$.
The definition is straightforwardly extended to the case where $b$ is a lepton or a photon, with the quark and gluon fields replaced by the corresponding lepton and photon fields.

The short-distance hard parts $\widehat{H}_{ia\to jbX}$ in eq.~(\ref{e.sidis-fac-co}) are infrared safe and perturbatively calculable in QCD and QED.
These are equal to the partonic cross section for the sub-process
    $i(k) + a(p) \to j(k') + b(p') + X(k+p-k'-p')$,
where all perturbative collinear divergences along the momentum directions of the active particles, $k$, $p$, $k'$, and $p'$, respectively, removed and resummed into the corresponding LDFs, LFFs, PDFs and FFs, respectively.
The factorization formalism in eq.~(\ref{e.sidis-fac-co}) also provides a prescription for evaluating the short-distance hard parts $\widehat{H}_{ia \to jbX}$ by applying the factorization formalism to lepton or parton states order-by-order in perturbation theory.
With the requirement that $\PTbar \sim \pTbar \gg \Lqcd$, the leading order contribution to $\widehat{H}_{ia\to jbX}$ is given by the $2 \to 3$ scattering processes. 
For example, by applying the factorization formalism in (\ref{e.sidis-fac-co}) to SIDIS with the nucleon $N$ and hadron $h$ each replaced by a quark, 
    $e(k) + q(p) \to e(k') + q(p') + g(k+p-k'-p')$
at the lowest order, one can derive $\widehat{H}^{(2,1)}_{eq \to eqg}$ at order ${\cal O}(\alpha^2\,\alpha_s)$.

% .......................................................................
\subsection{TMD factorization for semi-inclusive DIS with QED contributions}
\label{ss.sidis-fac-tmd}

When $\PTbar \gg \pTbar$, the transverse momentum imbalance between the observed lepton of momentum $\lp$ and hadron of momentum $P_h$ becomes sensitive to the infrared-sensitive collinear radiation from both QCD and QED.
In this case, the collinear factorization for the SIDIS cross section in this kinematic regime is no longer reliable. 
A TMD factorization is instead needed to take into account the transverse momentum dependence of the active particles (partons or leptons) probed by the hard collisions. 
The transverse momentum of a colliding particle (a parton or a lepton) is generated by the induced radiation of the hard collision plus the particle's intrinsic transverse momentum in the bound hadron state, if the particle is a parton.
Therefore, a TMD factorization for the SIDIS process should take into account the active particles' transverse momentum generated by both collision-induced QCD and QED showers (or radiation).

However, a full TMD QED and QCD factorization for the momentum imbalance of the observed lepton and hadron in SIDIS is likely to be violated~\cite{Collins:2007nk}. 
Instead of the full TMD factorization for all four observed particles (the two leptons and two hadrons) in SIDIS, in ref.~\cite{Liu:2020rvc}, we proposed a hybrid factorization approach, with the collinear factorization for the two leptons and TMD factorization for the two hadrons for SIDIS in the two-scale regime. 
The hybrid factorization approach was motivated by the observation that the momentum imbalance between the observed lepton and hadron in SIDIS is dominated by the transverse momentum of active partons from the two hadrons in all relevant collision energies. 
To justify this hybrid factorization for SIDIS, we demonstrate below that the transverse momentum broadening generated by the collision-induced radiation for a ``point-like'' lepton is much smaller than the typical transverse momentum of an active parton in a colliding hadron, and argue that such hybrid factorization should be valid up to power corrections.

\begin{figure}[t]
\centering
    \includegraphics[width=0.25\textwidth]{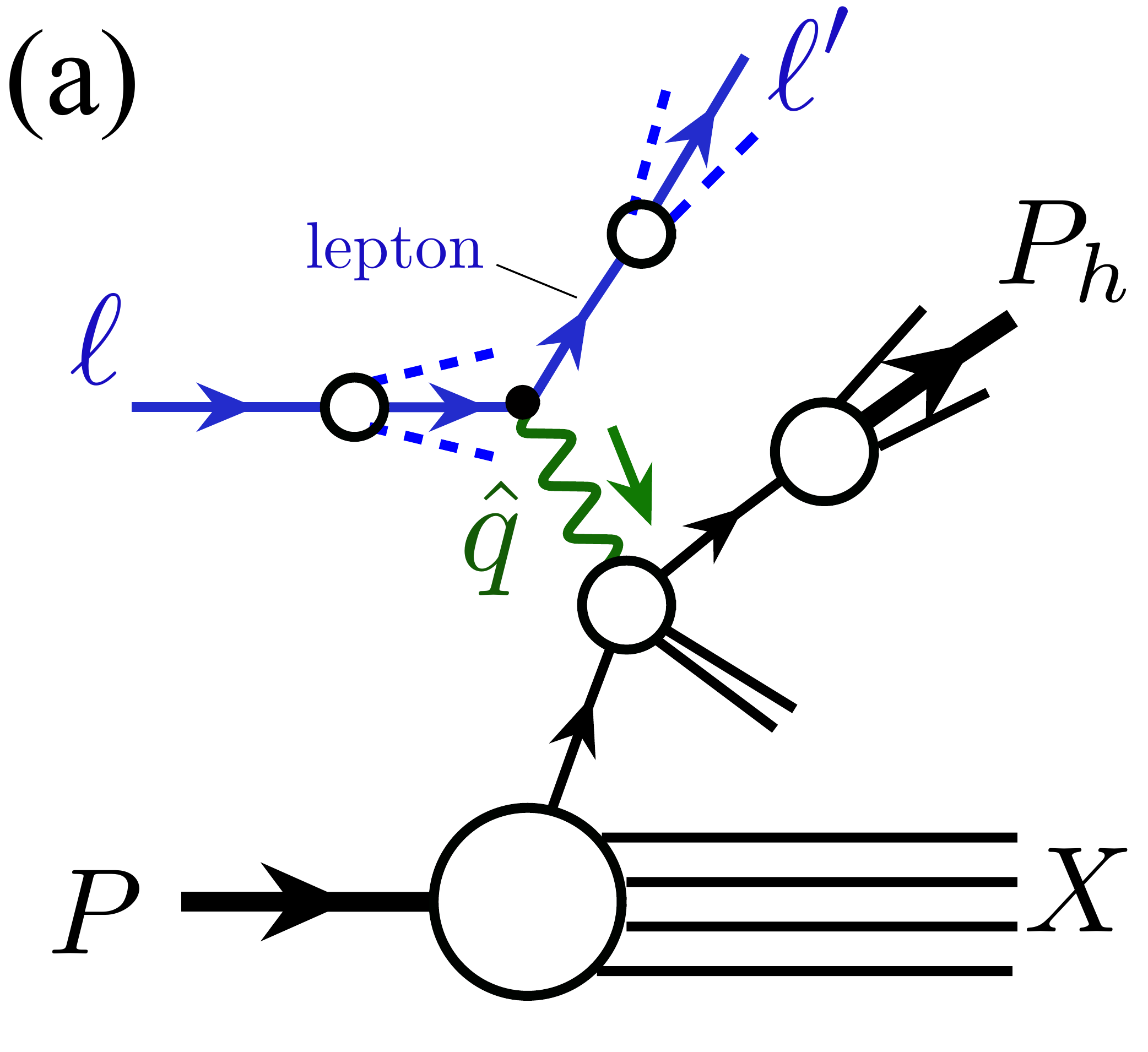}
    \hspace{1cm}
    \includegraphics[width=0.25\textwidth]{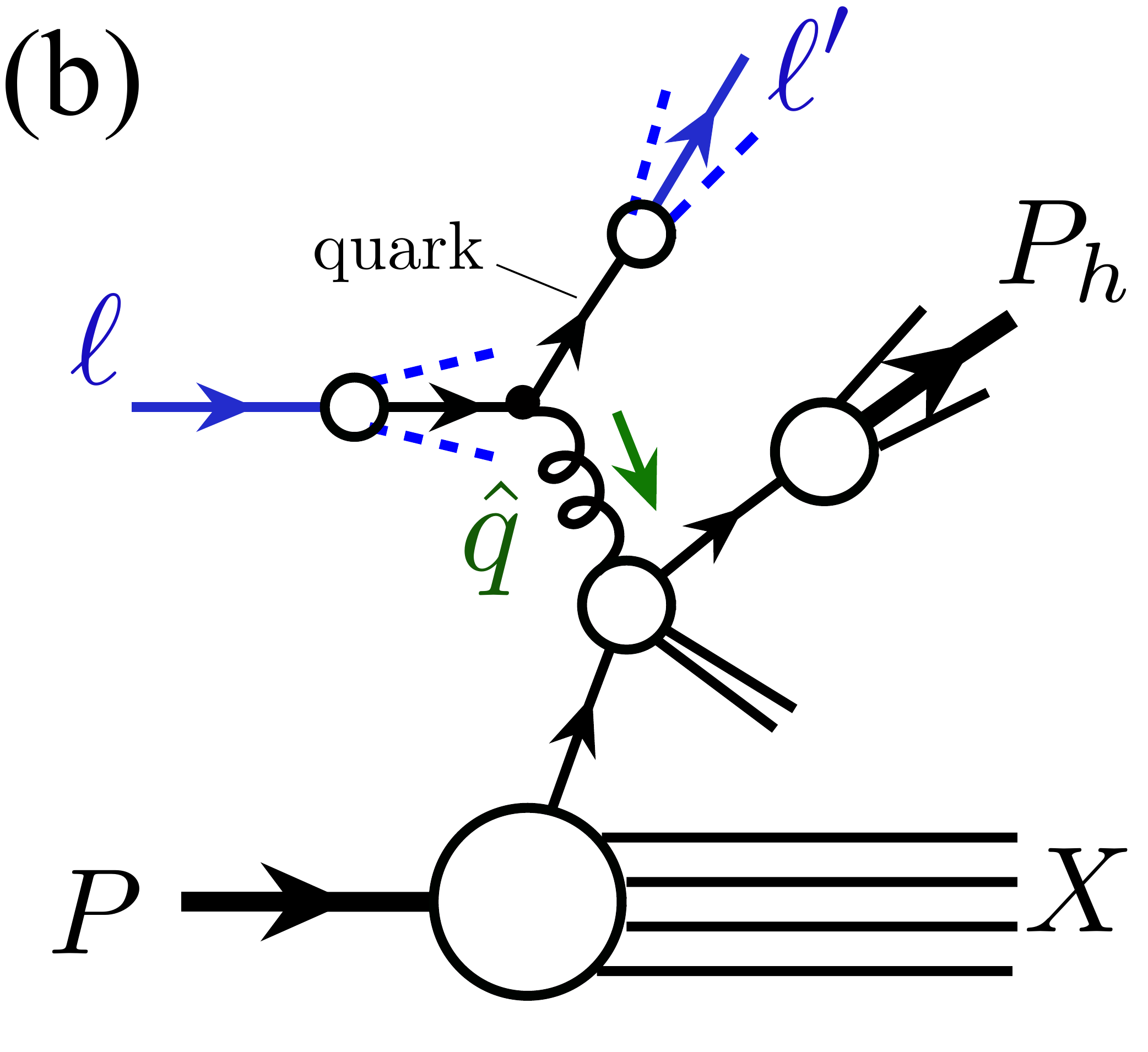} 
    \caption{Sketch for sample scattering amplitudes of SIDIS with {\bf (a)}~one-photon exchange, and~{\bf (b)}~one-gluon exchange.}
\label{f.sidis-amp}
\end{figure}

With a sufficiently large momentum transfer between the leptons and hadrons, the one-photon approximation, as shown in figure~\ref{f.sidis}(b), is often adopted for evaluating the SIDIS cross sections. 
To ensure this large momentum transfer, we require the transverse momentum of the scattered lepton $\lp$ and the observed hadron (or jet) $P_h$ in the lepton-nucleon frame to be the hard scales, with $\ell'_T \gg \Lqcd$ and $P_{hT} \gg \Lqcd$.
However, as an inclusive production of the lepton $\ell'$ and hadron $P_h$, this large momentum transfer could also be achieved by exchanging a virtual parton, such as a gluon, as in figure~\ref{f.sidis-amp}(b).
Here, the colliding lepton radiates a photon that turns into a quark-antiquark pair, and the quark then undergoes the hard scattering with the colliding hadron via the exchange of a virtual gluon.
As discussed in section~\ref{ss.ldf-lff} in connection with the contribution from the subprocess in figure~\ref{f.dis-amp}(b), the type of subprocess in figure~\ref{f.sidis-amp}(b) is likely to be further suppressed if we require the observed lepton of momentum $\ell'$ not to have strong hadronic activity around it.
In the rest of this paper, we take the one-photon approximation to include only the scattering amplitude in figure~\ref{f.sidis-amp}(a) for the SIDIS cross section, leaving the study of SIDIS beyond one-photon exchange for future work.

\begin{figure}[b]
\centering
    \includegraphics[width=0.35\textwidth]{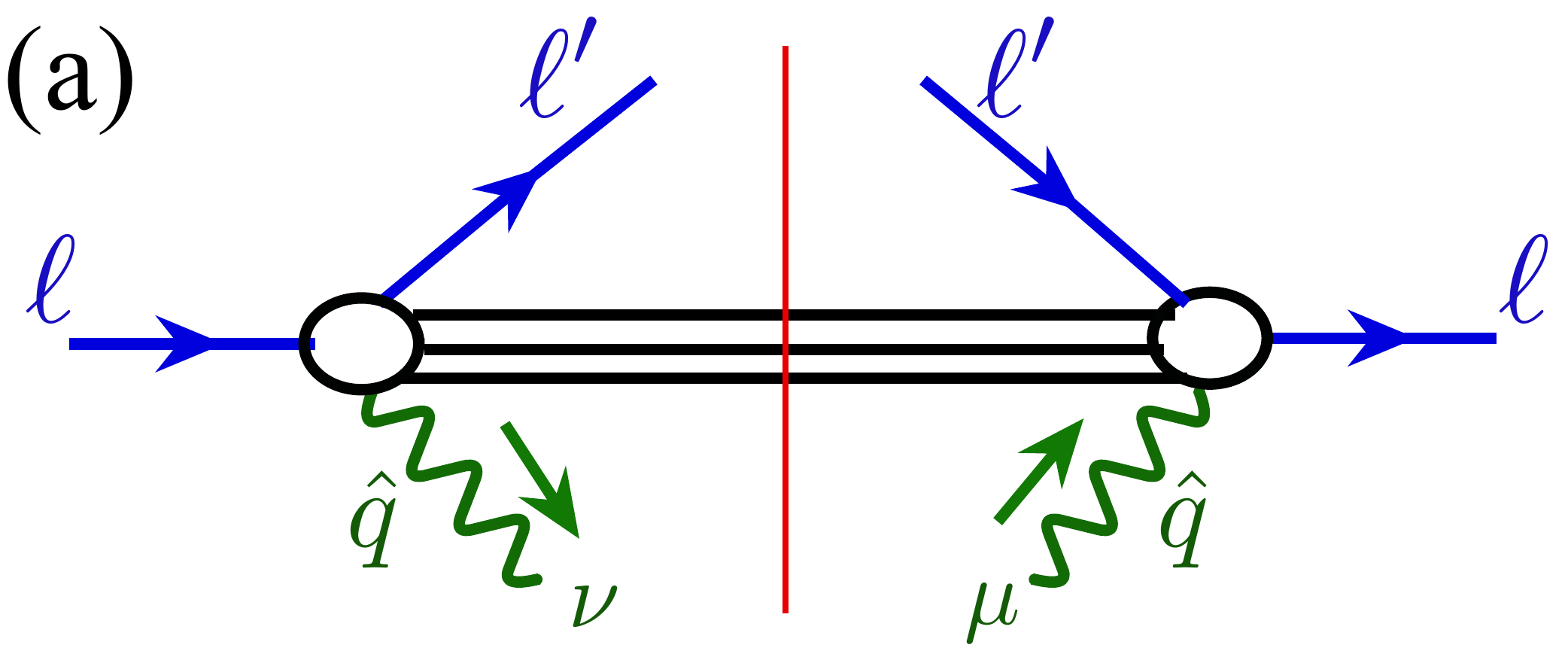}
    \hspace{1cm}
    \includegraphics[width=0.35\textwidth]{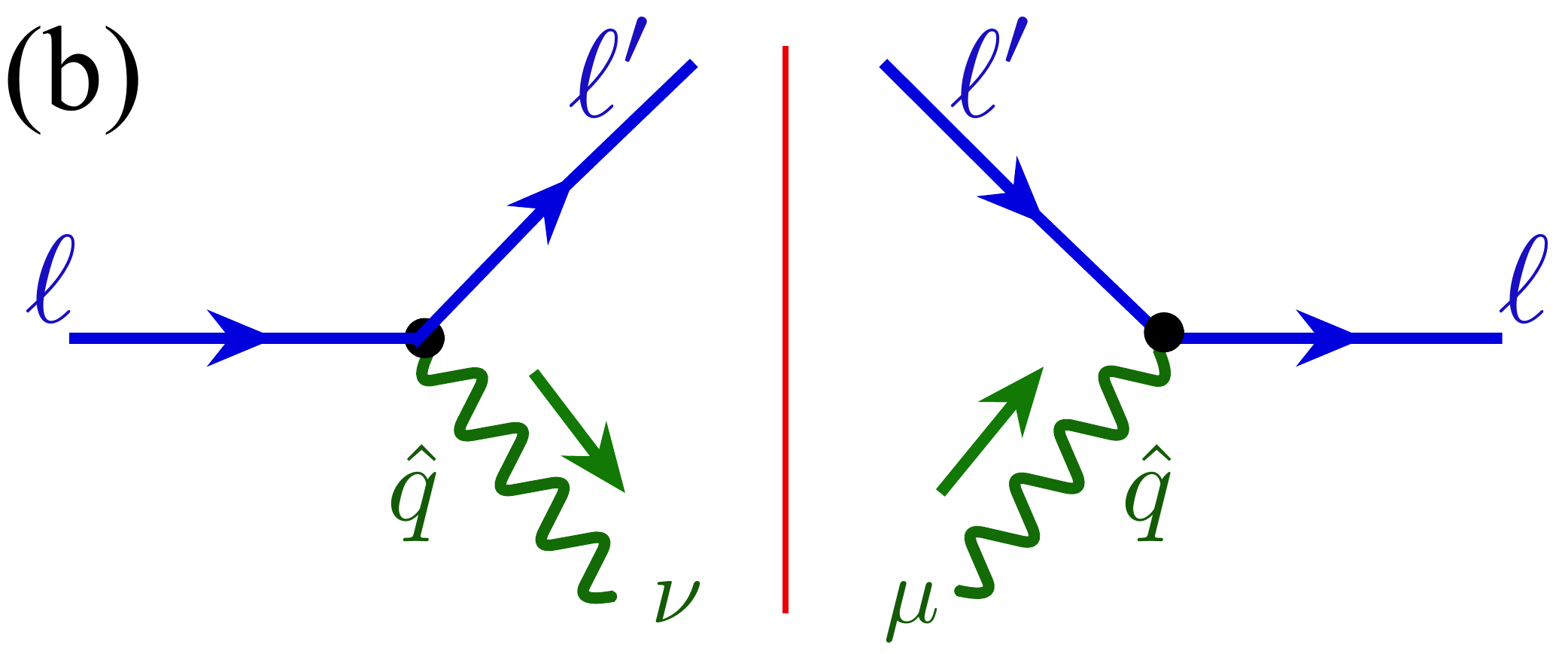} 
    \caption{Illustration of {\bf (a)}~the general leptonic tensor, $\widetilde{L}^{\mu\nu}(\l,\lp,\hat{q})$, and {\bf (b)} the lowest-order contribution to $\widetilde{L}^{\mu\nu}(\l,\lp,\hat{q})$.}
\label{f.sidis-lmn}
\end{figure}

With the approximation of one-photon exchange, we can write the SIDIS cross section in eq.~(\ref{e.sidis}) in terms of leptonic, $\widetilde{L}^{\mu\nu}$, and semi-inclusive hadronic, $\widetilde{W}_{\mu\nu}$, tensors,
\begin{eqnarray}
\label{e.sidis-invariant} 
E_{\lp} E_{P_h} 
\frac{\diff^6\sigma_{\l P\to \lp P_hX}}{\diff^3\lp\, \diff^3P_h}\,
\approx\,
\frac{\alpha^2}{2s} \int \diff^4\hat{q}
\left(\frac{1}{\hat{q}^2}\right)^2\
\widetilde{L}^{\mu\nu}(\l,\lp,\hat{q})\,
\widetilde{W}_{\mu\nu}(\hat{q}, P, P_h, S),
\end{eqnarray}
where $\hat{q}$ is the momentum carried by the exchanged virtual photon.
The leptonic tensor $\widetilde{L}^{\mu\nu}(\l,\lp,\hat{q})$, sketched in figure~\ref{f.sidis-lmn}(a), is defined as
\begin{eqnarray}
\label{e.sidis-lmn} 
\widetilde{L}^{\mu\nu}(\l,\lp,\hat{q})\,
&\equiv&\,
\sum_{X_L}\int 
\prod_{i\in X_L}\frac{\diff^3k_i}{(2\pi)^3 2E_i}\,
\delta^{(4)}\Big(\l-\lp-\hat{q}
-\sum_{i\in X_L} k_i \Big)
\nonumber\\
& &\times
\langle \l| j^\mu(0)|\lp X_L \rangle
\langle \lp X_L| j^\nu(0)|\l \rangle,
\end{eqnarray}
where the electromagnetic current $j^\mu(0)$ couples to leptons, and the sum over all final states $X_L$ includes radiation associated with the incoming 
and scattered leptons.

\begin{figure}[t]
\centering
    \includegraphics[width=0.35\textwidth]{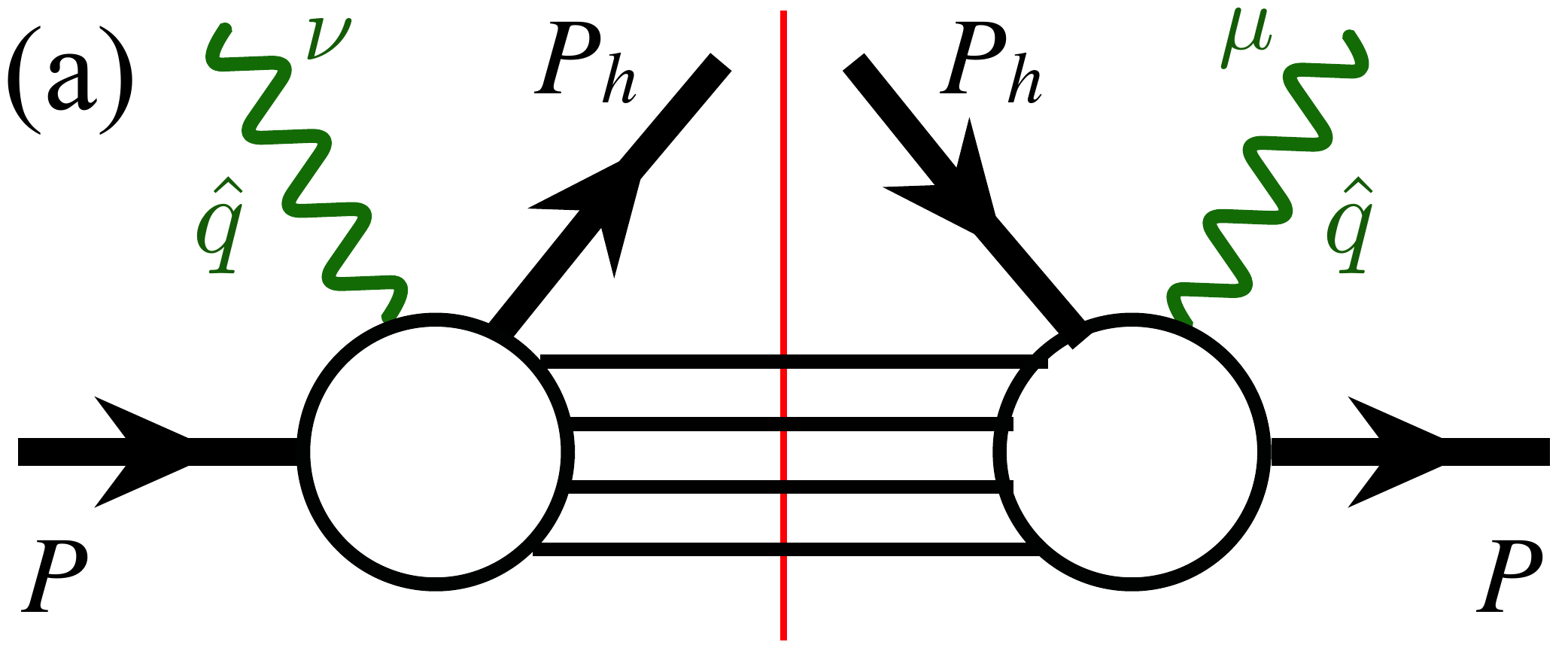}
    \hspace{1cm}
    \includegraphics[width=0.35\textwidth]{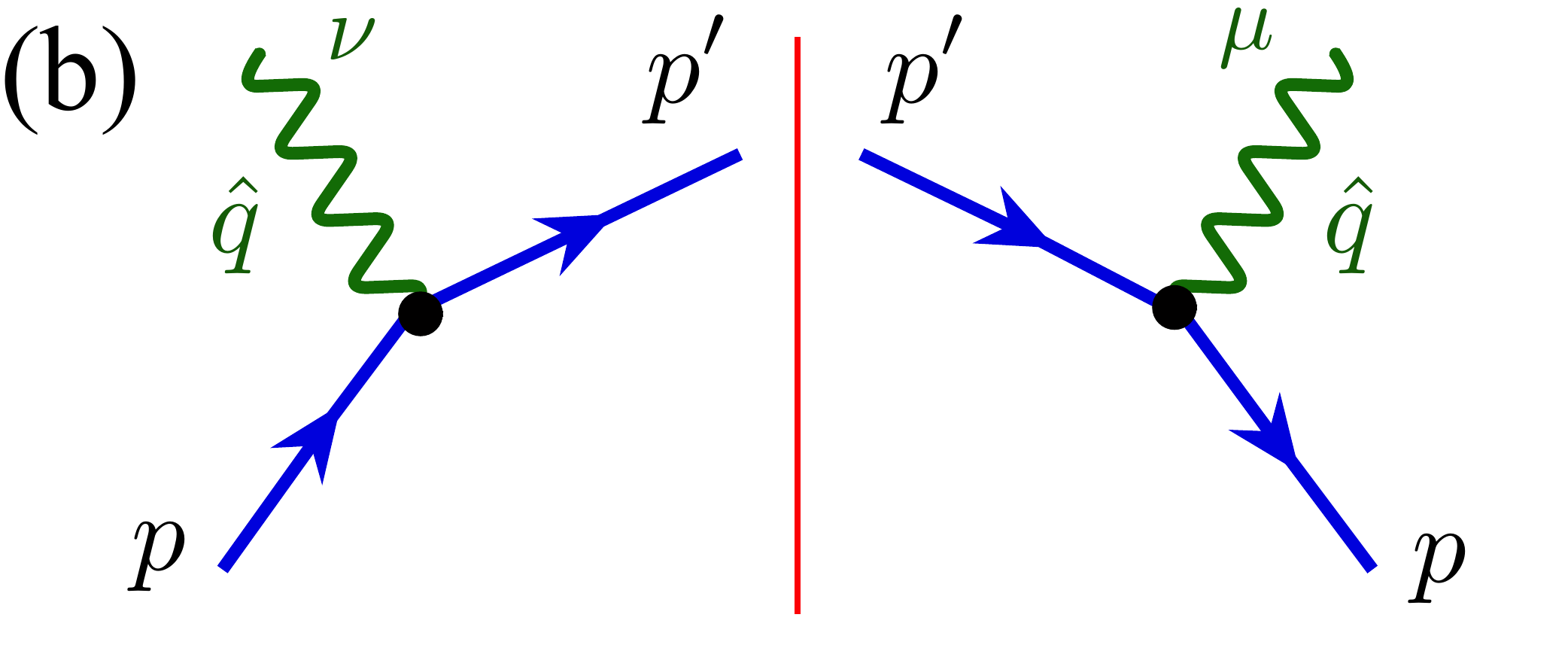} 
    \caption{Illustration of {\bf (a)}~the semi-inclusive hadronic tensor, $\widetilde{W}_{\mu\nu}(\hat{q},P,P_h,S)$, and {\bf (b)} the lowest-order contribution to $\widetilde{W}_{\mu\nu}(\hat{q},P,P_h,S)$.}
\label{f.sidis-wmn}
\end{figure}

The corresponding semi-inclusive hadronic tensor $\widetilde{W}_{\mu\nu}(\hat{q},P,P_h,S)$, representing the colliding nucleon of momentum $P$ and observed hadron of momentum $P_h$, is sketched in figure~\ref{f.sidis-wmn}(a).
It is defined similarly to the hadronic tensor for inclusive DIS in the one-photon approximation in eq.~(\ref{e.dis-wmn}),
\begin{eqnarray}
\label{e.sidis-wmn} 
\widetilde{W}_{\mu\nu}(\hat{q},P,P_h,S)
&=& \sum_{X_h}\int
\prod_{i\in X_h}\frac{\diff^3p_i}{(2\pi)^3 2E_i}\,
\delta^{(4)}\Big(\hat{q} + P - P_h - \sum_{i\in X_h} p_i\Big)
\nonumber\\
& & \times\
\langle P,S | J_\mu(0) | P_h X_h \rangle
\langle P_h X_h | J_\nu(0) | P,S \rangle,
\end{eqnarray}
where the electromagnetic current $J_\mu(0)$ couples to quarks (or charged leptons), and the sum is taken over all final states $X_h$.

The lowest order contribution to the leptonic tensor for an unpolarized lepton of momentum $\l$ in eq.~(\ref{e.sidis-lmn}), as sketched in figure~\ref{f.sidis-lmn}(b), is given by 
\begin{eqnarray}
\label{e.sidis-lmn0} 
\widetilde{L}^{\mu\nu(0)}(\l,\lp,\hat{q})
= 2\big( \l^\mu \lp^\nu + \lp^\mu\l^\nu - \l\cdot \lp g^{\mu\nu} \big)\,
\delta^{(4)}(\l-\lp-\hat{q})\, .
\end{eqnarray}
By substituting $\widetilde{L}^{\mu\nu(0)}(\l,\lp,\hat{q})$ into eq.~(\ref{e.sidis-invariant}), and using the $\delta^{(4)}(\l-\lp-\hat{q})$ function to remove the integration over $\diff^4\hat{q}$, we obtain the familiar expression for the SIDIS cross section in the Born QED approximation and a fully determined virtual photon momentum $q^\mu = \hat{q}^\mu$.
On the other hand, with QED radiation, the exchanged virtual photon momentum $\hat{q}^\mu$ cannot be fully determined without measuring all radiated final states $X_L$. 
In other words, there is no well-defined ``photon-hadron frame'' without having full control of the leptonic tensor $\widetilde{L}^{\mu\nu}(\l,\lp,\hat{q})$.

% .......................................................................
\subsection{Leptonic tensor and lepton structure functions}
\label{ss.lmn}

The leptonic tensor $\widetilde{L}^{\mu\nu}(\ell,\ell',\hat{q})$ in eq.~\eqref{e.sidis-lmn} is in general nonperturbative and cannot be calculated fully in QED and QCD perturbation theory. 
Since the QCD contribution to $\widetilde{L}^{\mu\nu}$ is from much higher order in $\alpha$, we will first calculate the leading QED contributions to this leptonic tensor and the transverse momentum broadening from the collision-induced QED radiation, which is perturbatively calculable.
We quantify and demonstrate the impact of the collision-induced QED radiation on the momentum change of the exchanged virtual photon from $q^\mu$ to $\hq^\mu$, and, in particular, the transverse momentum broadening of the $\hq^\mu$ in the lepton back-to-back frame where $\l$ is along the $+\bm{z}$ direction while $\lp$ is along the $-\bm{z}$ direction. 
The size of such transverse momentum broadening would directly impact the transverse momentum distribution of the observed final-state hadron in all SIDIS measurements.

% . . . . . . . . . . . . . . . . . . . . . . . . . . . . . . . . . . 
\subsubsection{Lepton structure functions}

In analogy to the decomposition of the hadronic tensor in eq.~(\ref{e.dis-wmnSFs}), we can express the leptonic tensor $\widetilde{L}^{\mu\nu}$ in terms of lepton structure functions, 
\begin{align}
    \widetilde{L}^{\mu\nu}(\l,\lp,\hq) 
    &= -\tg^{\mu\nu}(\hat{q})\, L_1
    + \frac{\tl^\mu\tl^\nu}{\l\cdot \lp} L_2
    + \frac{\tlp^\mu\tlp^\nu}{\l\cdot\lp} L_3
    + \frac{\tl^\mu\tlp^\nu+\tlp^\mu\tl^\nu}{2\l\cdot\lp} L_4,
\label{e.sidis-lmn-invariant}
\end{align}
where $\tg^{\mu\nu}(\hat{q})$ is given by eq.~(\ref{e.currentcon}) with $q$ replaced by $\hat{q}$, and we define
\begin{align}
\tl^\mu &= \tg^{\mu\nu}(\hq)\, \l_\nu,
\qquad
\tlp^\mu = \tg^{\mu\nu}(\hq)\, \lp_\nu,
\end{align}
such that $\hq_\mu \tl^\mu = \hq_\mu \tlp^\mu = 0$. 
In eq.~(\ref{e.sidis-lmn-invariant}), the lepton structure functions $L_i$ ($i=1,2,3,4$) depend on the four independent Lorentz scalars
    $Q^2$, 
    $\widehat{Q}^2$, 
and the ratios
\begin{align}
\xi_B &\equiv \frac{\hq\cdot\lp}{\l\cdot\lp}, 
\qquad
\frac{1}{\zeta_B} \equiv - \frac{\hq\cdot\l}{\l\cdot\lp}\, .
\label{e.xiBzetaB}
\end{align}
In the lepton back-to-back frame, in which
\begin{align}
    \l^\mu  & = (\l^+, 0, {\bf 0}_T),
    \qquad
    \lp^\mu\, = (0, \lp^-, {\bf 0}_T),
\end{align}
with $ \l^+ = \lp^- =Q/\sqrt{2}$, the exchanged virtual photon momentum can be written as
\begin{align}
    \hq^\mu 
    = \left( \hat{q}^+, \hat{q}^-,  \hat{\bm q}_T \right)
    = \Big( \xi_B \l^+, -\frac{1}{\zeta_B}\lp^-, \hat{\bm q}_T \Big),
\end{align}
where the transverse component is
\begin{align}
    \hat{\bm q}_T^2 = \widehat{Q}^2-\frac{\xi_B}{\zeta_B}Q^2.
\end{align}
We can also use $\hat{\bm q}_T^2$ to replace $\widehat{Q}^2$ or $Q^2$ as one of the four independent scalar variables for all lepton structure functions.
In this frame, the variable $\xi_B$ is effectively the momentum fraction of the incoming lepton carried by the active lepton at the hard collision, and $\zeta_B$ is the momentum fraction of the scattered lepton carried by the observed lepton in the final state.
The transverse momentum of the exchanged virtual photon $\hat{\bm q}_T$ is generated by the collision-induced radiation (mainly QED radiation, if we neglect the hadronic contribution to the leptonic tensor).
The amount of transverse momentum broadening $\hat{\bm q}_T$ from QED radiation will directly impact the direction of the exchanged virtual photon and the transverse momentum distribution of the extracted TMDs.

For the study of the $\hat{\bm q}_T$ dependence of the leptonic tensor $\widetilde{L}^{\mu\nu}(\ell,\ell',\hat{q})$ and the size of $\hat{\bm q}_T$ generated by QED radiation, it is convenient to express the tensor in eq.~(\ref{e.sidis-lmn-invariant}) in a helicity basis of the exchanged virtual photon,
\begin{align}
    L_{\rho\sigma}(\l,\lp,\hq) &\equiv \epsilon_{\rho}^\mu(\hq)\, 
    \widetilde{L}_{\mu\nu}(\l,\lp,\hq)\, \epsilon_{\sigma}^{*\nu}(\hq),
\label{e.sidis-lmn-helicity}
\end{align}
where the polarization vectors
    $\epsilon_{\rho}^\mu(\hq)$
and
    $\epsilon_{\sigma}^{*\nu}(\hq)$,
with polarization indices $\rho, \sigma=0,+,-$, depend on the reference frame and the coordinate system.
We construct the coordinate system by introducing the basis four-vectors $T$, $X$, $Y$, and $Z$, which satisfy the orthogonal and normalization relations,
\begin{subequations}
\begin{align}
    T\cdot X = T\cdot Y &= T\cdot Z = X\cdot Y = X\cdot Z = Y\cdot Z = 0,    \\
    T^2 &= 1, \qquad X^2 = Y^2 = Z^2 = -1 \, .
\end{align}
\end{subequations}
Since $\hat{q}^\mu$ is a space-like vector, we choose the basis vector $Z^\mu$ to be parallel to $\hat{q}^\mu$,
\begin{align}
    Z^\mu = \frac{1}{\widehat{Q}}\, \hq^\mu\, ,
\end{align}
with $\widehat{Q}\equiv \sqrt{\widehat{Q}^2} = \sqrt{- \hat{q}^2} > 0$.
If we choose the leptonic plane, defined by $\l$ and $\lp$, as the $X$--$Z$ plane, the other three basis vectors can be constructed from the conserved momenta, $\tl$ and $\tlp$, as
\begin{subequations}
\begin{align}
     T^\mu &= \frac{\sqrt{\xi_B \zeta_B}}{Q}\, \tl^\mu
     + \frac{1}{Q \sqrt{\xi_B \zeta_B}}\, \tlp^\mu,\\
     X^\mu &= -\frac{\widehat{Q}\sqrt{\xi_B \zeta_B}}{Q\, \hat{q}_T}\, \tl^\mu
     + \frac{\widehat{Q}}{Q\, \hat{q}_T \sqrt{\xi_B \zeta_B}}\, \tlp^\mu,\\
     Y^\mu &= \epsilon^{\mu\nu\rho\sigma} Z_\nu T_\rho X_\sigma,
\end{align}
\end{subequations}
where $\hat{q}_T$ is the magnitude of the photon transverse momentum, $\hat{q}_T \equiv \sqrt{\hat{\bm q}_T^2}$.
With the basis vectors defined, the three independent polarization vectors 
can then be written as
\begin{subequations}
\begin{align}
    \epsilon_0^\mu(\hq) &= T^\mu,
    \\
    \epsilon_\pm^\mu(\hq) &= \mp \frac{1}{\sqrt{2}}X^\mu - \frac{i}{\sqrt{2}} Y^\mu,
\end{align}
\end{subequations}
This ensures that the vectors $\epsilon^\mu_\rho(\hq)$ are orthogonal to $\hq$, $ \epsilon_\rho^\mu(\hq)\, \hq_\mu = \epsilon_\rho^{*\mu}(\hq)\, \hq_\mu = 0$, and orthogonal and normalized among themselves,
\begin{subequations}
\begin{align}
    \epsilon^*_0(\hq)\cdot \epsilon_\pm(\hq)
    &= \epsilon_\pm^*(\hq)\cdot \epsilon_\mp(\hq) = 0,
    \\
    \epsilon^*_0(\hq)\cdot \epsilon_0(\hq) = 1&, \quad
    \epsilon_\pm^*(\hq)\cdot \epsilon_\pm(\hq) = -1\, .
\end{align}
\end{subequations}
With the polarization vectors defined in the lepton back-to-back frame, we introduce the helicity-based lepton structure functions according to
\begin{subequations}
\begin{align}
    \widetilde{L}^{\mu\nu} &= 
    \epsilon_0^{*\mu}\epsilon_0^\nu\, L_{00}
    + (\epsilon_+^{*\mu}\epsilon_+^\nu + \epsilon_-^{*\mu}\epsilon_-^\nu)\, L_{++}
    + (\epsilon_+^{*\mu}\epsilon_-^\nu + \epsilon_-^{*\mu}\epsilon_+^\nu)\, L_{+-}
\nonumber\\
    &\quad
    - \epsilon_0^{*\mu}(\epsilon_+^\nu - \epsilon_-^\nu)\, L_{0+}
    - (\epsilon_+^{\mu} - \epsilon_-^{\mu})^*\epsilon_0^\nu\, L_{+0}
    \\
    &= T^\mu T^\nu\, L_{00} + (X^\mu X^\nu + Y^\mu Y^\nu)\, L_{TT}
\nonumber\\
    &\quad 
    + (T^\mu X^\nu + T^\nu X^\mu)\, L_{\Delta}
    + (Y^\mu Y^\nu - X^\mu X^\nu)\, L_{\Delta\Delta},
\label{e.sidis-lmn-helicity}
\end{align}
\end{subequations}
where 
$L_{TT} \equiv L_{++}$, 
$L_{\Delta} \equiv (L_{0+}+L_{+0})/\sqrt{2}$, and 
$L_{\Delta\Delta} \equiv L_{+-}$.
From these one can derive the relations between the original $L_i$ (\ref{e.sidis-lmn-invariant}) and helicity-based lepton structure functions,
\begin{subequations}
\label{e.Lij}
\begin{align}
    L_{00} &= \widetilde{L}_{\mu\nu}T^\mu T^\nu
    = -L_1 + \frac{1}{2\xi_B \zeta_B}\, L_2 + \frac{\xi_B \zeta_B}{2}\, L_3 + \frac12\, L_4, \\
    L_{TT} &= \frac12 \widetilde{L}_{\mu\nu}(X^\mu X^\nu + Y^\mu Y^\nu)
    = L_1 + \frac{1}{4\xi_B \zeta_B}\frac{\hat{\bm q}_T^2}{\widehat{Q}^2}\, L_2
    + \frac{\xi_B \zeta_B}{4}\frac{\hat{\bm q}_T^2}{\widehat{Q}^2}\, L_3
    -\frac14 \frac{\hat{\bm q}_T^2}{\widehat{Q}^2}\, L_4, \\
    L_{\Delta} &= -\frac12 \widetilde{L}_{\mu\nu}(T^\mu X^\nu + T^\nu X^\mu)
    = -\frac{1}{2\xi_B \zeta_B} \frac{\hat{q}_T}{\widehat{Q}}\, L_2 
    + \frac{\xi_B \zeta_B}{2}\frac{\hat{q}_T}{\widehat{Q}}\, L_3, \\
    L_{\Delta\Delta} &= \frac12 \widetilde{L}_{\mu\nu}(Y^\mu Y^\nu - X^\mu X^\nu) 
    = -\frac{1}{4\xi_B \zeta_B}\frac{\hat{\bm q}_T^2}{\widehat{Q}^2}\, L_2
    - \frac{\xi_B \zeta_B}{4}\frac{\hat{\bm q}_T^2}{\widehat{Q}^2}\, L_3
    +\frac14 \frac{\hat{\bm q}_T^2}{\widehat{Q}^2}\, L_4 
    \, .
\end{align}
\end{subequations}
As for the original lepton structure functions $L_i$ ($i=1,\ldots,4$), the helicity-based lepton structure functions are also functions of $\xi_B,\zeta_B,Q^2$, and $\hat{\bm q}_T^2$ (or equivalently, $\widehat{Q}^2$).

Without the collision-induced QED radiation, the transverse momentum, $\hat{\bm q}_T$, of the exchanged virtual photon in the lepton back-to-back frame should vanish. 
Any nonzero transverse momentum of the exchanged virtual photon in the lepton back-to-back frame is generated by the collision-induced radiations from both QED and QCD. 
If we can neglect the QCD contributions arising at higher orders in $\alpha$, the lepton structure functions could be perturbatively calculable in QED and expanded as a power series in $\alpha$,
\begin{align}
    L_{\rho\sigma}(\xi_B,\zeta_B,Q^2,\hat{\bm q}_T^2)
    = \sum_{N=0}^{\infty} \Big(\frac{\alpha}{\pi}\Big)^{\!N} 
    L^{(N)}_{\rho\sigma}(\xi_B,\zeta_B,Q^2,\hat{\bm q}_T^2),
    \label{eq:Lexpand}
\end{align}
where an overall factor $e^2$ (or $4\pi\alpha$) is factored out for our definition of the perturbative lepton structure functions.
From the lowest order leptonic tensor in eq.~(\ref{e.sidis-lmn0}), we derive the corresponding helicity-based lepton structure functions, 
\begin{subequations}
\begin{align}
    L_{TT}^{(0)} &=2\, \delta(\xi_B - 1)\, \delta\Big(\frac{1}{\zeta_B}-1\Big)\, \delta^{(2)}(\hat{\bm q}_T) 
    \label{e.lmn-tt-LO}, \\
    L_{00}^{(0)} &= 0, \qquad       % \label{e.lmn-00-LO}
    L_{\Delta}^{(0)} = 0, \qquad    % \label{e.lmn-d-LO}
    L_{\Delta\Delta}^{(0)} = 0.     % \label{e.lmn-dd-LO}
\end{align}
\end{subequations}
As expected from helicity conservation, apart from $L_{TT}$, all the other helicity-based lepton structure functions vanish at lowest order, and are suppressed by powers of $\hat{q}_T/\widehat{Q}$ at higher orders.
In the next subsection, we quantify the amount of photon transverse momentum $\hat{q}_T$ that can be generated by the collision-induced QED radiation in the relevant collision energies of SIDIS.

% . . . . . . . . . . . . . . . . . . . . . . . . . . . . . . . . . . 
\subsubsection{TMD factorization for lepton structure functions}
\label{sss.ltmd-fac}

The collision-induced QED radiation in SIDIS can generate nonvanishing $\hat{\bm q}_T$ for the leptonic tensor $\widetilde{L}^{\mu\nu}$.
As defined in eq.~(\ref{e.sidis-lmn}), the leptonic tensor has effectively the same operator definition as the corresponding hadronic tensor $\widetilde{W}_{\mu\nu}(\hat{q},P,P_h,S)$ in eq.~(\ref{e.sidis-wmn}), with the electromagnetic quark currents replaced by lepton currents, and hadronic states replaced by lepton states.
From studies of the factorization in SIDIS~\cite{Collins:2011zzd}, it is known that the hadronic tensor can be factorized in the hadron back-to-back frame in terms of QCD collinear factorization when $\hat{q}_T \sim \widehat{Q}$, and in terms of the TMD factorization when $\hat{q}_T \ll \widehat{Q}$.
Similarly, we can study the radiation induced $\hat{\bm q}_T$ dependence of the leptonic tensor $\widetilde{L}^{\mu\nu}$ in the lepton back-to-back frame by applying the same factorization in QED, with TMD factorization describing the low-$\hat{\bm q}_T$ region and collinear factorization for the high-$\hat{\bm q}_T$ region, along with a proper matching procedure for the phase space in between.

\begin{figure}[t]
\centering
    \includegraphics[width=0.85\textwidth]{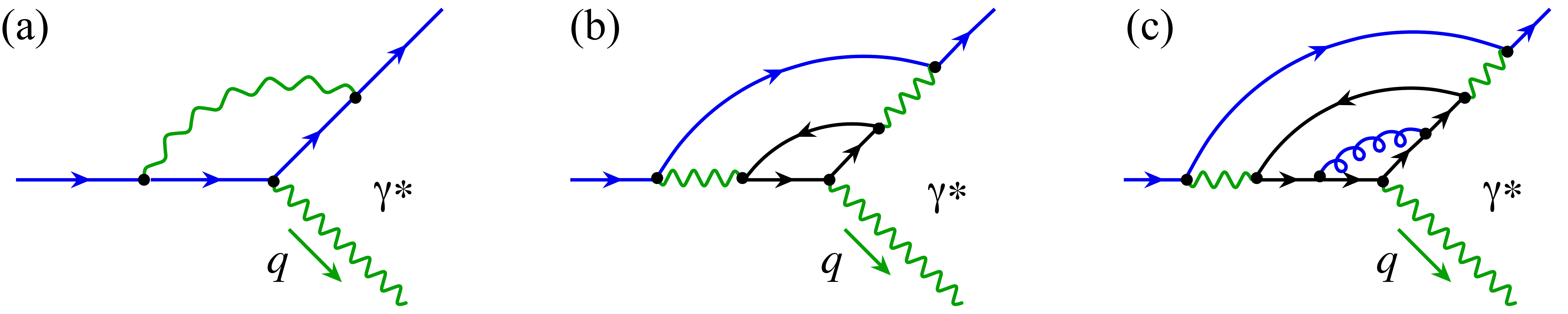}
    \caption{Sample diagrams, like the virtual diagram in figure~\ref{fig:nlo-v}, responsible for the leading double logarithmic contributions to the coefficient function $A$ of the lepton tensor in eq.~\eqref{eq:resumform} at {\bf (a)} LO ($\alpha$), {\bf (b)} NLO ($\alpha^2$), and {\bf (c)} NNLO ($\alpha^2\alpha_s$).}
    \label{fig:sud_ff}
\end{figure}

The collision-induced radiation dominates the low-$\hat{\bm q}_T$ region due to the logarithmic enhancement of the radiation in this regime.
Since the lepton structure functions $L_{00}$, $L_{\Delta}$, and $L_{\Delta\Delta}$ are power suppressed in $\hat{q}_T/\widehat{Q}$, we focus on the leading-power lepton structure function $L_{TT}$ in the following to study the size of $\hat{\bm q}_T$ generated by the induced photon shower.
In analogy with the QCD factorization of the SIDIS hadronic tensor~\cite{Collins:2011zzd}, we write $L_{TT}$ as
\begin{align}
    L_{TT}(\xi_B,\zeta_B,Q^2,\hat{\bm q}_T^2)
    &= W_{TT}(\xi_B,\zeta_B,Q^2,\hat{\bm q}_T^2)\,
    +\, Y_{TT}(\xi_B,\zeta_B,Q^2,\hat{\bm q}_T^2),
\label{eq:W+Y}
\end{align}
where the first term is given by a Fourier transform of an impact parameter distribution $\widetilde{W}_{TT}(\xi_B,\zeta_B,Q^2,b_T)$, with $b_T \equiv |\bm{b}_T|$ the magnitude of the impact parameter $\bm{b}_T$ conjugated to $\hat{\bm q}_T$,
\begin{align}
    W_{TT}(\xi_B,\zeta_B,Q^2,\hat{\bm q}_T^2)
    &= \int\frac{\diff^2{\bm b}_T}{(2\pi)^2}\,
    e^{i\hat{\bm q}_T\cdot {\bm b}_T}\,
    \widetilde{W}_{TT}(\xi_B,\zeta_B,Q^2,b_T).
\label{eq:WTT}
\end{align}
This term is mainly responsible for the region where $\hat{q}_T\ll \widehat{Q}$, including the resummation of large $\ln(\widehat{Q}^2/\hat{\bm q}_T^2)$-type logarithms associated with the radiations.
The second term in (\ref{eq:W+Y}), $Y_{TT}$, provides a smooth matching to the region of collinear factorization where $\hat{q}_T\sim\widehat{Q}$.
The formulation in eq.~(\ref{eq:W+Y}) is usually referred to as the ``W+Y'' formalism~\cite{Collins:1984kg}.
As for the hadronic case, the function $\widetilde{W}_{TT}$ for the leptonic structure function can be factorized as
\begin{align}
    \widetilde{W}_{TT}(\xi_B,\zeta_B,Q^2,b_T) 
    &=
    2 \int_{\zeta_B}^1\frac{d\zeta}{\zeta^2}  \int_{\xi_B}^1\frac{d\xi}{\xi}
    \bigg[ C_D\Big(\frac{\zeta_B}{\zeta},\alpha\Big) D(\zeta,\mu_b^2) \bigg] 
    \bigg[ C_f\Big(\frac{\xi_B}{\xi},\alpha\Big) f(\xi,\mu_b^2) \bigg] \nonumber\\
    &\quad \times
    \exp\bigg\{ \!-\!\int_{\mu_b^2}^{\mu_Q^2}\frac{d\mu'^2}{\mu'^2}
    \Big[ A\big(\alpha(\mu')\big) \ln\frac{\mu_Q^2}{\mu'^2}+B\big(\alpha(\mu')\big)
    \Big]
    \bigg\},
    \label{eq:resumform}
\end{align}
where we neglected the lepton flavor-changing contribution and kept only 
    $f_{e/e}=f$ and $D_{e/e}=D$. 
The factorized expression for the leading leptonic structure function in eq.~\eqref{eq:resumform} has effectively the same form as the $b_T$-space expression of the combined QED and QCD transverse momentum resummation for the transverse momentum ($q_T$) distribution of $Z^0$ bosons produced in hadronic collisions~\cite{Cieri:2018sfk,Cieri:2020ikq}, if we replace the leptons by hadrons, along with proper kinematic factors and adjustment from space-like to time-like hard processes.
For $Z^0$ production in hadronic collisions, both QED and QCD radiative contributions to the transverse momentum broadening, or to the coefficient functions $A$, $B$, $C_f$, and $C_D$ in eq.~\eqref{eq:resumform}, start at leading order in $\alpha$ and $\alpha_s$, respectively.
Since $\alpha \ll \alpha_s$, the transverse momentum broadening induced by the QED shower is much smaller than that generated by the QCD shower, and is of the order of a percent~\cite{Cieri:2018sfk, Cieri:2020ikq}.  
However, unlike the hadronic case, QCD radiative contributions to the transverse momentum broadening of the leptonic tensor in eq.~\eqref{eq:resumform} start at a much higher order in $\alpha$.
For example, as shown in figure~\ref{fig:sud_ff}, QCD contributions to the coefficient function, $A$, which is responsible for the resummation of the leading double logarithmic effects for the broadening, starts at ${\cal O}(\alpha^2\alpha_s)$, which is an order of $\alpha\alpha_s$ suppressed compared to the leading QED contribution at ${\cal O}(\alpha)$.
That is, the transverse momentum broadening caused by the collision-induced radiation from a ``point-like'' lepton is dominated by the QED radiation.

If we neglect the hadronic contribution to the lepton structure function $\widetilde{W}_{TT}$, all the coefficient functions $A$, $B$, $C_f$, and $C_D$, as well as the LDF and LFF, are perturbatively calculable in QED and valid at a scale $\mu_b$ that is much lower than the order of GeV in hadronic collisions.
Generally, we can expand them in a power series in $\alpha$,
\begin{subequations}
\begin{align}
A   &= \sum_{N=1}^{\infty} \Big(\frac{\alpha}{\pi}\Big)^N A^{(N)},\quad
B\   = \sum_{N=1}^{\infty} \Big(\frac{\alpha}{\pi}\Big)^N B^{(N)}, \\ 
C_f &= \sum_{N=0}^{\infty} \Big(\frac{\alpha}{\pi}\Big)^N C_f^{(N)},\quad
C_D\ = \sum_{N=0}^{\infty} \Big(\frac{\alpha}{\pi}\Big)^N C_D^{(N)}.
\end{align}
\end{subequations}
To estimate the size of $\hat{\bm q}_T$ broadening from induced radiation, we choose the convention to define the factorization scales in eq.~(\ref{eq:resumform}) as
    $\mu_b = 2 e^{-\gamma_E}/b_T$
and 
    $\mu_Q = \sqrt{\xi_B/\zeta_B}\, Q$~\cite{Collins:1984kg}.
In principle, we could introduce two proportional constants of ${\cal O}(1)$ to test the uncertainty associated with the scale choice.

With the leading order $L_{TT}^{(0)}$ given in eq.~(\ref{e.lmn-tt-LO}), together with $f^{(0)}(\xi)=\delta(1-\xi)$ and $D^{(0)}(\xi)=\delta(1-\zeta)$, we have for the hard coefficients
\begin{align}
    C_f^{(0)}(\lambda) &= \delta(\lambda-1),
    \qquad
    C_D^{(0)}(\eta) = \delta\bigg(\frac{1}{\eta}-1\bigg),
\end{align}
with $\lambda=\xi_B/\xi$ and $\eta=\zeta_B/\zeta$.
As presented in appendix~\ref{s.app-nlo}, at ${\cal O}(\alpha)$ in the $\overline{\rm MS}$ scheme we find the following coefficient functions,
\begin{subequations}
\begin{align}
    C_f^{(1)}(\lambda) &= \frac12 (1-\lambda) 
    - \frac12 \left[\frac{1+\lambda^2}{1-\lambda}\right]_{+}
    \ln\frac{\mu_{\overline{\rm MS}}^2}{\mu_b^2}
    - 2\delta(1-\lambda), \\
    C_D^{(1)}(\eta) &= \frac{1}{2\eta}(\eta-1)
    -\frac{1}{2\eta}\left[\frac{1+\eta^2}{1-\eta}\right]_{+}
    \ln\frac{\mu_{\overline{\rm MS}}^2}{\mu_b^2}
    - \frac{2}{\eta}\delta(\frac{1}{\eta}-1), \\
    A^{(1)} &= 1, \\
    B^{(1)} &= -\frac32,
\end{align}
\end{subequations}
as well as for the ``Y'' term,
\begin{align}
    \widehat{Y}_{TT}^{(1)} 
    &=
    \frac{1}{2\pi\mu_Q^2}
    \bigg[-\frac{2(\hat{u}^2+\hat{v}^2)+4\hat{t}(\hat{t}+\hat{u}+\hat{v})}{\hat{u}\hat{v}}
    +\frac{1+\lambda^2\eta^2}{\lambda\eta}
    \bigg]
    \delta\Big(\frac{1}{\lambda}(1-\lambda)(1-\eta)-\frac{\hat{\bm q}_T^2}{\mu_Q^2}\Big)
    \nonumber\\
    &\quad
    -\frac{1}{\pi\hat{\bm q}_T^2}
    \bigg[
    \frac{1+\lambda^2}{(1-\lambda)_+}\delta(1-\eta)
    +\frac{1}{\eta}\frac{1+\eta^2}{(1-\eta)_+}\delta(1-\lambda)
    -2\delta(1-\lambda)\delta(1-\eta)
    \ln\frac{\hat{\bm q}_T^2}{\mu_Q^2}
    \bigg],
    \label{e.YTThat(1)}
\end{align}
with the variables defined as
$\hat{t}=(k-k')^2=-(\xi/\zeta) Q^2$,
$\hat{u} = (k-\hat{q})^2 = [(\xi-\xi_B)/\zeta_B]Q^2 - \hat{\bm q}_T^2$, 
and 
\mbox{$\hat{v} = (k'+\hat{q})^2 = [\xi_B(\zeta_B-\zeta)/(\zeta \zeta_B)]Q^2 - \hat{\bm q}_T^2$}.
The expression in eq.~\eqref{e.YTThat(1)} is presented as a difference between the NLO perturbative contribution to the leptonic tensor and the asymptotic piece of this contribution when $\hat{\bm q}_T^2 \to 0$.

\begin{figure}[t]
    \centering
    \includegraphics[width=0.9\textwidth]{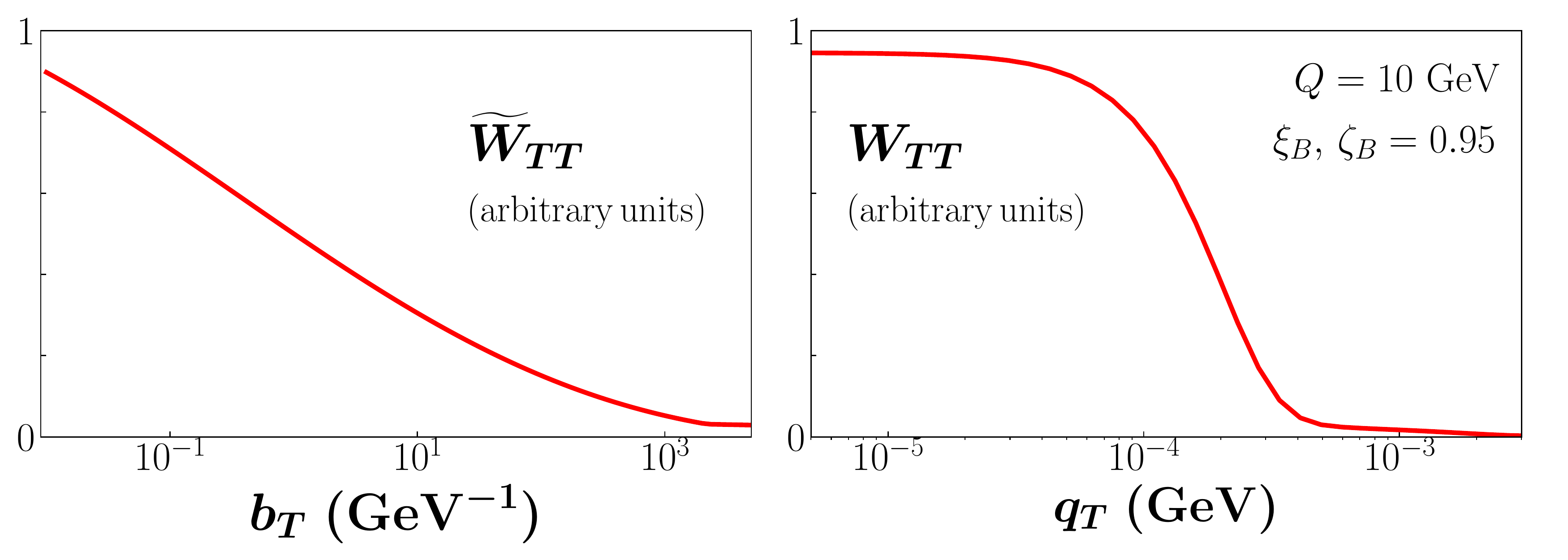}
    \caption{Shape of the lepton structure function $\widetilde{W}_{TT}(\xi_B,\zeta_B,Q^2,b_T)$ (in arbitrary units) in impact parameter space {\bf (left)} and the corresponding Fourier transform in transverse momentum space $W_{TT}(\xi_B,\zeta_B,Q^2,\hat{\bm q}_T^2)$ {\bf (right)}, evaluated at fixed $Q=10$~GeV and $\xi_B=\zeta_B=0.95$.}
    \label{fig:leptonTMD}
\end{figure}

In figure~\ref{fig:leptonTMD} we plot the impact parameter distribution of the lepton structure function, $\widetilde{W}_{TT}(\xi_B,\zeta_B,Q^2,b_T)$, along with its Fourier transform $W_{TT}(\xi_B,\zeta_B,Q^2,\hat{\bm q}_T^2)$ in conjugate $\hat{\bm q}_T$ space, at fixed values of $Q=10$~GeV and $\xi_B=\zeta_B=0.95$.
The $W_{TT}$ function, which is generated by resumming the leading logarithmic enhanced QED radiation, has a very steep and narrow peak at $\hat{\bm q}_T^2=0$.
Although the logarithms due to the hard collisions could be large, the QED fine structure constant $\alpha \sim 1/137$ is much smaller than $\alpha_s$ in the same kinematic regime, which effectively makes the $\hat{\bm q}_T$-broadening from QED radiation negligible compared with the typical transverse momentum broadening of QCD radiation.  
As discussed following eq.~\eqref{eq:resumform}, QCD contributions to the same $\hat{\bm q}_T$-broadening are even smaller than what can be generated from QED showers from the leptons, although they are much more important, even dominant, for hadronic $Z^0$ production~\cite{Cieri:2018sfk,Cieri:2020ikq}.  One can therefore safely use the collinear factorization approach to account for the leading power radiative contributions from the leptons of the SIDIS cross section as a controllable approximation.

% .......................................................................
\subsection{SIDIS cross section with collinearly factorized QED contributions}
\label{ss.sidis-xsec}

As demonstrated quantitatively in the previous subsection, the transverse momentum imbalance generated by the collision-induced QED and QCD radiation from the leptons for SIDIS at relevant collision energies is much smaller than the typical intrinsic parton transverse momentum of hadrons.
This fact ensures that contributions to the momentum imbalance $\pTbar$ between the observed lepton of momentum $\lp$ and hadron of momentum $P_h$ are completely dominated by the active parton's transverse momentum generated by the collision-induced QCD shower on top of parton's intrinsic transverse momentum from the hadrons.
The leading collision-induced radiative contributions to the SIDIS cross section from the two leptons can therefore be systematically accounted for in terms of the collinear factorization approach.
In this case, the SIDIS cross section for a colliding lepton of momentum $\l$ and helicity $\lambda_\l$ and a nucleon of momentum $P$ and spin $S$ can be factorized as~\cite{Qiu:1990xy}
\begin{align}
\label{e.sidis-fac-qed} 
    E_{\ell'} E_{P_h} 
    \frac{\diff^6\sigma_{\l(\lambda_\l) P(S)\to \ell' P_hX}}
         {\diff^3\ell'\, \diff^3P_h}
    &\approx\, \sum_{i j \lambda_k}
    \int_{\zeta_{\rm min}}^1\frac{\diff\zeta}{\zeta^2} \, D_{e/j}(\zeta)
    \int_{\xi_{\rm min}}^1 \diff\xi\, f_{i(\lambda_k)/e(\lambda_\l)}(\xi)
    \nonumber\\
    &\quad
    \times
    \left[
    E_{k'} E_{P_h} 
    \frac{\diff^{6}{\hat{\sigma}}_{k(\lambda_k) P(S)\to k' P_hX}}
         {\diff^3k'\, \diff^3P_h}
    \right]_{k=\xi\l, k'=\lp/\zeta},
\end{align}
where the integration limits $\zeta_{\rm min}$ and $\xi_{\rm min}$ are given in eq.~(\ref{e.minvalues}), and $\hat{\sigma}_{k(\lambda_k) P(S)\to k' P_hX}$ is infrared-safe as $m_e \to 0$, with all infrared sensitive collinear QED radiative contributions to the cross section resummed into the LDFs and LFFs.
(For ease of notation the dependence on the factorization scale in (\ref{e.sidis-fac-qed}) is suppressed.)
With the one-photon approximation, we set $i=j=e$ in eq.~(\ref{e.sidis-fac-qed}).

In the Born approximation in QED (which is the LO contribution in $\alpha$), the QED infrared-safe cross section 
    $\hat{\sigma}_{k(\lambda_k) P(S) \to k' P_h X}$ 
in eq.~(\ref{e.sidis-fac-qed}) further simplifies to
\begin{align}
\label{e.sidis-fac-qed0} 
    E_{k'} E_{P_h}
    \frac{\diff^{6}\hat{\sigma}^{(0)}_{k(\lambda_k) P(S) \to k' P_h X}}
         {\diff^3 k'\, \diff^3 P_h}
    &= \frac{\alpha^2}{2\hat{s}} \bigg( \frac{1}{\hat{q}^2} \bigg)^2
    \widehat{L}_{\mu\nu}^{(0)}(k,k',\lambda_k)\,
    \widetilde{W}^{\mu\nu}(\hq,P,P_h,S),
\end{align}
where the $0^{\rm th}$-order leptonic tensor hard part is
\begin{align}
    \widehat{L}_{\mu\nu}^{(0)}(k,k',\lambda_k) = 2\big( k_\mu k'_\nu + k_\nu k'_\mu - k\cdot k' g_{\mu\nu} + i \lambda_k \epsilon_{\mu\nu\rho\sigma} k^\rho k'^\sigma\big),
\end{align} 
and the hadronic tensor $\widetilde{W}^{\mu\nu}(\hq,P,P_h,S)$ is defined in eq.~(\ref{e.sidis-wmn}). 
Our factorization formula for SIDIS in eqs.~(\ref{e.sidis-fac-qed}) and (\ref{e.sidis-fac-qed0}) indicates that the impact of QED radiative contribution to the SIDIS cross section is not only from the change of the exchanged virtual photon momentum $q^\mu \to \hq^\mu$, weighted by the convolution over the LDFs and LFFs, but also from non-logarithmic and infrared-safe higher-order QED corrections to $\hat{\sigma}_{k(\lambda_k) P(S)\to k' P_hX}$. 
Without QED radiation, the probing scale for the hard collision --- the momentum of the exchanged virtual photon $\hat{q}^\mu = q^\mu$ --- is uniquely determined from the experimental measurement of the colliding and scattered lepton momenta, $\l^\mu$ and $\lp^\mu$.
On the other hand, with QED radiation the momentum of the exchanged virtual photon $\hq^\mu$ is no longer determined by direct experimental measurement, but instead is a function of the lepton momentum fractions $\xi$ and $\zeta$.

\begin{figure}[t]
\centering
    \includegraphics[width=0.6\textwidth]{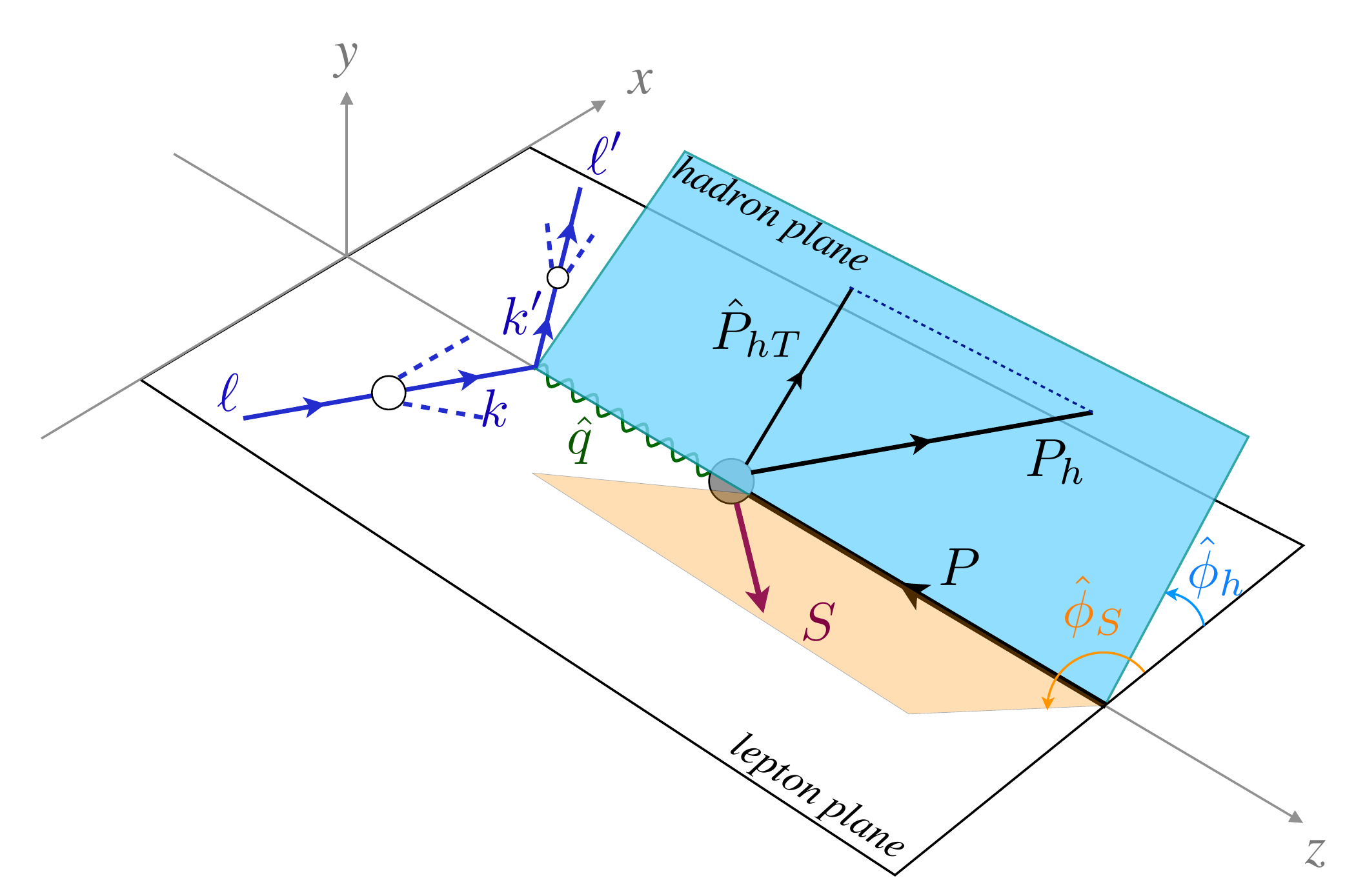}
    \caption{Sketch for the ``photon-hadron frame'' for SIDIS according to the Trento convention~\cite{Bacchetta:2004jz}.}
\label{f.sidis-frame}
\end{figure}

To proceed with the application of QED and QCD factorization to SIDIS hadron production experiments, we convert the differential cross section in eq.~(\ref{e.sidis-fac-qed0}) to more standard kinematic variables consistent with the Trento convention~\cite{Bacchetta:2004jz}.
In particular, we use the change of variables 
\begin{align}
\label{e.sidis-trento} 
    E_{k'} E_{P_h}
    \frac{\diff^6\hat{\sigma}^{(0)}_{k(\lambda_k) P(S) \to k' P_hX}}
         {\diff^3 k'\, \diff^3 P_h}
    = \bigg( \frac{4\hxb}{\widehat{Q}^2} 
      \sqrt{\hat{z}_h^2-(\hat{\gamma} \widehat{P}_{hT}/\widehat{Q})^2}
      \bigg)
      \frac{\diff^6\hat{\sigma}^{(0)}_{k(\lambda_k) P(S)\to k' P_hX}}
           {\diff\hxb \diff\hat{y}\, \diff{\hat\psi}\, \diff\hat{z}_h\, \diff\hat{\phi}_h \diff \widehat{P}_{hT}^2},
\end{align}
where $\hat{z}_h=P\cdot \widehat{P}_h/P\cdot \hat{q}$ and $\hat{\phi}_h$ is the angle from the leptonic plane to the hadronic plane defined in the virtual ``photon-nucleon frame'', as shown in figure~\ref{f.sidis-frame}. 
The angle $\hat{\psi}$ is the azimuthal angle of $k'$ around $k$ with respect to the transverse spin of the nucleon. 
In DIS kinematics, one has 
    $\diff\hat{\psi}\approx \diff \hat{\phi}_S$~\cite{Diehl:2005pc},
with $\hat{\phi}_S$ the angle from the leptonic plane to the spin plane, as shown in figure~\ref{f.sidis-frame}.
Parametrizing the one-photon exchange cross section in terms of the usual 18 SIDIS structure functions,
    $F_n^h(\hxb,\widehat{Q}^2,\hat{z}_h,\widehat{P}_{hT}^2)$
    ($n=1,\ldots,18$)~\cite{Bacchetta:2006tn},
weighted by factors $\hat{w}_n$ that are functions of the kinematic variables, we can write the differential SIDIS cross section in the presence of QED effects as
\begin{align}
    &\frac{\diff^6\sigma_{\l(\lambda_\l) P(S) \to \lp P_h X}}
         {\diff\xb \diff y\, \diff \psi\, \diff z_h\, \diff \phi_h \diff P_{hT}^2}
    = \sum_{i j \lambda_k}
       \int_{\zeta_{\rm min}}^1 \frac{\diff\zeta}{\zeta^2}
       \int_{\xi_{\rm min}}^1 \frac{\diff\xi}{\xi}\, f_{i(\lambda_k)/e(\lambda_\l)}(\xi)\, D_{e/j}(\zeta)
    \notag\\
    & \hspace{2.8cm}
    \times
    \frac{\hxb}{\xb \xi \zeta}
    \bigg[ \frac{\alpha^2}{\hxb\, \hat{y}\, \widehat{Q}^2} 
        \frac{\hat{y}^2}{2(1-\hat{\varepsilon})}
    \left(1+\frac{\hat{\gamma}^2}{2\hxb}\right)
    \sum_n \hat{w}_n F_n^h(\hxb,\widehat{Q}^2,\hat{z}_h,\widehat{P}_{hT}^2)
    \bigg],
\label{e.sidisRC}
\end{align}
where the kinematic variables with carets in the factorized expression can be written in terms of momentum fractions $\xi$, $\zeta$ and the measured variables without carets.

Our strategy to evaluate the lepton-nucleon SIDIS cross section with QED contributions, as in eq.~(\ref{e.sidis-fac-qed}), is as follows:
\begin{enumerate}
\item[(1)]
Evaluate the SIDIS cross section without QED radiation in the ``photon-nucleon frame'' (defined by the exchanged virtual photon of momentum $\hq$ and colliding nucleon of momentum $P$ for a given set of momentum fractions $(\xi,\zeta)$) in terms of TMD factorization and the corresponding momentum variables
    $\{ \hq$, $P$, $\widehat{P}_h \}$ 
if $\widehat{P}_{hT} \ll \widehat{Q}$,
and in terms of collinear factorization if $\widehat{P}_{hT} \sim \widehat{Q}$, along with a matching prescription between these two regimes.
\item[(2)]
Apply a $(\xi,\zeta)$-dependent Lorentz transformation to change all variables of the calculated SIDIS cross section, composed of
    $\{ \hq$, $P$, $\widehat{P}_h \}$
and the spin vectors in figure~\ref{f.sidis-frame} for polarized lepton-nucleon SIDIS, to the corresponding variables in a frame suitable for comparison with experiment, such as the lepton-hadron frame, or the experimentally defined photon-nucleon Breit frame.
\end{enumerate}
This $(\xi,\zeta)$-dependent Lorentz transformation changes $\widehat{P}_{hT}$ and $\hat{\phi}_h$, the angle between the leptonic plane and the hadronic plane, in a virtual ``photon-nucleon frame'', as shown in figure~\ref{f.sidis-frame}, to be functions of $\xi$, $\zeta$, and experimentally measured kinematic variables $\xb$, $Q^2$, $z_h$, $P_{hT}$ and $\phi_h$, in a frame where the calculated SIDIS cross section is compared with experimental data.
This leads to a strong impact on the extraction of TMDs from SIDIS cross section data, as we discuss in more detail in the next section.

%%%%%%%%%%%%%%%%%%%%%%%%%%%%%%%%%%%%%%%%%%%%%%%%%%%%%%%%%%
\section{Numerical impact of QED in semi-inclusive DIS}
\label{s.pheno}

Having derived the SIDIS cross section formulas in our combined QED + QCD factorized approach, in this section we discuss the numerical impact of the QED radiation on the extraction of SIDIS structure functions and asymmetries.
As illustrative examples, we consider the unpolarized SIDIS structure function, as well as the angular modulations for scattering unpolarized leptons from transversely polarized nucleons, such as those associated with the Collins and Sivers asymmetries.

For the numerical implementation of the radiative effects on the SIDIS calculation, we use LDFs and LFFs evolved in Mellin space and computed in momentum space via a numerical inverse Mellin transform.
One problem encountered in implementing \eref{sidisRC} numerically is the accuracy of the calculation in the vicinity of the end-point regions when $\xi$, $\zeta \to 1$.
These regions contribute maximally to the cross sections, reflecting the presence of peaks in the LDFs and LFFs.
However, the end-point region contributions are numerically inaccurate if one naively evaluates them via the inverse Mellin transform directly.
Instead, one can make use of a subtraction trick, whereby the differential cross section is first written in the form
\begin{align}
\label{e.d6sig}
    \frac{\diff^6\sigma_{{ \l(\lambda_{\l})} P(S) \to { \lp} P_h X}}
         {\diff\xb \diff y\, \diff \psi\, \diff z_h\, \diff \phi_h \diff P_{hT}^2}
    &= \sum_{ij} 
       \int_{\zeta_{\rm min}}^1 \diff\zeta
       \int_{\xi_{\rm min}(\zeta)}^1 \diff\xi\,
       f_{i/e}(\xi)\, D_{e/j}(\zeta)\, {\cal H}_{ij}(\xi,\zeta),
\end{align}
where ${\cal H}_{ij}(\xi,\zeta)$ contains all other factors in the integrand of \eref{sidisRC} that are not contained in the LDF and LFF.
The right-hand-side of (\ref{e.d6sig}) can then be written as
\begin{align}
\frac{\diff^6\sigma_{{ \l(\lambda_{\l})} P(S) \to { \lp} P_h X}}
     {\diff\xb \diff y\, \diff \psi\, \diff z_h\, \diff \phi_h \diff P_{hT}^2}
&= \sum_{ij}
    \bigg[\!\int_{\zeta_{\rm min}}^1 \!\diff\zeta\,
    D_{e/j}(\zeta)\, 
    \big[ g_{ij}(\zeta)-g_{ij}(1) \big]
    \nonumber\\
&\qquad\quad
 +  g_{ij}(1) \frac{\zeta_{\rm min}}{2\pi i} \int \!\diff N\, 
    \zeta_{\rm min}^{-N} \frac{D^N_j}{N-1}
    \bigg],
\end{align}
where the function $g_{ij}$ is defined as
\begin{align}
g_{ij}(\zeta)
= 
\int_{\xi_{{\rm min}(\zeta)}}^1 \!\diff\xi
    \big[ {\cal H}_{ij}(\xi,\zeta)-{\cal H}_{ij}(1,\zeta) \big]\,
+\, {\cal H}_{ij}(1,\zeta)
    \frac{\xi_{\rm min}(\zeta)}{2\pi i} 
    \int\!\diff M\, \xi^{-M}_{\rm min}(\zeta) \frac{F^M_i}{M-1}
    ,
\end{align}
and $F^M_i$ and $D^N_j$ are the Mellin moments of the LDFs and LFFs, respectively,
\begin{subequations}
\begin{align}
F^M_i &= \int \diff \xi\, \xi^{M-1}\, f_{i/e}(\xi), \\
D^N_j &= \int \diff \zeta\, \zeta^{N-1}\, D_{e/j}(\zeta).
\end{align}
\end{subequations}
The subtraction trick allows us to remove the numerically problematic region and evaluate the end-point contributions accurately through convolution of the LDF and LFF moments with simple factors $1/(M-1)$ and $1/(N-1)$, respectively.

With the numerical strategy in place, we proceed to quantify the radiative effects for the SIDIS process for various channels. 
In our earlier work~\cite{Liu:2020rvc}, we demonstrated the impact of QED effects on the unpolarized SIDIS cross section differential in the outgoing hadron's transverse momentum, $P_{hT}$, in the Breit frame.
The associated unpolarized SIDIS structure function $F_{UU,T}^h$ was modeled by a factorized Gaussian {\it ansatz} in the TMD framework~\cite{Anselmino:2013lza},
\begin{eqnarray}
F_{UU,T}^h(\xb,Q^2,z_h,P_{hT})
= \sum_q e_q^2\, f_{q/N}(\xb,Q^2)\, D_{h/q}(z_h,Q^2)\,
  \frac{\exp(-P_{hT}^2/\langle P_{hT}^2 \rangle)}
        {\pi \langle P_{hT}^2 \rangle},
\label{e.FUU}
\end{eqnarray}
where we adopt the notation of ref.~\cite{Bacchetta:2006tn} in which the first two subscripts of the structure functions denote the polarization states of the lepton and nucleon, respectively, while the third indicates the polarization of the virtual photon.

Using the fitted parameters from ref.~\cite{Anselmino:2013lza}, the $P_{hT}$ spectrum was found~\cite{Liu:2020rvc} to be significantly modified in the presence of QED effects.
Since the fitted Gaussian {\it ansatz} for $F_{UU,T}^h$ is only valid for small transverse momenta, it is instructive to see how the QED effects depend on its shape in the large transverse momentum region, where the Gaussian behavior is expected to transform into a power law-like dependence.
To explore this transition, we augment the original function $F_{UU,T}^h$ by modifying its large-$P_{hT}$ behavior,
\begin{align}
    F_{UU,T}^h\ \to\ 
    F_{UU,T}^{h\, {\rm (mod)}}\, =\, F_{UU,T}^h\, R\, +\, (1-R) F_{\rm tail},
\label{e.FUUtail}
\end{align}
where   
\begin{align}
    R = \exp\bigg[ -N\Big(\frac{q_T}{Q}\Big)^3 \bigg],
    \qquad
    F_{\rm tail} = \frac{C_{\rm tail}}{q_T^2}, 
\label{e.FUUtailRF}
\end{align}
with $q_T = P_{hT}/z$, and the parameters set to $N = 20$ and $C_{\rm tail} = 0.01$~GeV$^2$. 
The modification mimics the enhancement of the structure function at large $P_{hT}$ stemming from hard QCD radiation, which overwhelms the effects from intrinsic transverse momentum in this region.

\begin{figure}[t]
\centering
    \includegraphics[width=0.55\textwidth]{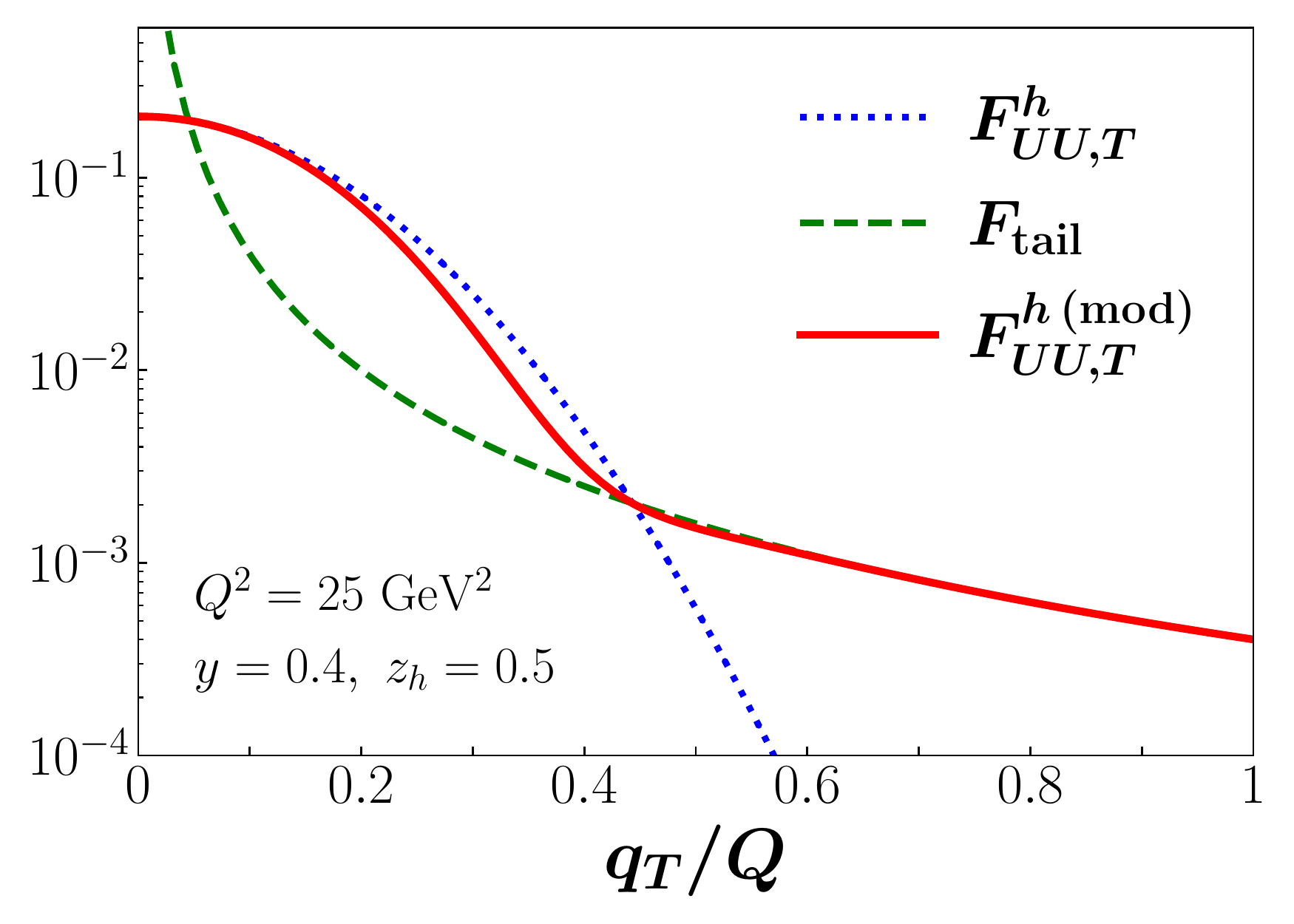}\vspace*{-0.5cm}
    \caption{Unpolarized SIDIS structure function $F_{UU,T}^h$ as a function $q_T/Q$, where $q_T = P_{hT}/z$, at fixed values of $Q^2=25$~GeV$^2$, $y=0.4$, and $z_h=0.5$. The unmodified function (dotted blue line) is taken from the JAM3D20 global QCD analysis~\cite{Cammarota:2020qcw}, while the additional power-law tail contribution (dashed green line) distorts the region $q_T/Q>0.5$ by enhancing the modified $F_{UU,T}^{h\, {\rm (mod)}}$ (solid red line) to mimic QCD radiation effects in collinear factorization.}
\label{f.FUU}
\end{figure}

In \fref{FUU} we illustrate our modification to $F_{UU,T}^h$, showing the dependence on $q_T/Q$ for fixed values of
    $Q^2 = 25$~GeV$^2$, $y = 0.3$, and $z_h = 0.5$,
using for the unmodified $F_{UU,T}^h$ structure function the result extracted from the recent JAM3D20 global analysis in the TMD framework~\cite{Cammarota:2020qcw}.
The specific parameters used for the modification are simply illustrative, but chosen to approximate a typical scenario for the large-$P_{hT}$ region within collinear factorization. 
Note that when implementing the QED effects described in the previous sections,  eqs.~(\ref{e.FUU})--(\ref{e.FUUtailRF}) are utilized by replacing the arguments of $F_{UU,T}^h$ with the corresponding variables $\hxb$, $\widehat{Q}^2$, $\hat{z}_h$ and $\widehat{P}_{hT}$.

\begin{figure}[t]
\centering
    \includegraphics[width=0.9\textwidth]{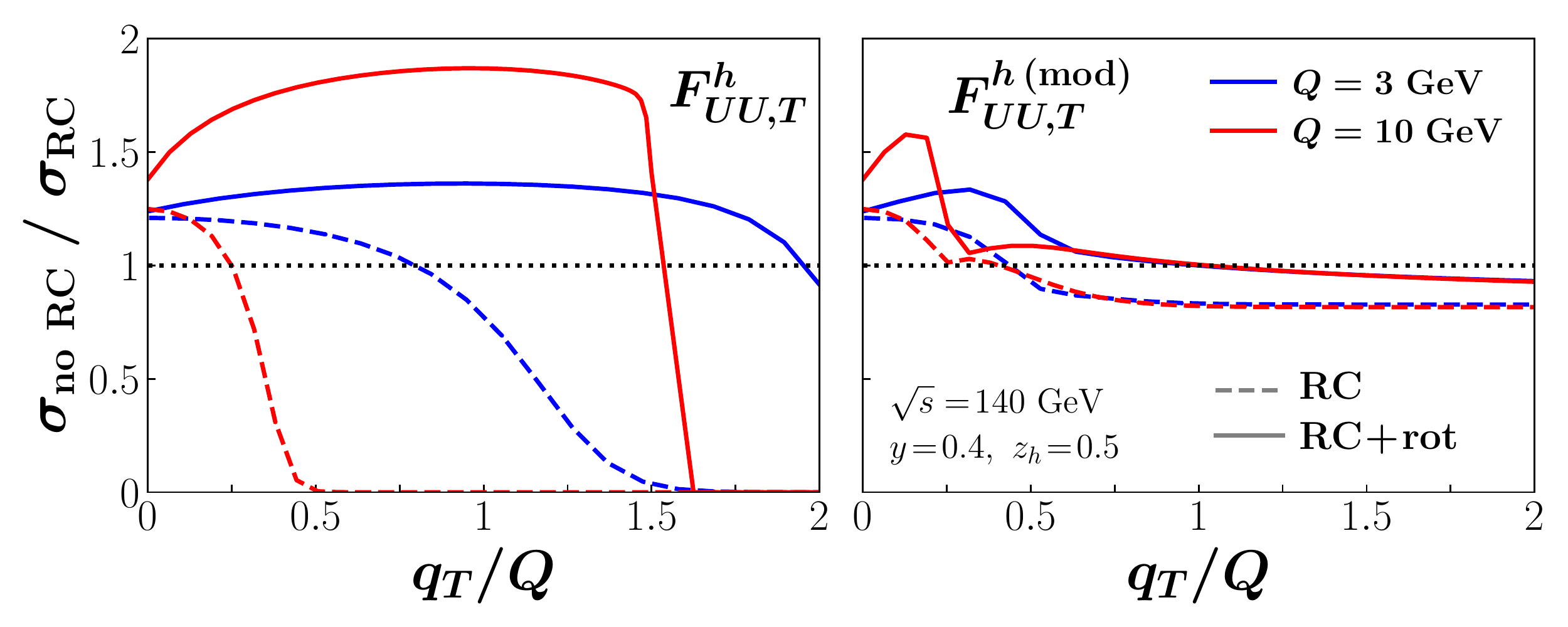}\vspace*{-0.5cm}
    \caption{Ratios of unpolarized SIDIS cross sections without radiation (``no RC'') to those including QED effects, as a function of $q_T/Q$, for the $F_{UU,T}^h$ structure function from ref.~\cite{Cammarota:2020qcw} using the Gaussian {\it ansatz} in the TMD framework (left) and with the modified $F_{UU,T}^{h\, \rm (mod)}$ as in eq.~(\ref{e.FUUtail}) (right), for $\sqrt{s} = 140$~GeV, $y = 0.4$ and $z_h = 0.5$, at $Q=3$~GeV (blue lines) and 10~GeV (red lines). The full calculation with QED radiation (``RC+rot'', solid lines) is compared with that removing the QED rotational effects induced to the transverse momentum in Breit frame (``RC'', dashed lines).}
\label{f.sidisUU}
\end{figure}

In \fref{sidisUU} we show the impact of the QED radiative effects on the ratios of unpolarized SIDIS cross sections, calculated at the Born level and with RCs, as a function of $q_T/Q$ at fixed values of
    $\sqrt{s} = 140$~GeV, 
    $y = 0.4$ and 
    $z_h = 0.5$, 
for $Q=3$ and 10~GeV, typical of those expected at the EIC.
The QED radiative effects are observed to be stronger in the absence of hard QCD radiation enhancements in $F_{UU,T}^h$ at large $P_{hT}$, and relatively mild otherwise.
To isolate the rotational effects induced by the QED radiation in relating the true Breit frame transverse momentum and the one computed with external kinematics, we set $\widehat{P}_{hT} \to P_{hT}$, but keep the other ($\xi$, $\zeta$)-dependent variables unmodified.
This effectively removes the rotational effect, and reveals its suppressed role for the power-law enhanced $F_{UU,T}^{h\, {\rm (mod)}}$ structure function compared with the unmodified function.
The striking dependence of the QED radiative effects on the specific behavior of $F_{UU,T}^h$ indicates the difficulty in establishing a universal QED correction that can be applied to extract the pure QED,  ``free'' SIDIS structure function from the data.
Since the corrections depends on the behavior of $F_{UU,T}^h$ itself, one is confronted with an inverse problem that can only be solved within a QCD analysis framework that incorporates QED effects simultaneously.

Turning now to the QED radiative effects on the leading-twist spin modulations in SIDIS, we note that for scattering of unpolarized leptons  ($U$) from nucleons with transverse ($T$) polarization $\bm{S}_T$ there are three contributions that enter in the sum $\sum_n \hat{w}_n F_n^h$ in eq.~(\ref{e.sidisRC}).
These $UT$ modulations depend on the relative angles $\hat{\phi}_h$ and $\hat{\phi}_S$ in the combinations given by~\cite{Bacchetta:2006tn}
\begin{align}
\sum_n 
\hat{w}_n F_n^h(\hxb,\widehat{Q}^2,\hat{z}_h,\widehat{P}_{hT}^2)\Big|_{UT}
&= |\bm{S}_T|
\Big[ 
  \sin(\hat{\phi}_h-\hat{\phi}_S)
  F_{UT{,T}}^{\, \sin({\phi}_h-{\phi}_S)}
+ \sin(\hat{\phi}_h+\hat{\phi}_S)
  F_{UT}^{\, \sin({\phi}_h+{\phi}_S)}
\notag\\
& \qquad\quad
+ \sin(3\hat{\phi}_h-\hat{\phi}_S)
  F_{UT}^{\, \sin(3{\phi}_h-{\phi}_S)}
\Big],
\label{e.F_UT}
\end{align}
where the first and second terms correspond to the Sivers and Collins asymmetries, respectively, and the third term contains the pretzelosity TMD function in the small-$P_{hT}$ region.

\begin{figure}[t]
\centering
    \includegraphics[width=0.87\textwidth]{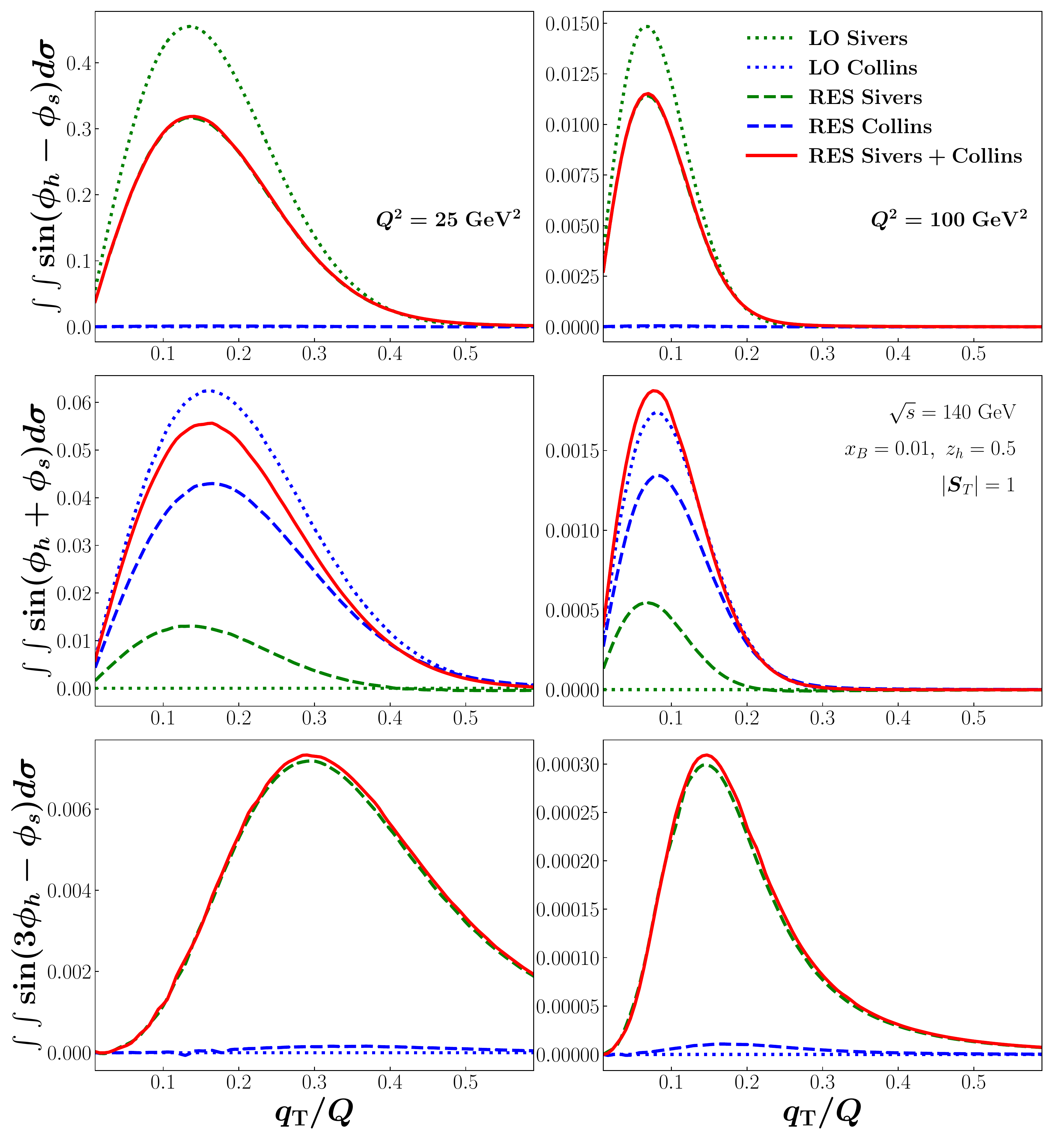}\vspace*{-0.5cm}
    \caption{QED radiation effects for
    $\sin(\phi_h-\phi_S)$ {\bf (top)},
    $\sin(\phi_h+\phi_S)$ {\bf (middle)} and
    $\sin(3\phi_h-\phi_S)$ {\bf (bottom)} SIDIS $UT$ spin modulations versus $q_T/Q$ at $\sqrt{s}=140$~GeV, $\xb=0.01$, $z_h=0.5$, and $Q^2=25$~GeV$^2$ {\bf (left)} and 100~GeV$^2$ {\bf (right)}, with $|\bm{S}_T|=1$. The cross sections with no QED effects (``LO'', dotted lines) are compared with the QED resummed cross sections (``RES'', dashed lines) for the Sivers (green lines) and Collins (blue lines) asymmetries.
    The total spin modulations (solid red lines) include the full QED contribution along with leakage effects.}
\label{f.sidisUT}
\end{figure}

Typically, the measured differential cross sections are integrated over the physical angles $\phi_h$ and $\phi_S$.
In the absence of QED radiative effects, the Sivers asymmetry, for instance, would be isolated via the external $\sin(\phi_h-\phi_S)$ projecting phase,
\begin{align}
\frac{\diff^6\sigma_{\l P(S_T) \to \lp P_h X}}
     {\diff\xb \diff y\, \diff \psi\, \diff z_h\, \diff P_{hT}^2}
    \Bigg|_{UT,T}^{\sin(\phi_h-\phi_S)}\,
= \int\diff\phi_h\, \diff\phi_S\, \sin(\phi_h-\phi_S)\,
\frac{\diff^6\sigma_{\l P(S_T) \to \lp P_h X}}
       {\diff\xb \diff y\, \diff \psi\, \diff z_h\, \diff \phi_h \diff P_{hT}^2},
\label{e.sivers_xsec}
\end{align}
which at the structure function level is equivalent to the identity
\begin{align}
F_{UT{,T}}^{\, \sin(\phi_h-\phi_S)}\ 
&\stackrel{\rm no\; \mbox{\tiny QED}}{=}\, 
\int\diff\phi_h\, \diff\phi_S\, \sin(\phi_h-\phi_S)
\bigg[
  \sin(\phi_h-\phi_S) F_{UT{,T}}^{\, \sin(\phi_h-\phi_S)}
\notag\\
& \qquad\quad
+ \sin(\phi_h+\phi_S) F_{UT}^{\, \sin(\phi_h+\phi_S)}
+ \sin(3\phi_h-\phi_S) F_{UT}^{\, \sin(3\phi_h-\phi_S)}
\bigg].
\label{e.sivers_stf}
\end{align}
In contrast, with QED radiation \eref{sivers_stf} no longer holds, as the projecting phases are not orthogonal with the ``internal'' phases,
\begin{align}
\int \diff\phi_h\, \diff\phi_S\,
\sin(\phi_h-\phi_S) \sin(\hat{\phi}_h+\hat{\phi}_S)\, \neq\, 0
\end{align}
for $\xi$, $\zeta \neq 1$.
The external projecting phases will then not uniquely isolate the desired structure function, but instead receive ``leakage'' from other modulations.

In \fref{sidisUT} we illustrate this phenomenon by integrating the $UT$ cross section (with $|\bm{S}_T|=1$) over the three different modulations in eq.~(\ref{e.F_UT}), calculated at some typical EIC kinematics versus $q_T/Q$. 
The structure functions $F_{UT,T}^{\, \sin(\phi_h-\phi_S)}$ and $F_{UT}^{\sin(\phi_h+\phi_S)}$ are taken from the JAM3D20 analysis~\cite{Cammarota:2020qcw} and $F_{UT}^{\sin(3\phi_h-\phi_S)}$ is set to zero.
As expected, the lowest order QED calculations isolate only the structure function associated with the relevant phase.
In the presence of radiation, however, the modulations are no longer orthogonal, and identification of the desired signal requires more care.
The largest effect is seen for the $\sin(\phi_h-\phi_S)$ modulation, where the cross section decreases uniformly across all $q_T/Q$, with no visible leakage from other modulations. 
For the $\sin(\phi_h+\phi_S)$ modulation, a similar depletion is found, but is partly compensated by leakage from the Sivers contribution.
Finally, the $\sin(3\phi_h-\phi_S)$ modulation, which in the ``true'' Breit frame (at LO) is set to zero, acquires a sizable contribution due to leakage from the Sivers $\sin(\phi_h-\phi_S)$ effect, with a small effect from the Collins $\sin(\phi_h+\phi_S)$ modulation.

As for the unpolarized SIDIS cross section, the QED radiative effects for the spin modulations generally depend on the shape of the transverse momentum distribution of the structure functions.
The presence of the inverse problem makes it impossible in practice to establish a universal set of QED corrections for SIDIS.
For the spin-dependent cross sections the problem is further aggravated since the radiative effects are not even universal within a specific type of modulation due to leakage effects.
The direct and simultaneous inclusion of QED radiation, along with QCD frameworks such as collinear or TMD factorization, is therefore indispensable for a meaningful QCD global analysis involving SIDIS data.

%%%%%%%%%%%%%%%%%%%%%%%%%%%%%%%%%%%%%%%%%%%%%%%%%%%%%%%%%%
\section{Conclusion and Outlook}
\label{s.conclusion}

In this paper we have proposed a QCD-like factorization approach to take into account the collision-induced QED radiative contributions to the experimentally measured cross sections of both inclusive and semi-inclusive lepton-nucleon DIS.
In this new hybrid factorized approach, based on the perturbative sensitivity in the limit when the lepton mass $m_e/Q \to 0$, all-order collision-induced QED contributions to the lepton-nucleon cross sections are organized into three groups: 
    infrared sensitive, 
    infrared safe, and 
    power suppressed.
In the limit when $m_e/Q \to 0$, the infrared sensitive contributions diverge logarithmically in powers of $\ln(Q^2/m_e^2)$, while the infrared safe contributions are independent of $m_e$ and can be calculated order-by-order in powers of~$\alpha$.
The power suppressed contributions (in powers of $m_e/Q$) are typically very small and can be safely neglected.

Taking advantage of the fact that the logarithmically enhanced and infrared sensitive contributions from the induced radiation are process independent, we collect them into universal LDFs and LFFs, and resum the logarithms to all orders in $\alpha$ by solving corresponding renormalization group equations.
Since QED contributions are in principle perturbatively calculable in the energy regime of relevant experiments, our factorization approach to the collision-induced radiation provides a consistent and perturbatively stable method to include all order QED contributions to both inclusive and semi-inclusive DIS cross sections, up to the power corrections in $m_e/Q$.
This provides excellent predictive power for lepton-nucleon SIDIS cross sections, given the universality of LDFs and LFFs, our ability to calculate the infrared-safe contributions perturbatively to all orders in $\alpha$, and the fact that $m_e/Q$ is a small number. 
Although we can calculate LDFs and LFFs in QED, they can be further improved by performing global analyses of lepton-nucleon scattering data to include the nonperturbative QCD contributions to these lepton distributions.

We have demonstrated that the traditional approach to handling the contributions from collision-induced QED radiation by imposing a ``radiative correction" factor to the ``Born" cross section with no QED radiation will not work for semi-inclusive lepton-nucleon processes when a final-state hadron or jet of momentum $P_h$ is measured in addition to the scattered lepton of momentum $\lp$.
Furthermore, without being able to account for all radiation, the photon-nucleon frame, where the TMD factorization was proven for SIDIS, is not well-defined.
Consequently, there is no unique connection between the produced hadron's transverse momentum $P_{hT}$ defined in the TMD factorization formalism and the measured hadron's $P_{hT}$, either in the laboratory frame, where the lepton and nucleon collide head-on, or in the ``Breit"-frame defined experimentally without taking into account the collision-induced radiation.
In addition, without knowing the ``true'' photon-nucleon frame, we cannot uniquely determine the hadronic plane, and will lose all advantages of extracting different TMDs from the different modulations of angle distribution between the leptonic plane and the hadronic plane.
Numerically, we found significant ``leakage" between different angular modulations, which could impact the precision with which TMDs can be extracted in practice.

Our factorization approach to the inclusive and semi-inclusive lepton-nucleon DIS naturally goes beyond the ``one-photon approximation".
We define the inclusive DIS as an inclusive production of a scattered lepton of momentum $\lp$ with $\ell'_T \gg \Lqcd$, and the semi-inclusive DIS as an inclusive production of a scattered lepton of momentum $\lp$ plus an observed hadron of momentum $P_h$ with both $\ell'_T$ and $P_{hT} \gg \Lqcd$ in the lepton-nucleon frame. 
We demonstrated quantitatively that the collision-induced QED and QCD radiative contributions to SIDIS from the observed leptons can be consistently treated in terms of collinear factorization, which allows a uniform treatment of the infrared-sensitive part of induced QED and QCD radiative contributions for both DIS and SIDIS, by resumming them into universal collinear LDFs and LFFs.
This factorization framework therefore provides excellent predictive power for QED and QCD radiative contributions.

With the collinear factorization approach to the induced QED and QCD radiation for the leptons, and the ``one-photon approximation'', we can define a ``virtual photon-nucleon'' frame for a given combination of lepton momentum fractions ($\xi, \zeta$), and take advantage of all factorization formalisms, including collinear and TMD factorization, to evaluate the semi-inclusive virtual-photon cross sections with the observed hadronic variables defined in the ``virtual photon-nucleon'' frame.
A ($\xi,\zeta$)-dependent Lorentz transformation can then be applied to transfer these variables from the ``virtual photon-nucleon'' frame to a frame (either laboratory or experimentally defined Breit frame), where the hadronic variables are measured.  
Finally, the integration over ($\xi,\zeta$), weighted by LDFs and LFFs, sums up the total impact of the induced QED radiation on the SIDIS cross sections.

We stress that even though $\alpha$ is very small, the logarithmic enhanced QED radiation could significantly alter the momentum transfer to the colliding nucleon, including the invariant mass (which defines the hard scale), as well as the direction that impacts on the angular distributions between the leptonic and hadronic planes, and the precision of extracting the TMDs from lepton-nucleon  scattering.
Our new and renormalization improved factorization approach allows the systematic resummation of the logarithmically enhanced collision-induced radiative effects from observed leptons into factorized LDFs and LFFs that are universal for all final states, applicable for DIS, SIDIS, as well as for $e^+ e^-$ annihilation and Drell-Yan lepton-pair production processes, leaving the fixed-order QED corrections completely infrared-safe and stable in the limit as $m_e \to 0$.
Our hybrid factorization approach goes beyond the ``one photon-approximation'', and provides a new paradigm for a uniform treatment of QED radiation in the extraction of PDFs, TMDs and other partonic correlation functions.
This will have important implications for the future analyses of hard scattering process at the EIC, in the quest to map the nucleon's three-dimensional structure in momentum space from lepton-nucleon collision data.

%%%%%%%%%%%%%%%%%%%%%%%%%%%%%%%%%%%%%%%%%%%%%%%%%%%%%%%%%%
%\begin{acknowledgements}
\acknowledgments
We thank B.~Badelek, P.~Bosted, H.~Gao, and D.~Gaskell for valuable comments.
We thank the participants of the {\it Theory-Experiment Dialogue} at Jefferson Lab, including A.~Accardi, H.~Avakian, \mbox{J.-P.} Chen, R.~Ent, C.~E.~Keppel, A.~Prokudin, T.~C.~Rogers and P.~Rossi, for helpful discussions.
This work is supported by the U.S. Department of Energy contract DE-AC05-06OR23177, under which Jefferson Science Associates, LLC, manages and operates Jefferson Lab, and within the framework of the TMD Topical Collaboration.
The work of TL is supported in part by National Natural Science Foundation of China under Contract No. 12175117.
The work of NS was supported by the DOE, Office of Science, Office of Nuclear Physics in the Early Career Program.

\newpage
\appendix
%%%%%%%%%%%%%%%%%%%%%%%%%%%%%%%%%%%%%%%%%%%%%%%%%%%%%%%%%%
\section{Perturbative coefficients of the leptonic tensor at NLO}
\label{s.app-nlo}

In this appendix, we derive the NLO perturbative coefficients $A^{(1)}$, $B^{(1)}$, $C_f^{(1)}$, and $C_D^{(1)}$ of the W-term $\widetilde{W}_{TT}$
and the first nontrivial Y-term $Y_{TT}$ in eq.~(\ref{eq:W+Y}) of section~\ref{sss.ltmd-fac}.
Among the four helicity basis lepton structure functions $L_{00}$, $L_{TT}$, $L_{\Delta}$, and $L_{\Delta\Delta}$ in eqs.~(\ref{e.Lij}), only $L_{TT}$ has a nonzero contribution at LO.
With $L_{TT}^{(0)}$ given in eq.~(\ref{e.lmn-tt-LO}) and the LO LDF and LFF given by
\begin{subequations}
\begin{align}
    f^{(0)}_{i/e}(\xi)   &= \delta_{ie}\, \delta(1-\xi),     \\
    D^{(0)}_{e/j}(\zeta) &= \delta_{ej}\, \delta(1-\zeta),
\end{align}
\end{subequations}
one can obtain the hard part,
\begin{align}
    \widehat{L}^{(0)}_{TT} 
    &= 2\delta(1-\lambda)\, \delta\Big(\frac{1}{\eta}-1\Big)\, \delta^{(2)}(\hat{\bm q}_T),
\end{align}
and the $C$ functions,
\begin{align}
    C_f^{(0)}(\lambda)\, C_D^{(0)}(\eta)
    = \delta(1-\lambda)\, \delta(1-\eta).
\end{align}
As a natural choice, we set
\begin{subequations}
\begin{align}
    C_f^{(0)}(\lambda) &= \delta(1-\lambda),\\
    C_D^{(0)}(\eta) &= \delta(1-\eta). 
\end{align}
\end{subequations}

\begin{figure}[t]
    \centering
    \includegraphics[width=0.6\textwidth]{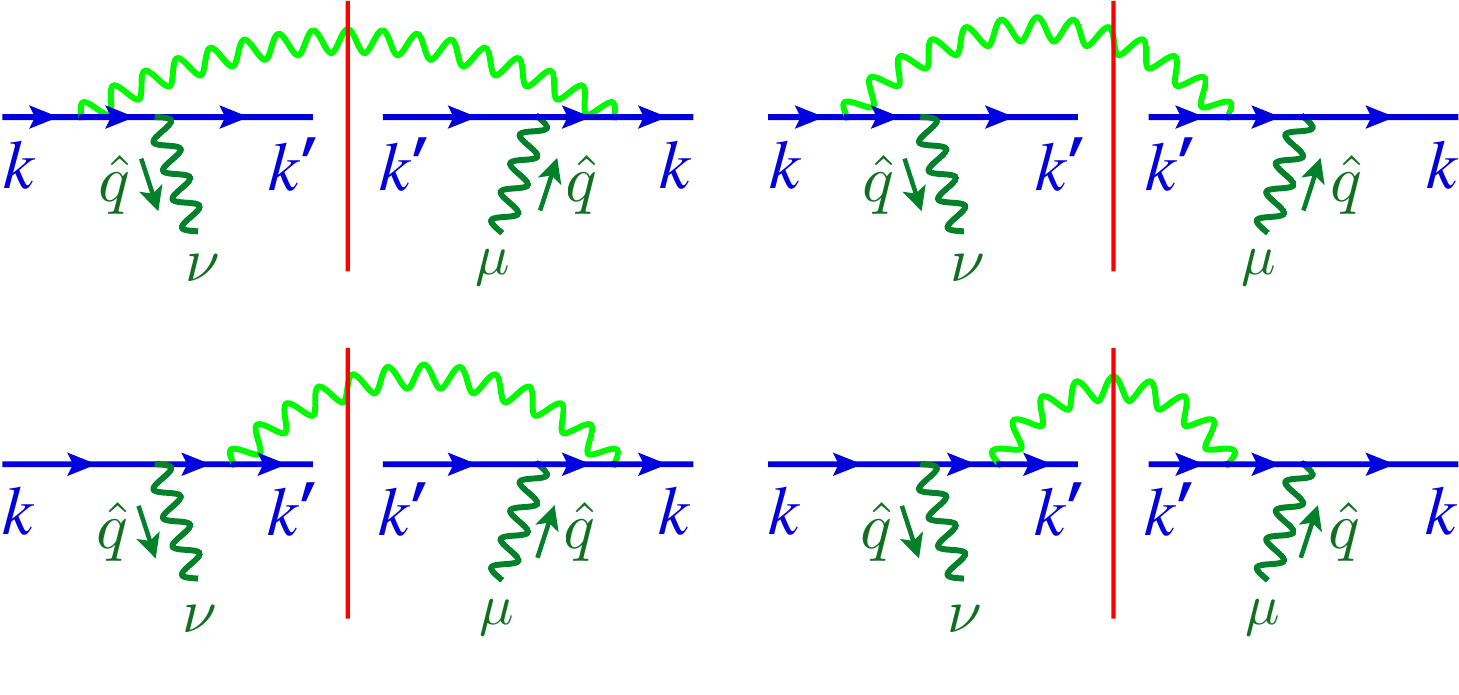}\vspace*{-0.5cm}
    \caption{Real photon emission diagrams for the calculation of the leptonic tensor at NLO.}
    \label{fig:nlo-r}
\end{figure}

\begin{figure}
    \centering
    \includegraphics[width=0.3\textwidth]{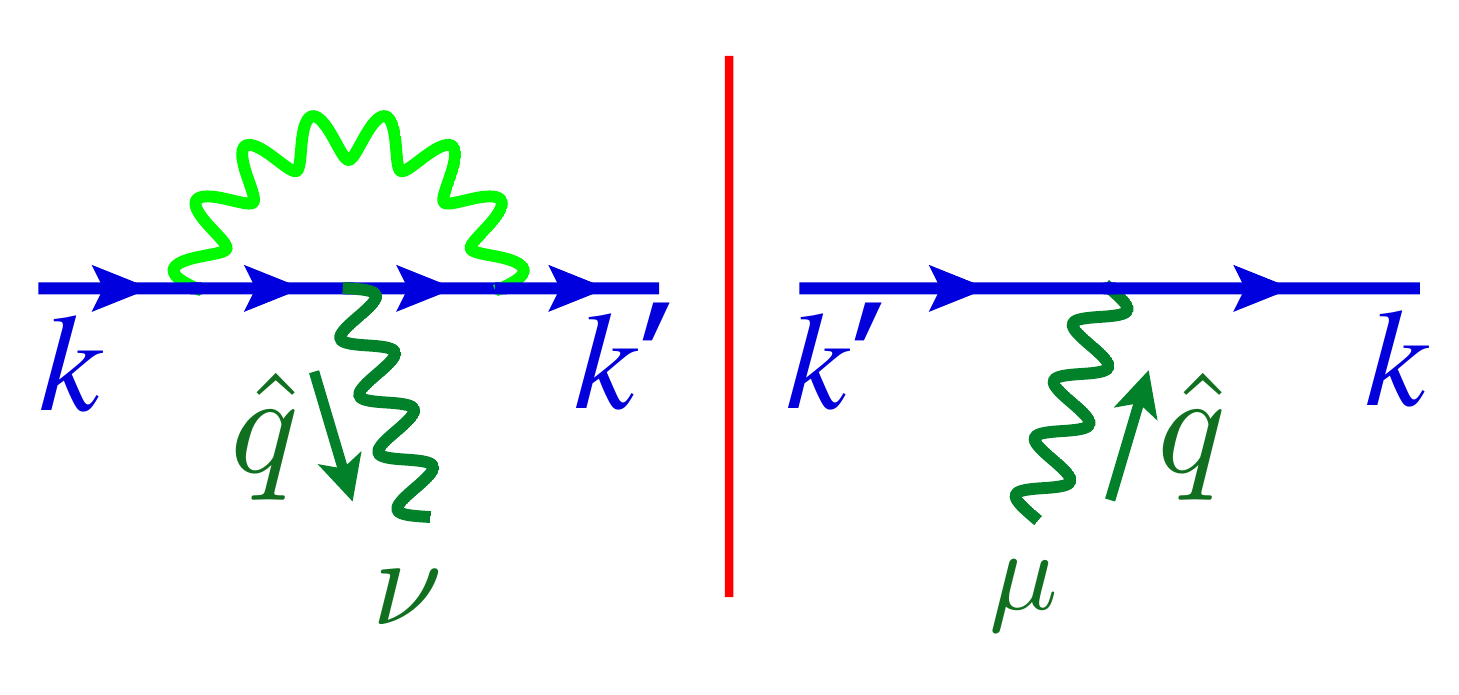}\vspace*{-0.5cm}
    \caption{Virtual correction diagram for the calculation of the leptonic tensor at NLO. The corresponding Hermitian conjugate diagram (omitted here) is also included in the calculation.}
    \label{fig:nlo-v}
\end{figure}

The NLO coefficients can be extracted from the NLO lepton structure function $L_{TT}^{(1)}$, which can be derived by calculating the real and virtual diagrams in figures~\ref{fig:nlo-r} and~\ref{fig:nlo-v}, respectively.
To extract the helicity basis lepton structure function $L_{TT}^{(1)}$, we contract these diagrams with $(X^\mu X^\nu+Y^\mu Y^\nu)/2$, and perform our calculation in $D=4-2\epsilon$ dimension to regulate all perturbative divergences.
Computing the amplitude squared of the real diagrams (figure~\ref{fig:nlo-r}), we obtain the real contribution to $L_{TT}^{(1)}$,
\begin{align}
\widehat{R}_{TT}^{(1)}
&= \frac{\left(4 \pi^2 \mu^2\right)^\epsilon}{2 \pi Q^2}
\left[ -\frac{2(1-\epsilon)^2 \left( \hat{u}^2+\hat{v}^2\right) + 4(1-\epsilon) \hat{t} (\hat{t}+\hat{u}+\hat{v})}{\hat{u} \hat{v}}\right.
\nonumber\\
&\quad
\left.+ \frac{1+\lambda^{2} \eta^{2}-\epsilon(1-\lambda \eta)^{2}}{\lambda \eta}-4 \epsilon^{2}\right]
\delta\Big( \frac{\xi_B}{\lambda \zeta_B}(1-\lambda)(1-\eta) - \frac{\hat{\bm q}_T^2}{Q^2} \Big),
\label{e.RTT(1)}
\end{align}
where the phase space factor
$ (2\pi)^{2\epsilon}/\big[ (2\pi)^3 Q^2 \big]\,
  \delta\Big( \big[\xi_B/\lambda \zeta_B\big] (1-\lambda)(1-\eta)-\hat{\bm q}_T^2/Q^2 \Big)
$
has been included, and the corresponding virtual contribution (figure~\ref{fig:nlo-v}),
\begin{align}
\widehat{V}_{TT}^{(1)}
&= - (1-\epsilon)\,
2 \delta(1-\lambda)\, \delta(1-\eta)\, \delta^{(2-2 \epsilon)} 
\left(\hat{\bm q}_T\right) 
\left[
\frac{1}{\epsilon^2}
+ \frac{2}{\epsilon} 
+ \frac{1}{\epsilon} \Big(\ln \frac{4 \pi \mu^2}{-\hat{t}}-\gamma_{E}\Big)
\right.
\notag\\
&\quad
\left.
+ \frac12\Big(\ln \frac{4 \pi \mu^2}{-\hat{t}}-\gamma_E\Big)^2
+ \frac32\Big(\ln \frac{4 \pi \mu^2}{-\hat{t}}-\gamma_E\Big)
- \frac{\pi^2}{12}
+ 4
\right].
\end{align}%
Note that in defining $\widehat{R}_{TT}^{(1)}$ and $\widehat{V}_{TT}^{(1)}$ we follow the same convention as for the leptonic tensor in eq.~(\ref{eq:Lexpand}), and the variables, $\hat{t}, \hat{u}$ and $\hat{v}$ are the same as those defined below eq.~(\ref{e.YTThat(1)}). 
In terms of the ``W+Y'' formalism in eq.~(\ref{eq:W+Y}), one can write the NLO lepton structure function as the sum
\begin{align}
    \widehat{L}_{TT}^{(1)} = \widehat{W}_{TT}^{(1)} + \widehat{Y}_{TT}^{(1)},
    \label{e.NLO-w+y}
\end{align}
where the Y-term is a regular part of $\widehat{L}_{TT}^{(1)}$ as $\hat{\bm q}_T\to 0$. 
Since the virtual contribution is proportional to
    $\delta^{(2-2 \epsilon)} \left(\hat{\bm q}_{T}\right)$,
and is singular as $\hat{\bm q}_T\to 0$, we only need the real diagrams to obtain the Y-term as 
    $Y_{TT}^{(1)} = \widehat{R}_{TT}^{(1)}
                  - \widehat{R}_{TT}^{(1)}\big|_{\hat{\bm q}_{T}\to 0}$,
where the subtraction term is known as the asymptotic term. 
Taking $\epsilon\to 0$ for the real-term in~\eqref{e.RTT(1)}, we obtain the Y-term as
\begin{align}
    \widehat{Y}_{TT}^{(1)} 
    &=
    \frac{1}{2\pi\mu_Q^2}
    \bigg[-\frac{2(\hat{u}^2+\hat{v}^2)+4\hat{t}(\hat{t}+\hat{u}+\hat{v})}{\hat{u}\hat{v}}
    +\frac{1+\lambda^2\eta^2}{\lambda\eta}
    \bigg]
    \delta\Big(\frac{1}{\lambda}(1-\lambda)(1-\eta)-\frac{\hat{\bm q}_T^2}{\mu_Q^2}\Big)
    \nonumber\\
    &\quad
    -\frac{1}{\pi \hat{\bm q}_T^2}
    \bigg[
    \frac{1+\lambda^2}{(1-\lambda)_+}\delta(1-\eta)
    +\frac{1}{\eta}\frac{1+\eta^2}{(1-\eta)_+}\delta(1-\lambda)
    -2\delta(1-\lambda)\delta(1-\eta)
    \ln\frac{\hat{\bm q}_T^2}{\mu_Q^2}
    \bigg],
\end{align}
where the asymptotic second term has been derived using the identity in eq.~(\ref{e.+int}) of Appendix~\ref{s.app-int}.

To extract the NLO coefficients $A^{(1)}$, $B^{(1)}$, $C_f^{(1)}$, and $C_D^{(1)}$, we expand the resummed expression for the W-term in~\eqref{eq:resumform} to ${\cal O}(\alpha)$ and remove the term proportional to $f^{(1)}$ and $D^{(1)}$ to isolate the perturbative part of the W-term in eq.~(\ref{e.NLO-w+y}), 
\begin{align}
    \widehat{\widetilde{W}}_{TT}^{(1)}
    &= C_f^{(1)}C_D^{(0)}S^{(0)}
    + C_f^{(0)}C_D^{(1)}S^{(0)}
    + C_f^{(0)}C_D^{(0)}S^{(1)},
    \label{e.resumexpand}
\end{align}
where $S^{(0)}$ and $S^{(1)}$ are the first two terms in the expansion of
\begin{align}
S&= \exp
    \bigg\{ -\int_{\mu_b^2}^{\mu_Q^2} \frac{\diff\mu'^2}{\mu'^2}
    \Big[ A\big(\alpha(\mu')\big)
          \ln\frac{\mu_Q^2}{\mu'^2} + B\big(\alpha(\mu')\big)
    \Big]
    \bigg\}
    \notag\\
 &=
    1 - \frac{\alpha}{\pi}
        \left[
        \frac12 A^{(1)} \ln^2\frac{\mu_Q^2}{\mu_b^2}
        + B^{(1)} \ln \frac{\mu_Q^2}{\mu_b^2} 
        \right]
    + {\cal O}(\alpha^2).
\end{align}
To derive $\widehat{\widetilde{W}}_{TT}^{(1)}$, we perform a Fourier transform of $\widehat{L}_{TT}^{(1)}$ to $b_T$ space using the relevant integral formulas in Appendix~\ref{s.app-int}.
The $1/\epsilon^2$ terms from the soft radiation exactly cancel between the real and virtual diagrams of figures~\ref{fig:nlo-r} and \ref{fig:nlo-v}.
The $1/\epsilon$ terms from collinear radiation can be absorbed into the LDF and LFF. 
The NLO perturbative part of the W-term can then be written
\begin{align}
    \widehat{\widetilde{W}}_{TT}^{(1)}
    &=
    2 \delta(1-\lambda) \delta(1-\eta)
    \bigg[ - \frac12 \ln^2\frac{\mu_Q^2}{\mu_b^2} 
           + \frac32 \ln\frac{\mu_Q^2}{\mu_b^2}
    \bigg]
    \nonumber\\
    &\quad
    -2\ln\frac{\mu_{\overline{\rm MS}}}{\mu_b}
    \bigg[\Big(\frac{1+\lambda^2}{1-\lambda}\Big)_+ \delta(1-\eta)
    +\frac{1}{\eta}\Big(\frac{1+\eta^2}{1-\eta}\Big)_+ \delta(1-\lambda)\bigg]
    \nonumber\\
    &\quad
    +2\bigg[ \frac12 (1-\lambda)\delta(1-\eta) +\frac{1}{2\eta}(1-\eta)\delta(1-\lambda)
    -4\delta(1-\lambda)\delta(1-\eta)
     \bigg].
    \label{e.WTT(1)}
\end{align}
By comparing eqs.~(\ref{e.resumexpand}) and (\ref{e.WTT(1)}), we determine the NLO perturbative coefficients,
\begin{subequations}
\begin{align}
    C_f^{(1)}(\lambda) 
    &= \frac12 (1-\lambda) 
    - \bigg(\frac{1+\lambda^2}{1-\lambda}\bigg)_+
      \ln\frac{\mu_{\overline{\rm MS}}}{\mu_b}
    - 2\delta(1-\lambda),
    \\
    C_D^{(1)}(\eta)
    &= \frac{1}{2\eta}(1-\eta)
    - \frac{1}{\eta} \bigg(\frac{1+\eta^2}{1-\eta}\bigg)_+
      \ln\frac{\mu_{\overline{\rm MS}}}{\mu_b}
    - 2\delta(1-\eta),
    \\
    A^{(1)} &= 1,
    \\
    B^{(1)} &= -\frac32.
\end{align}
\end{subequations}

%%%%%%%%%%%%%%%%%%%%%%%%%%%%%%%%%%%
\section{Some useful formulas}
\label{s.app-int}

In this appendix we collect some integral formulas which are relevant to the calculation of the NLO leptonic tensor in Appendix~\ref{s.app-nlo}.
First, we consider the integral
\begin{align}
    I&\equiv
    \int_{\xi_B}^1 \frac{\diff \xi}{\xi}
    \int_{\zeta_B}^1 \frac{\diff \zeta}{\zeta^2}\,
    f(\xi)\, D(\zeta)\, F(\lambda, \eta)\,
    \delta\bigg(
    \frac{\xi_B}{\lambda \zeta_B}(1-\lambda)(1-\eta)-\frac{\hat{\bm q}_T^2}{Q^2}
    \bigg)
    \nonumber\\
    &=\int_{\xi_B}^1\frac{\diff\lambda}{\lambda}
    \int_{\zeta_B}^1 \frac{\diff\eta}{\xi_B}\,
    f\Big(\frac{\xi_B}{\lambda}\Big)\,
    D\Big(\frac{\zeta_B}{\eta}\Big)\,
    F(\lambda,\eta)\,
    \delta\bigg( \Big(\frac{1}{\lambda}-1\Big) (1-\eta)-\chi^2 \bigg),
\end{align}
where 
    $\chi^2 \equiv (\zeta_B \hat{\bm q}_T^2)/(\xi_B Q^2)$
and $F(\lambda,\eta)$ is a smooth function of $\lambda$ and $\eta$. 
Taking the limit $\hat{\bm q}_T\to 0$, which corresponds to $\chi\to 0_+$, we have
\begin{align}
    I&= 
    \int_{\xi_B}^{1-\chi}\frac{\diff\lambda}{\lambda}
    \int_{2-1/\lambda}^1 \diff\eta\,
    \frac{1}{\xi_B}
    f\Big(\frac{\xi_B}{\lambda}\Big)\,
    D\Big(\frac{\zeta_B}{\eta}\Big)\,
    F(\lambda,\eta)
    \frac{\lambda}{1-\lambda}
    \delta(1-\eta)
    \nonumber\\
    &\quad +
    \int_{\zeta_B}^{1-\chi} \diff\eta
    \int_{1/(2-\eta)}^1\frac{\diff\lambda}{\lambda}
    \frac{1}{\xi_B}
    f\Big(\frac{\xi_B}{\lambda}\Big)\,
    D\Big(\frac{\zeta_B}{\eta}\Big)\,
    F(\lambda,\eta)
    \frac{\lambda}{1-\eta}
    \delta(1-\lambda)
    +{\cal O}(\chi^2)
    \nonumber\\
    &=
    \int_{\xi_B}^1 
    \frac{\diff\lambda}{(1-\lambda)_+}
    \frac{1}{\xi_B} 
    f\Big(\frac{\xi_B}{\lambda}\Big)
    D(\zeta_B)
    F(\lambda, 1)
    + \int_{\zeta_B}^1
    \frac{\diff\eta}{(1-\eta)_+}
    \frac{1}{\xi_B}\,
    f(\xi_B)\,
    D\Big(\frac{\zeta_B}{\eta}\Big)\,
    F(1, \eta)
    \nonumber\\
    &\quad
    -\frac{1}{\xi_B}\, f(\xi_B)\, D(\zeta_B)\, F(1,1) \ln\chi^2
    +{\cal O}(\chi^2)
    \nonumber\\
    &=\int_{\xi_B}^1 \frac{\diff\xi}{\xi} 
    \int_{\zeta_B}^1 \frac{\diff\zeta}{\zeta^2}\,
    f(\xi)\, D(\zeta)\, F(\lambda,\eta)\, 
    \frac{\lambda \zeta_B}{\xi_B}
    \bigg[
      \frac{1}{(1-\lambda)_+} 
      \delta(1-\eta)
    + \frac{1}{(1-\eta)_+}
      \delta(1-\lambda)
    \notag\\
    &\quad
    - \delta(1-\lambda)\, \delta(1-\eta) \,
      \ln \frac{\zeta_B\hat{\bm q}_T^2}{\xi_B Q^2}
    \bigg]
    + {\cal O}(\chi^2).
    \label{e.+int}
\end{align}
When computing the Fourier transform, one needs to calculate integrals of the type
\begin{align}
    I(a, k) \equiv
    \big(\mu^2\big)^{\epsilon+a-1} \int \diff^{2-2 \epsilon} \hat{\bm q}_T\,
    e^{-i \hat{\bm q}_T \cdot {\bm b}} \frac{1}{(\hat{\bm q}_T^2)^{a}} \ln ^k \frac{q_T^2}{\mu^2}.
\end{align}
For $k=0$, the integral can be evaluated directly,
\begin{align}
    I(a, 0) = \pi^{1-\epsilon} \left(\frac{\mu^2 b^2}{4}\right)^{\epsilon+\alpha-1}\, \frac{\Gamma(1-\epsilon-\alpha)}{\Gamma(\alpha)}.
\end{align}
For positive integer values of $k$, one can evaluate the integral by calculating the derivative of $I(a,0)$,
\begin{align}
    I(a, k) = \left( \frac{\diff}{\diff \delta} \right)^k\, I(a-\delta, 0) \bigg|_{\delta=0}.
\end{align}
Applying these results, we arrive at the following expressions for the $I(a,k)$ integrals utilized in Appendix~\ref{s.app-nlo},
\begin{subequations}
\begin{align}
I(1,0)
    &= \pi^{1-\epsilon}
    \Big[ -\frac{1}{\epsilon}
        - \gamma_E 
        - \ln \frac{\mu^2 b^2}{4} 
        + O(\epsilon)
    \Big], 
    \\
I(1,1)
    &= \pi^{1-\epsilon}
    \Big[ -\frac{1}{\epsilon^2}
        + \frac{1}{\epsilon} \gamma_E 
        + \frac12 \ln ^2 \frac{\mu^2 b^2}{4} 
        + 2\gamma_E \ln \frac{\mu^2 b^2}{4} 
        + \frac{\pi^2}{12} 
        + \frac32 \gamma_E^2 
        + O(\epsilon)
    \Big], 
    \\
I\big(\tfrac12, 0\big)
    &= \pi^{1-\epsilon} 
        \frac{2}{\mu b} + O(\epsilon),
    \\
I\big(\tfrac12, 1\big)
    &=-\pi^{1-\epsilon} \frac{2}{\mu b}
    \Big[2 \gamma_E 
        + 2 \ln 4 
        + \ln \frac{\mu^2 b^2}{4} 
        + O(\epsilon)
    \Big].
\end{align}
\end{subequations}

\newpage
%================================================================
\section{QED dependent SIDIS kinematic variables }
\label{s.sidis-kin}

In computing  SIDIS cross sections including QED effects, it is convenient to express the Breit frame kinematical variables, such as the angular phases and transverse momenta, in terms of lab frame variables that are directly accessible experimentally.
For notations and definitions, we follow the Trento convention as set out in ref.~\cite{Bacchetta:2006tn}.

We define symmetric and antisymmetric tensors for projections perpendicular to the direction of the virtual photon of momentum $q$
in the Breit frame scattering from an initial nucleon of momentum $P$,
\begin{align}
g_T^{\mu\nu}
    &= g^{\mu\nu}-\frac{P^\mu q^\nu + P^\nu q^\mu}{(1+\gamma^2)\, P\cdot q} 
    - \frac{\gamma^2}{1+\gamma^2}
    \bigg( \frac{P^\mu P^\nu}{M^2} - \frac{q^\mu q^\nu}{Q^2} \bigg),
\\
\epsilon_T^{\mu\nu}
    &= \frac{1}{\sqrt{1+\gamma^2}\, P\cdot q}\,
    \epsilon^{\mu\nu\rho\sigma} P_\rho\, q_\sigma.
\end{align}
Here by ``virtual photon'' we refer to the external momentum $q = \l-\lp$, which does not necessary coincide with the true photon momentum that enters in the hard scattering. 
With these transverse projectors we can write covariant expressions for the Breit frame transverse momenta,
\begin{align}
\l_T^\mu  &= g_T^{\mu\nu} \l_\nu,
\qquad
P_{hT}^\mu = g_T^{\mu\nu} P_{h\nu}.  
\label{e.transerse_mom}
\end{align}
Similarly, the angular dependence for the outgoing hadron ($\phi_h$) and the initial state spin vector ($\phi_S$) are given by  
\begin{align}
\cos\phi_h 
    &= -\frac{1}{\sqrt{\l^2_T P_{hT}^2}}\,
    \l_\mu P_{h\nu}\, g_T^{\mu\nu},
    \qquad
\sin\phi_h
    = -\frac{1}{\sqrt{\l^2_T P^2_{hT}}}\,
    \l_\mu P_{h\nu}\, \epsilon^{\mu\nu}_T,
    \\
\cos\phi_S
    &= -\frac{1}{\sqrt{\l^2_T S_T^2}}\,
    \l_\mu S_\nu\, g_T^{\mu\nu},
    \qquad\ \ \,
\sin\phi_S
    = -\frac{1}{\sqrt{\l^2_T S^2_T}}\,
    \l_\mu S_\nu\, \epsilon^{\mu\nu}_T,
\end{align}
where the spin vector $S$ of the initial nucleon is decomposed into longitudinal and transverse components,
\begin{subequations}
\label{e.spinvector}
\begin{align}
S^\mu 
    &= S_L\, 
    \frac{P^\mu-q^\mu M^2/P\cdot q}{M\sqrt{1+\gamma^2}}
    + S_T^\mu,
\\
S_L
    &= \frac{M}{\sqrt{1+\gamma^2}} \frac{S\cdot q}{P\cdot q},
\qquad
S_T^\mu
    = g_T^{\mu\nu}\, S_\nu.
\end{align}
\end{subequations}
Note that while the expressions in eqs.~(\ref{e.transerse_mom})--(\ref{e.spinvector}) are written in a covariant way, their interpretation in terms of longitudinal and transverse momentum directions is true only in the Breit frame.

The expressions above, along with the variables $\xb$, $z_h$ and $Q^2$, provide the full set of variables that characterize SIDIS for any  configuration of initial state particles. 
With these it is then possible to build the corresponding internal invariant variables that depend on the momentum fractions $\xi$ and $\zeta$.
To that end, it is important to note that only the external momenta $\l$ and $\lp$ are directly connected with the QED momentum fractions via $k=\xi \l$ and $k'=\lp/\zeta$, while the other external hadronic vectors, $P$, $P_h$ and $S$, do not depend on $\xi$ or $\zeta$.
It is only when the latter are decomposed in the Breit frame using the covariant projectors that the QED momentum fractions play a role.
For instance, in \eref{transerse_mom} the vector $P_{hT}^\mu$ becomes sensitive to the QED momentum fractions because the projector $g_T^{\mu\nu}$ depends on $\l$, $\lp$ or $q$, rather than because the original $P_h^\mu$ vector is sensitive to these.

The strategy then is to simply express all the relevant invariants in terms of the scalar products $\l\cdot V$ and $\lp\cdot V$, with $V=P$, $S$ or $P_h$, and include QED radiative effects via the substitutions
    $\l  \cdot V \to \xi\, \l \cdot V$ and
    $\lp \cdot V \to \lp \cdot V/\zeta$. 
For this purpose we form the scalar products from the external kinematics through the invariants $\xb$, $z_h$, $Q^2$, $y$ and $P_{hT}$, the angles $\phi_h$ and $\phi_S$, and the spin projections $S_L$ and $|\bm{S}_T|$.
We utilize both the sine and cosine of the phases to keep track of the signs of the modulations. 
Defining ${\cal Q}^2 \equiv Q^2 + 4M^2 \xb^2$, it is then straightforward to verify that      
\begin{subequations}
\begin{align}
\l_T^2 
    &= \frac{Q^2 \big(Q^2(1-y) - M^2 \xb^2 y^2\big)}{y^2 {\cal Q}^2},
\label{e.lperp}
\\
q\cdot P_h
    &= \frac{Q}{4 M^2 \xb^2}
    \Big( Q^3 z_h 
    - \sqrt{{\cal Q}^2 \big(Q^4 z_h^2 - 4M^2 \xb^2\, (m_h^2 + P_{hT}^2) \big)}
    \Big),
\\
\l\cdot P_h
    &= -\frac{1}{y {\cal Q}^2}
    \Big({\cal Q}^2\, P_{hT}\, 
    [\l_T]\, y\, \cos\phi_h 
        - \big(Q^2 + 2 M^2 \xb^2\, y\big)\, [q \cdot P_h]
        + Q^4 z_h \big(1 + \tfrac12 y\big)
    \Big),
\\
\lp\cdot P_h
    &= [\l\cdot P_h] - [q\cdot P_h].
\end{align}
\end{subequations}
For convenience, in the expressions for $\l\cdot P_h$ and $\lp\cdot P_h$ we have kept explicit the dependence on $\l_T$ and $q\cdot P_h$, highlighted by the brackets ``$[...]$.''
The QED effects on the various scalar products can be implemented in a covariant way through the substitutions
$\l \cdot P_h \to k \cdot P_h = \xi\, [\l \cdot P_h]$ and 
$\lp\cdot P_h \to k'\cdot P_h = [l'\cdot P_h]/\zeta$. 
Proceeding next to scalar products involving the spin vector $S$, we have
\begin{subequations}
\begin{align}
q\cdot S &=
    -\frac{Q\, {\cal Q}}{2M\xb} \sqrt{1 - \bm{S}_T^2},
\\
\l\cdot S
    &= -\frac{1}{y {\cal Q}^2}
    \Big({\cal Q}^2\, |\bm{S}_T|\, [\l_T]\, y\, \cos\phi_S 
    - (Q^2 + 2 M^2 \xb^2\, y)\, [q\cdot S]
    \Big),
\\
\lp\cdot S
    &= [\l\cdot S] - [q\cdot S].
\end{align}
\end{subequations}
As for the case of $P_h$, we can implement the scalar products involving the spin vector $S$ in a covariant way by making the replacements
$\l \cdot S \to k \cdot S =\xi\, [\l\cdot S]$ and
$\lp\cdot S \to k'\cdot S = [\lp\cdot S]/\zeta $.

For the sinusoidal phases involving contractions of the Levi-Civita tensor with $\l$ and $\lp$, we have the relations
\begin{subequations}
\begin{align}
&\epsilon_{\mu\nu\rho\sigma} P^\mu \l^\nu \lp^\rho P_h^\sigma
    = -\frac{P_{hT}\, Q\, {\cal Q}\, [\l_T]}{2x} \sin\phi_h,
\\
&\epsilon_{\mu\nu\rho\sigma} P^\mu \l^\nu \lp^\rho S^\sigma
    = -\frac{|\bm{S}_T|\, Q\, {\cal Q}\, [\l_T]}{2x} \sin\phi_S.
\end{align}
\end{subequations}
These also allow us to implement the $\xi$ and $\zeta$ dependence of the sinusoidal modulations via the substitutions
\begin{subequations}
\begin{align}
\epsilon_{\mu\nu\rho\sigma} P^\mu \l^\nu \lp^\rho P_h^\sigma
\to
\epsilon_{\mu\nu\rho\sigma} P^\mu k^\nu k'^\rho P_h^\sigma
&= \frac{\xi}{\zeta}\,
[\epsilon_{\mu\nu\rho\sigma} P^\mu \l^\nu \lp^\rho P_h^\sigma],
\\
\epsilon_{\mu\nu\rho\sigma} P^\mu \l^\nu \lp^\rho S^\sigma
\to
\epsilon_{\mu\nu\rho\sigma} P^\mu k^\nu k'^\rho S^\sigma
&= \frac{\xi}{\zeta}\,
[\epsilon_{\mu\nu\rho\sigma} P^\mu \l^\nu \lp^\rho S^\sigma].
\end{align}
\end{subequations}

We next consider the SIDIS invariants with $\xi$ and $\zeta$ dependence, which for the simplest variables are
\begin{subequations}
\begin{align}
\hxb = &\frac{\xb\, \xi\, y}{\xi \zeta + y - 1},
\quad
\hat{y} = \frac{\xi \zeta + y - 1}{\xi \zeta},
\quad
\hat{z}_h = \frac{y z_h\, \zeta}{\xi \zeta + y - 1},
\\
& \qquad\qquad
\widehat{Q}^2 = \frac{\xi}{\zeta}\, Q^2, 
\quad
\hat{\gamma}^2 = \frac{(2 M \hxb)^2}{\widehat{Q}^2}.
\end{align}
\end{subequations}
Defining again the shorthand notation
    $\widehat{{\cal Q}}^2 
    \equiv \widehat{Q}^2 + 4 M^2 \hxb^2 
         = \widehat{Q}^2 (1 + \hat{\gamma}^2)$,
from \eref{lperp} we can write the transverse momentum of the incoming lepton that enters the hard scattering in the true Breit frame as
\begin{align}
k_T^2
&= \frac{\widehat{Q}^2 
         \big(\widehat{Q}^2 (1-y)-M^2\, \hxb^2\, \hat{y}^2\big)}
        {\hat{y}^2 \widehat{{\cal Q}}^2}.
\end{align}
For the corresponding scalar products involving the internal $\hat{q}$ vector, we can write
\begin{align}
\hat{q}\cdot P_h 
= \llbrack k\cdot P_h\rrbrack - \llbrack k'\cdot P_h \rrbrack,
\qquad
\hat{q} \cdot S  = \llbrack k\cdot S\rrbrack
- \llbrack k'\cdot S\rrbrack.
\end{align}
Here we use the shorthand notation ``$\llbrack ...\rrbrack$'' for the ($\xi, \zeta$)-dependent scalar products to distinguish them from the ($\xi, \zeta)$-independent quantities above, with the understanding that the former are functions of ($\xi, \zeta)$ and of the latter scalar products, which we can represent schematically as
    ``$\llbrack...\rrbrack = \llbrack ...\rrbrack(\xi,\zeta,[...])$''.
The transverse momentum of the produced hadron can be computed using  \eref{transerse_mom} in terms of the ($\xi, \zeta$)-dependent variables as 
\begin{align}
\widehat{P}_{hT}^2 
= - \frac{1}{\widehat{Q}^2\, \widehat{{\cal Q}}^2}
  \Big(
    m_h^2\, \widehat{Q}^2\, \widehat{{\cal Q}}^2
    - \widehat{Q}^6 \hat{z}_h^2
    + 2 \big(\widehat{Q}^4 \hat{z}_h + 2 M^2 \hxb^2\big)\, \llbrack \hat{q}\cdot P_h \rrbrack
  \Big).
\end{align}
Similarly, the decomposition of the spin vector $S$ in the true Breit frame is given by 
\begin{subequations}
\begin{align}
\widehat{S}_L
&= \frac{\hat{\gamma}}{\widehat{\cal Q}}\, \llbrack \hat{q}\cdot S \rrbrack,
\\
|\widehat{\bm{S}}_T|^2
&= 1 - \frac{\hat{\gamma}^2}{\widehat{\cal Q}^2}\, \llbrack \hat{q}\cdot S \rrbrack^2
=\, 1 - \widehat{S}_L^2.
\end{align}
\end{subequations}
Finally, the angular phases can be expressed in terms of the scalar products as
\begin{subequations}
\begin{align}
\cos\hat{\phi}_h
&= \frac{\widehat{Q}^4\, \big( 2-\hat{y} \big) \hat{z}_h
            + 2 \big( \widehat{Q}^2 + 2 M^2 \hxb^2\, \hat{y} \big)\,
             \llbrack \hat{q} \cdot P_h\rrbrack
            - 2\widehat{\cal Q}^2\, \hat{y}\, \llbrack k\cdot P_h\rrbrack}
        {2{\widehat{\cal Q}^2\, \hat{y}\,  \llbrack \widehat{P}_{hT}\rrbrack \llbrack k_T\rrbrack}},
\\
& \notag\\
\sin\hat{\phi}_h
&= - \frac{2\hxb\, {\llbrack \epsilon_{\mu\nu\rho\sigma} 
            P^\mu k^\nu k'^\rho P_h^\sigma \rrbrack}}
          {\widehat{Q}\, \widehat{\cal Q}\,
            \llbrack \widehat{P}_{hT}\rrbrack \llbrack k_T\rrbrack},
\end{align}
\end{subequations}
and
\begin{subequations}
\begin{align}
\cos\hat{\phi}_S
&= \frac{\big( \widehat{Q}^2 + 2 M^2 \hxb^2\, \hat{y} \big)\,
            \llbrack\hat{q}\cdot S\rrbrack
            - \widehat{\cal Q}^2\, \hat{y}\, \llbrack k\cdot S\rrbrack}
        {\widehat{\cal Q}^2\, \hat{y}\, \llbrack|\widehat{\bm{S}}_T|\rrbrack \llbrack k_T\rrbrack},
\\
& \notag\\
\sin\hat{\phi}_S
&= - \frac{2\hxb\, \llbrack \epsilon_{\mu\nu\rho\sigma} 
            P^\mu k^\nu k'^\rho S^\sigma\rrbrack}
          {\widehat{Q}\, \widehat{\cal Q}\,
            \llbrack|\widehat{\bm{S}}_T|\rrbrack \llbrack k_T\rrbrack}.
\\
& \notag
\end{align}
\end{subequations}
With these expressions, we have now completed the translation of the SIDIS invariants expressed in terms of external degrees of freedom into their corresponding internal ones that depend on $\xi$ and $\zeta$.

%%%%%%%%%%%%%%%%%%%%%%%%%%%%%%%%%%%%%%%%%%%%%%%%%%%%%%%%%%

\end{document}